\documentclass[journal, 10pt]{IEEEtran}

\usepackage{cite}
\usepackage{amsmath,amssymb,amsfonts,amsthm, empheq}
\usepackage{algorithm}
\usepackage{algorithmic}
\usepackage{setspace}
\usepackage{graphicx}
\usepackage{textcomp}
\usepackage{xcolor}
\usepackage{mathtools}
\usepackage{graphicx}
\usepackage{booktabs}
\usepackage{makecell}
\usepackage[font={small}]{caption}
\usepackage{subcaption}
\usepackage{cite}
\usepackage{breqn}
\usepackage{mathrsfs}
\usepackage{accents}
\usepackage{acronym}
\usepackage{multirow}
\usepackage{soul}
\usepackage{bm}
\usepackage{url}
\usepackage{wrapfig}
\usepackage{todonotes}
\usepackage{verbatim}
\usepackage{pdfpages}
\usepackage{siunitx}
\usepackage{adjustbox}

\usepackage{pgfplots}
\pgfplotsset{compat=1.3}
\usepgfplotslibrary{statistics}
\usepackage{tikz}
\usepackage{pgfplotstable}
\usetikzlibrary{shapes, shapes.misc, arrows, positioning, calc, backgrounds, angles, quotes, patterns, pgfplots.groupplots}

\renewcommand{\thetable}{\arabic{table}}

\DeclareMathOperator*{\argmin}{arg\,min}

\graphicspath{{./figures/}}
\acrodef{prop}[\textit{MIMORPH}]{MIMO Radio Platform for Heterogeneous wireless systems}

\acrodef{abft}[A-BFT]{Association Beamforming Training}
\acrodef{ack}[ACK]{Acknowledge}
\acrodef{adc}[ADC]{Analog-to-Digital Converter}
\acrodef{aoa}[AoA]{Angle of Arrival}
\acrodef{aod}[AoD]{Angle of Departure}
\acrodef{ap}[AP]{Access Point}
\acrodef{amc}[AMC]{Advanced Mezzanine Card}
\acrodef{awv}[AWV]{Antenna Wave Vector}
\acrodef{axi}[AXI]{Advanced eXtensible Interface}
\acrodef{adc}[ADC]{Analog-to-Digital Converter}
\acrodef{ber}[BER]{Bit Error Rate}
\acrodef{bft}[BFT]{Beamforming Training}
\acrodef{bp}[BP]{Beam Pattern}
\acrodef{brp}[BRP]{Beam Refinement Phase}
\acrodef{bss}[BSS]{Blind Source Separation}
\acrodef{cacc}[CACC]{Cross-Antenna Cross-Correlation}
\acrodef{casr}[CASR]{Cross-Antenna Signal Ratio}
\acrodef{cs}[CS]{Compressed Sensing}
\acrodef{cdf}[CDF]{Cumulative Distribution Function}
\acrodef{cef}[CEF]{Channel Estimation Field}
\acrodef{cfo}[CFO]{Carrier Frequency Offset}
\acrodef{cir}[CIR]{Channel Impulse Response}
\acrodef{cfr}[CFR]{Channel Frequency Response}
\acrodef{crb}[CRB]{Cramér-Rao Bound}
\acrodef{crlb}[CRLB]{Cramér-Rao Lower Bound}
\acrodef{csi}[CSI]{Channel State Information}
\acrodef{csirs}[CSI-RS]{CSI-Reference Signal}
\acrodef{cs}[CS]{Compressed Sensing}
\acrodef{cv}[CV]{Constant Velocity}
\acrodef{cnn}[CNN]{Convolutional Neural Network}
\acrodef{cots}[COTS]{Commercial-Off-The-Shelf}
\acrodef{dac}[DAC]{Digital-to-Analog Converter}
\acrodef{dft}[DFT]{Discrete Fourier Transform}
\acrodef{dl}[DL]{Deep Learning}
\acrodef{dma}[DMA]{Direct Memory Access}
\acrodef{dmg}[DMG]{Directional Multi Gigabit}
\acrodef{dmrs}[DMRS]{Demodulation Reference Signal}
\acrodef{srs}[SRS]{Sounding Reference Signal}
\acrodef{dti}[DTI]{Data Transfer Interval}
\acrodef{edmg}[EDMG]{Enhanced Directional Multi Gigabit}
\acrodef{ekf}[EKF]{Extended Kalman Filter}
\acrodef{elu}[ELU]{Exponential-Linear Unit}
\acrodef{fft}[FFT]{Fast Fourier Transform}
\acrodef{fmcw}[FMCW]{Frequency-Modulated Continuous-Wave}
\acrodef{fim}[FIM]{Fisher Information Matrix}
\acrodef{fov}[FOV]{Field-of-View}
\acrodef{ft}[FT]{Fourier Transform}
\acrodef{fr2}[FR2]{Frequency Range 2}
\acrodef{fpga}[FPGA]{Field Programmable Gate Array}
\acrodef{gpio}[GPIO]{General Purpose Input/Output}
\acrodef{gsps}[GSPS]{Giga-Samples per Second}
\acrodef{har}[HAR]{Human Activity Recognition}
\acrodef{ht}[HT]{High Throughput}
\acrodef{ica}[ICA]{Independent Component Analysis}
\acrodef{idft}[IDFT]{Inverse Discrete Fourier Transform}
\acrodef{if}[IF]{Intermediate Frequency}
\acrodef{ifft}[IFFT]{Inverse Fast Fourier Transform}
\acrodef{ifs}[IFS]{Inter-Frame Spacing}
\acrodef{iht}[IHT]{Iterative Hard Thresholding}
\acrodef{ista}[ISTA]{Iterative Shrinkage-Thresholding Algorithm}
\acrodef{isac}[ISAC]{Integrated Sensing And Communication}
\acrodef{iot}[IoT]{Internet of Things}
\acrodef{imu}[IMU]{Inertial Measurement Unit}
\acrodef{jcs}[JCS]{Joint Communication and Sensing}
\acrodef{jm}[JM]{JUMP-MUSIC}
\acrodef{jpdaf}[JPDAF]{Joint Probabilistic Data Association Filter}
\acrodef{lo}[LO]{Local Oscillator}
\acrodef{los}[LoS]{Line-of-Sight}
\acrodef{lbm}[LBM]{Loop-Back Memory}
\acrodef{lora}[LoRa]{Long-Range wide area}
\acrodef{ls}[LS]{Least-Squares}
\acrodef{mae}[MAE]{Mean Absolute Error}
\acrodef{mcs}[MCS]{Modulation and Coding Scheme}
\acrodef{md}[$\mu$D]{micro-Doppler}
\acrodef{mimo}[MIMO]{Multiple Input Multiple Output}
\acrodef{mmwave}[mmWave]{Millimeter-Wave}
\acrodef{msps}[MSPS]{Mega-Samples per Second}
\acrodef{mu}[MU]{Multiple User}
\acrodef{music}[MUSIC]{MUltiple SIgnal Classification}
\acrodef{nac}[NAC]{Normalized Auto Correlation}
\acrodef{nco}[NCO]{Numerical Controlled Oscillator}
\acrodef{nlos}[NLoS]{Non-Line-of-Sight}
\acrodef{nn}[NN]{Neural Network}
\acrodef{nls}[NLS]{Nonlinear Least-Squares}
\acrodef{ofdm}[OFDM]{Orthogonal Frequency Division Multiplexing}
\acrodef{omp}[OMP]{Orthogonal Matching Pursuit}
\acrodef{per}[PER]{Packet Error Rate}
\acrodef{pdf}[p.d.f.]{Probability Density Function}
\acrodef{phy}[PHY]{Physical Layer}
\acrodef{pl}[PL]{Programmable Logic}
\acrodef{pov}[POV]{Point-of-View}
\acrodef{ps}[PS]{Processing System}
\acrodef{po}[PO]{Phase Offset}
\acrodef{qam}[QAM]{Quadrature Amplitude Modulation}
\acrodef{ransac}[RANSAC]{Random Sample Consensus}
\acrodef{rf}[RF]{Radio Frequency}
\acrodef{rfsoc}[RFSoC]{Radio Frequency System on a Chip}
\acrodef{rcs}[RCS]{Radar Cross-Section}
\acrodef{rmse}[RMSE]{Root Mean Square Error}
\acrodef{rss}[RSS]{Received Signal Strength}
\acrodef{rom}[ROM]{Read Only Memories}
\acrodef{rx}[RX]{receiver}
\acrodef{sc}[SC]{Single Carrier}
\acrodef{sdr}[SDR]{Software Defined Radio}
\acrodef{sinr}[SINR]{Signal-to-Interference-plus-Noise Ratio}
\acrodef{siso}[SISO]{Single Input Single Output}
\acrodef{sls}[SLS]{Sector Level Sweep}
\acrodef{snr}[SNR]{Signal-to-Noise Ratio}
\acrodef{soc}[SoC]{System on a Chip}
\acrodef{spb}[SPB]{Signal Processing Blocks}
\acrodef{srrc}[SRRC]{Square-Root-Raised-Cosine}
\acrodef{ssb}[SSB]{Synchronization Signal Block}
\acrodef{ssr}[SSR]{Super Sample Rate}
\acrodef{sta}[STA]{Station}
\acrodef{std}[STD]{Standard Deviation}
\acrodef{stf}[STF]{Short Training Field}
\acrodef{stft}[STFT]{Short Time Fourier Transform}
\acrodef{su}[SU]{Single User}
\acrodef{tf}[TF]{Time-Frequency}
\acrodef{to}[TO]{Timing Offset}
\acrodef{toa}[ToA]{Time of Arrival}
\acrodef{tx}[TX]{transmitter}
\acrodef{ue}[UE]{User Equipment}
\acrodef{ula}[ULA]{Uniform Linear Array}
\acrodef{usrp}[USRP]{Universal Software Radio Peripheral}
\acrodef{vht}[VHT]{Very High Throughput}
\acrodef{wlan}[WLAN]{Wireless Local Area Network}
\acrodef{wnalp}[WNALP]{Weighted Normalized Auto-correlation Linear Predictor}

\newcounter{mytempeqncnt}

\newcommand{\eq}[1]{Eq.~\eqref{#1}}
\newcommand{\eqs}[2]{Eqs.~(\ref{#1}-\ref{#2})}
\newcommand{\fig}[1]{Fig.~\ref{#1}}
\newcommand{\tab}[1]{Tab.~\ref{#1}}
\newcommand{\secref}[1]{Section~\ref{#1}}

\newcommand{\alg}[1]{Algorithm~\ref{#1}}

\newcommand{\rev}[1]{{\color{black}#1}}

\newcommand{\mytexttilde}{{\raise.17ex\hbox{$\scriptstyle\mathtt{\sim}$}}}

\newcommand{\ourname}{AsyMov}

\IEEEaftertitletext{\vspace{-2.\baselineskip}}

\begin{document}
\bstctlcite{IEEEexample:BSTcontrol}
\pagenumbering{gobble}

\title{AsyMov: Integrated Sensing and Communications with Asynchronous Moving Devices}

\author{Gianmaria Ventura$^{\dag}$,~\IEEEmembership{Graduate Student Member,~IEEE}, Michele Rossi$^{\dag *}$~\IEEEmembership{Senior Member,~IEEE}, \\Jacopo Pegoraro$^{\dag}$,~\IEEEmembership{Member,~IEEE}
\thanks{$^{\dag}$These authors are with the University of Padova, Dept. of Information Engineering. $^{*}$This author is with the University of Padova, Dept. of Mathematics.
Corresponding author email: \texttt{jacopo.pegoraro@unipd.it}. 
This work was partially supported by the European Union under the Italian National Recovery and Resilience Plan (NRRP) Mission 4, Component 2, Investment 1.3, CUP C93C22005250001, partnership on “Telecommunications of the Future” (PE00000001 - program “RESTART”).
This work was partially supported by the Smart Networks and Services Joint Undertaking (SNS JU) under the European Union's Horizon Europe research and innovation programme, project MultiX (Grant Agreement No 101192521).
}
}

\maketitle

\begin{abstract}
Estimating the Doppler frequency shift caused by moving targets is one of the key objectives of \ac{isac} systems, as it enables applications such as target classification, human activity recognition, and gait analysis. In practical scenarios, Doppler estimation is hindered by the movement of transmitter and receiver devices, and by the phase offsets caused by their clock asynchrony. Existing approaches have \textit{separately} addressed these two aspects, either assuming clock-synchronous moving devices or asynchronous static ones. In fact, jointly tackling device motion and clock asynchrony is extremely challenging, as the Doppler shift from device movement differs for each propagation path and the phase offsets are time-varying. 
In this work, we present \ourname{}, a method to estimate the bistatic Doppler frequency of a target and its velocity in \ac{isac} setups featuring \textit{mobile and asynchronous} devices.
It leverages the channel impulse response at the receiver, by originally exploiting the invariance of phase offsets across propagation paths and the bistatic geometry, where the target Doppler and the device velocity are jointly estimated by a newly proposed alternating minimization algorithm. Moreover, it can be seamlessly integrated with device velocity measurements obtained from onboard sensors (if available), for enhanced reliability.
Here, \ourname{} is thoroughly characterized by way of theory (Cram\'er-Rao bound), simulation, and experiments, implementing it on an IEEE~802.11ay \rev{testbed and testing it on multiple setups in the 60~GHz and 28~GHz bands, including moving human subjects.} Numerical and experimental results show superior performance against state-of-the-art methods and are on par with scenarios featuring \textit{static} \ac{isac} devices.
\end{abstract}


\IEEEpeerreviewmaketitle

\section{Introduction}\label{sec:intro}

\acf{isac} is expected to be a cornerstone of future wireless communication systems~\cite{liu2023seventy}. 
One of the key uses of \ac{isac} concerns the estimation of the Doppler frequency caused by a moving target, which also reveals its velocity. On the one hand, Doppler estimation enhances the sensing resolution of the system, as it allows distinguishing targets moving at different speeds~\cite{richards2010principles}. On the other hand, the Doppler effect can be leveraged by a wide range of fine-grained cellular and Wi-Fi sensing applications, such as person identification based on gait~\cite{vandersmissen2018indoor}, fall detection~\cite{chu2023deep}, drone monitoring~\cite{fang2021rotor}, and human movement classification for smart homes and remote healthcare~\cite{zhang2021enabling, singh2021multi}. 

Compared to indoor radar systems, Doppler estimation in \ac{isac} poses several challenges. First, \ac{isac} transmitter (TX) and receiver (RX) devices are spatially separated (bistatic), so they rely on \textit{asynchronous} \acp{lo} to generate the carrier signal, which is consequently affected by a time-varying relative drift~\cite{zhang2022integration}. 
This causes \ac{to}, \ac{cfo}, and a random \ac{po} across different transmissions. \ac{cfo} and \ac{po} prevent the coherent processing of channel measurements across time, which is required for Doppler estimation~\cite{wu2024sensing}. 

\begin{figure}[t!]
    \centering
    \includegraphics[width=\linewidth]{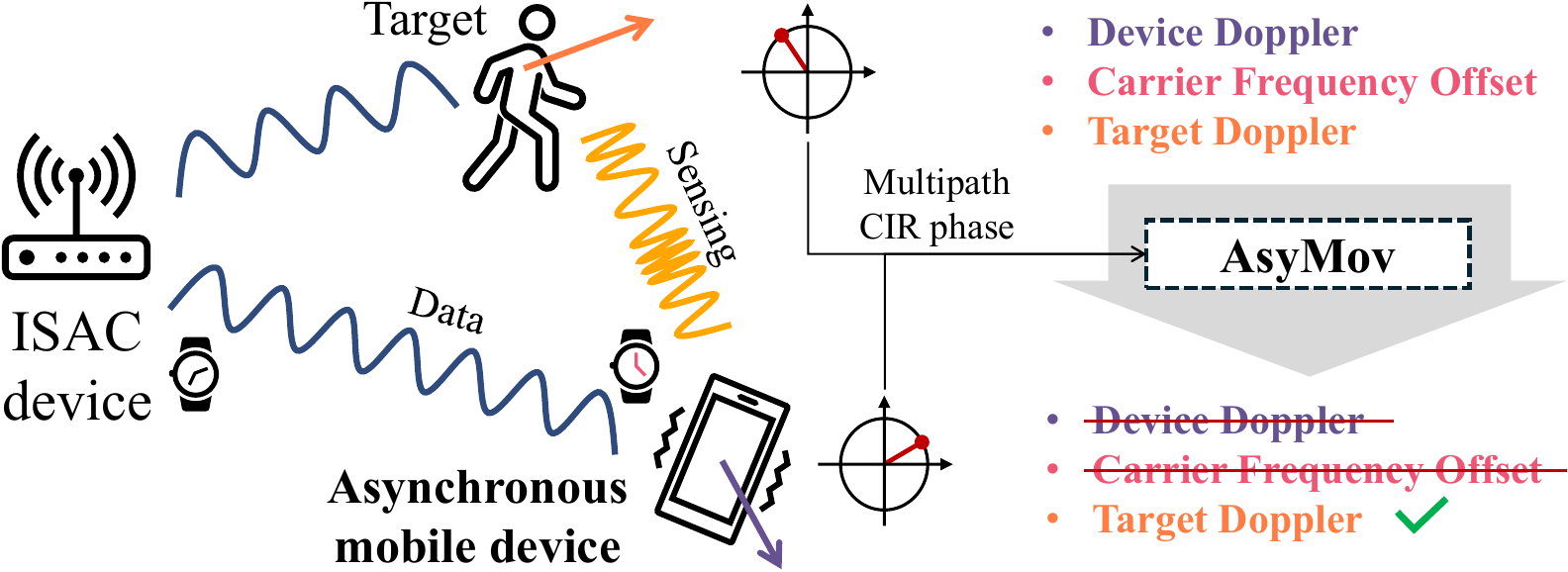}
    \caption{Overview of AsyMov.}
    \label{fig:concept}
\end{figure}

The reference scenario that we consider in the present work is depicted in Fig.~\ref{fig:concept}. It features a TX/RX pair where one of these two devices is mobile, and the target is also moving. For this setup, differently from what is commonly assumed in \ac{isac} research, an \textit{additional} Doppler frequency contribution appears on the reflected signal due to the movement of the TX or RX device.
The key challenge in {\it jointly} addressing clock asynchrony and moving TX/RX devices descends from the different properties of \ac{cfo}/\ac{po} and from the motion-induced Doppler component of the TX/RX device, which adds up to the Doppler caused by the target. In fact, the noise due to \ac{cfo}/\ac{po} is the same for all propagation paths~\cite{pegoraro2024jump} (spatial domain), but changes across subsequent communication packets (time domain), whereas the device Doppler changes slowly over time but causes a different frequency shift on each propagation path. 
We underline that existing techniques based on cross-antenna processing,~\cite{zhang2022integration}, or reference paths,~\cite{pegoraro2024jump, meneghello2022sharp}, cannot concurrently  cope with these different spatial and temporal dynamics. In fact, existing solutions only address the following scenarios: (i)~\textit{moving} monostatic radar systems, which are not affected by \ac{cfo} and \ac{po}~\cite{zhang2022mobi2sense, chang2024msense, liu2023towards} and only need to compensate for the device Doppler shift, or (ii)~asynchronous \ac{isac} systems with \textit{static} devices~\cite{zhao2024multiple, pegoraro2024jump, wu2024sensing}, which are only affected by \ac{cfo} and \ac{po}. 
Either case poses stringent assumptions to the application scenario, limiting the range of use of previous solutions. 

In this work, we propose \ourname{}, a method to estimate the Doppler frequency of a target and the device velocity in bistatic mobile \ac{isac} setups using the sole wireless signal. 
To this end, we leverage the growing trend in next-generation wireless systems of adopting a large communication bandwidth~\cite{garcia2021ieee}. This yields a good multipath resolution, which can be exploited to extract {\it path-specific signal features}. 

Our approach (see Fig.~\ref{fig:concept}) obtains phase measurements at the RX from different multipath reflections across time in the \acf{cir}. Hence, it removes \ac{cfo} and \ac{po} by leveraging their invariance across the received propagation paths. Path-specific phase terms are subsequently removed via time-domain phase differencing, thanks to their slow varying nature over short time windows. 
Most importantly, each multipath component is \textit{differently} affected by the TX or RX motion, and we exploit this as a source of \textit{diversity} to reduce the number of unknowns in the phase measurements model. As a result, we can successfully cast the Doppler frequency estimation of the target as a non-linear system of equations, that can be solved if at least two multipath components from static scatterers are available. In such a case, the target's Doppler and the \ac{isac} device velocity are jointly estimated via a newly proposed and efficient alternating minimization algorithm that operates on time-averaged phase measurements.
Remarkably, \ourname{} is robust to \textit{irregular} sampling of the \ac{cir}. \rev{This means that it can be implemented in \ac{isac} systems with minimal overhead, by opportunistically exploiting the transmissions of data packets, and without requiring the transmission of additional signals for sensing purposes or applying sparse channel reconstruction methods~\cite{pegoraro2022sparcs}.}
As an additional feature, \ourname{} can seamlessly integrate measurements of any moving device velocity obtained with an onboard \ac{imu}, which may be available on mobile devices. As we show in Section~\ref{sec:exp-results}, our experimental results reveal that the integration of an onboard \ac{imu} leads to less accurate results than using our approach with the sole wireless signal, but increases the estimation reliability in case fewer than two static multipath components are available.

\ourname{} is extensively evaluated along two lines. First, its performance limits are analyzed by obtaining the \ac{crb} for the target's Doppler frequency estimation and via numerical simulations. Second, it is evaluated on real measurements obtained with an IEEE~802.11ay \ac{isac} system operating in \rev{the $60$ or $28$~GHz bands}. \ourname{} is finally compared against standard Doppler estimation techniques based on \ac{music}~\cite{schmidt1986multiple} and to the case where both TX and RX are static.

The contributions of this paper are summarized as follows.

\begin{enumerate}
\item We solve the problem of estimating the Doppler frequency of a target and the TX/RX device velocity in bistatic \ac{isac} with asynchronous \textit{moving} devices. Our solution uniquely relies on the analysis of the received wireless signal by means of an original \ac{cir} phase measurements differencing step and a new alternating minimization algorithm.

\item We derive the \ac{crb} for Doppler frequency estimation based on our measurement model and propose a more practical, approximate \ac{crb} expression. Using numerical simulations, we show that \ourname{} achieves accurate Doppler frequency estimation with different carrier frequencies and \acp{snr}.

\item We evaluate \ourname{} performance on real measurements, collected with an \rev{IEEE 802.11ay testbed at $60$ and $28$ GHz,} in different setups. Our results show that \ourname{} achieves median estimation errors as low as $3$~Hz for the Doppler frequency, $3^{\circ}$ for device movement direction, and $2$~cm/s for the device speed.

\item We demonstrate improvements over existing \ac{music}-based methods~\cite{schmidt1986multiple}, combined with JUMP~\cite{pegoraro2024jump} for asynchrony compensation. Moreover, \ourname{} is more accurate than a benchmark algorithm that estimates the device movement with an onboard \ac{imu} sensor, and it can be easily integrated with \ac{imu} measurements when the number of multipath components in the \ac{cir} is insufficient.
\end{enumerate}

The paper is organized as follows. The related work is discussed in \secref{sec:related}. Reference scenario and system model are introduced in \secref{sec:sysmodel}. \ourname{} is presented in \secref{sec:method}. \secref{sec:num-results} contains its analysis, with the derivation of the \ac{crb} and related numerical results. Experimental results are shown in \secref{sec:exp-results}, to demonstrate the effectiveness of our approach against state-of-the-art methods. Concluding remarks are given in \secref{sec:conclusion}.

\section{Related Work}\label{sec:related}

\textbf{Synchronous \ac{isac}.} Monostatic full-duplex technology for \ac{isac} has been investigated in many research works. In~\cite{barneto2021full, heino2021design}, beamforming, waveform optimization, and phased array architectures have been utilized to enable full-duplex \ac{isac} in \ac{mmwave} 5G systems. The authors of\cite{pegoraro2024rapid, pegoraro2022sparcs} focused on micro-Doppler spectrogram estimation for indoor human sensing, using a monostatic full-duplex IEEE~802.11ay system. However, full-duplex technology suffers from strong self-interference which remains a challenging and still open issue~\cite{barneto2021full}. As a solution, the transmitted power can be reduced to avoid saturation at the receiver, but this is only feasible in small indoor scenarios and reduces the communication \ac{snr}~\cite{pegoraro2024rapid}.
Other works~\cite{kanhere2021target, pucci2022performance, gao2023integrated} have addressed bistatic Doppler estimation through linear methods in \ac{mmwave} \ac{isac}, assuming synchronized static devices. 

Since synchronous transceivers are unaffected by \ac{cfo} and \ac{po}, the above works apply standard techniques from time-frequency analysis to obtain the target's Doppler shift. However, with asynchronous time systems and moving devices, these methods fail due to the superposition of the phase offsets, the device movement, and the target's Doppler frequency.  

\textbf{Bistatic asynchronous \ac{isac}.} Existing works on bistatic \ac{isac} which tackled the clock asynchronism between TX and RX can be categorized into two groups. 

The first group exploits the fact that the clock offsets across multiple antennas at the RX are the same due to the shared \ac{lo}. In~\cite{ni2021uplink}, the authors propose the \ac{cacc} method to remove \ac{to} and \ac{cfo}. 
Similarly, \cite{li2022CSI, zhang2020exploring, zeng2020multisense} use the \ac{casr} method, which removes phase offsets by obtaining the ratio of the channel state information between antennas and enables Doppler estimation. 
Authors in~\cite{zhang2020exploring} analyze \ac{casr} using the Mobius transform to estimate the Doppler frequency shift with the \ac{lora} technology. In~\cite{zeng2020multisense}, instead, multi-person respiration sensing is modeled as a blind source separation problem and efficiently solved by independent component analysis. 
Cross-antenna methods require multiple antennas at the receiver, which becomes costly at higher frequencies (such as in \ac{mmwave} systems), where typically phased arrays are preferred. Moreover, they add complexity to the estimation of the sensing parameters by introducing non-linearity. 

The second group of works exploits a reference propagation path.
\cite{yu2024efficient} proposes a new low-complexity method relying on the correlation between \ac{los} and \ac{nlos} paths to compensate for the \ac{cfo}. After the compensation, the Doppler frequency is estimated through \ac{dft}.
\cite{pegoraro2024jump} proposes a correlation-based algorithm to compensate for the \ac{cfo} which is robust to multitarget and \ac{nlos} scenarios to compute the micro-Doppler spectrum via a short-time \ac{dft}.

The above bistatic \ac{isac} methods all assume \textit{static} TX and RX, which poses severe limitations to real-world \ac{isac} applications. The device movement introduces an additional Doppler frequency shift, which \textit{differs} for each propagation path and can not be tackled by the mentioned methods.

More recently, \cite{wang2024clutter, wang2024windowing} have proposed a fingerprint-based \ac{to} and \ac{cfo} removal method for vehicular scenarios. However, the Doppler introduced by a moving TX or RX is not considered as a disturbance factor for the target Doppler estimation, so these works inherit the above limitations.

\textbf{Mobile transceiver \ac{isac}.} Existing solutions for wireless sensing with moving devices mostly consider dedicated monostatic \ac{mmwave} radars. This avoids self-interference, \ac{cfo}, and \ac{po}, and makes the problem significantly simpler.
In~\cite{zhang2022mobi2sense} and~\cite{chang2024msense}, the Doppler phase information is extracted by removing the effect of device motion using reflections on background static objects. Recently, the authors of~\cite{liu2023towards} introduced the dynamic Fresnel zone model to link the variation in the received signal and the receiver movement. However, this work does not provide a method to estimate the Doppler frequency of the target and focuses instead on activity recognition.

To the best of our knowledge, \ac{isac} with asynchronous and mobile devices has only been considered in our previous work~\cite{ventura2024bistatic}, since other works all consider \textit{static} \ac{isac} devices. \rev{This work presents important novel contributions compared to~\cite{ventura2024bistatic}, as detailed in the following.
\begin{itemize}
\item \cite{ventura2024bistatic} relies on standard \ac{nls}, whereas we propose an original, more efficient solution based on alternating minimization and analyze its identifiability and convergence.
\item We provide a theoretical analysis of the sensitivity of the Doppler estimation to noise and multipath channel parameters, which is missing in \cite{ventura2024bistatic}.
\item We perform an extensive experimental campaign in several scenarios using an IEEE~802.11ay \ac{rfsoc} \ac{isac} system, validating our algorithm on real data at $60$ and $28$~GHz, whereas \cite{ventura2024bistatic} relies on simulations.
\item \cite{ventura2024bistatic} assumes a constant channel estimation period, which is unrealistic in \ac{isac}. Conversely, \ourname{} is robust to irregular channel estimation (sampling) times.
\end{itemize}
}

\section{Mobile ISAC System Model}\label{sec:sysmodel}

In this section, we introduce the mobile \ac{isac} system model. This includes the reference scenario, the channel model, and the expressions of the phase measurements.
\vspace{-0.2cm}
\subsection{Reference scenario}
\label{sec:scenario}
Consider two \ac{isac} devices acting, respectively, as transmitter~(TX) and receiver~(RX), located on a $2$-dimensional ($2$D), horizontal Cartesian plane. 
Denote by $t$ the continuous time variable, and refer to $\mathbf{x}^{\rm tx} (t)$ and $\mathbf{x}^{\rm rx} (t)$ as the time-varying locations of the TX and RX devices, respectively. 
In the following, we consider the TX to be mobile, while the RX is static, i.e., $\mathbf{x}^{\rm rx} (t) = \mathbf{x}^{\rm rx}$. We call $v^{\rm tx}(t)$ and $\eta(t)$ the TX speed and movement direction with respect to the \ac{los} segment between TX and RX.
Note that this assumption is not restrictive, since the symmetric case in which the RX moves and the TX is static is analogous, as further discussed in \secref{sec:cir-model}, and leads to similar derivations. 
Note also that our model still holds if the TX and RX are both mobile. 
In this case, we consider $v^{\rm tx}(t), \eta(t)$ to be the \textit{relative} velocity magnitude and the direction between the TX and the RX.

We consider $Q$ static or mobile scatterers in the environment, indexed by $q=1,\dots, Q$, each at location $\mathbf{x}_{q}(t)$. Mobile scatterers have a velocity $v_q(t)$ that forms an angle $\gamma_q(t)$ with the bisector line of the angle formed by the TX, the scatterer, and the RX, denoted by $\beta_q(t)$ (\textit{bistatic angle}), as shown in \fig{fig:geometric_disp}.
We aim to estimate the bistatic Doppler frequency caused by a moving scatterer of interest, called \textit{target} using channel estimates obtained by the RX as part of normal communication and \ac{aod} estimates. In the symmetric case where the RX moves, the \ac{aod} would be replaced by the \ac{aoa} (see  \secref{sec:cir-model}).

\subsection{Channel model}\label{sec:cir-model}

Consider the wireless signal transmitted by the TX device. We call \mbox{$\lambda = c/f_{\rm c}$} the wavelength of the transmitted signal carrier, with $c$ being the speed of light and $f_{\rm c}$ the carrier frequency. 
We consider a multipath channel including $Q$ delayed, Doppler-shifted, and attenuated copies of the transmitted signal, caused by the $Q$ scatterers in the environment. 
As customary in the \ac{isac} literature, we only consider first-order signal reflections and the \ac{los} propagation path, since these are significantly stronger than second-order reflections~\cite{zhang2021overview}. 

Next, we present the Doppler frequency model and the continuous-time channel model underlying the considered scenario. Then, we derive the practical discrete-time channel estimated by the \ac{isac} system.

\subsubsection{Doppler frequency model}  
The Doppler frequency on a generic path $q$ is obtained by differentiating its path length with respect to time as~\cite{willis2005bistatic}
\begin{align}
	\frac{1} {\lambda}\frac{\partial} {\partial t}& \left(\|\mathbf{x}^{\rm rx} - \mathbf{x}_q(t))\| + \|\mathbf{x}_q(t) - \mathbf{x}^{\rm tx} (t))\|\right)\approx\\
	&\approx  \underbrace{\frac{v^{\rm tx} (t)}{ \lambda}  \cos \xi_q (t)}_{f_{{\rm D},q}^{\mathrm{tx}}(t)} +  \underbrace{\frac{2 v_q (t)}{\lambda} \cos \gamma_q(t) \cos \frac{\beta_q (t)}{2}}_{f_{{\rm D},q}(t)},\label{eq:insta-doppler}
\end{align}
where $\xi_q(t)$ is the angle between the segment connecting the $q$-th scatterer to the TX and the TX velocity vector.
In \eq{eq:insta-doppler}, the total Doppler shift is decomposed into the contributions of the TX movement, $f_{{\rm D},q}^{\mathrm{tx}}(t)$, and that of the target, $f_{{\rm D}, q}(t)$.  
The approximation in \eq{eq:insta-doppler} holds for short processing intervals, during which we consider the movement of the TX to be well approximated by a constant-velocity model~\cite{willis2005bistatic}.
We remark that the Doppler due to the movement of the TX or RX is not modeled in recent works that tackle asynchronous bistatic \ac{isac} systems, e.g.,  \cite{pegoraro2024jump, wang2024clutter}.

\eq{eq:insta-doppler} shows that the target-induced bistatic Doppler frequency shift depends on the direction of motion with respect to the bistatic system.
\rev{In this paper, we do not estimate the moving velocity vector of the target, given by the pair $v_q(t), \gamma_q(t)$ (or in Cartesian coordinates), but only the aggregate Doppler shift $f_{{\rm D},q}(t)$.
Hence, we do not require the knowledge of $\gamma_q(t), \beta_q (t)$, nor estimate them.
The estimation of the velocity vector is, in general, ambiguous using a single bistatic TX and RX pair~\cite{willis2005bistatic}, and is solved for multistatic scenarios~\cite{ma2021target}.
We focus instead on the estimation of the Doppler frequency in the unsolved case of moving TX-RX \textit{and} presence of \ac{cfo}, so the ambiguity does not arise in our model that considers $f_{{\rm D},q}(t)$ \textit{as a whole}.
In multistatic settings, the Doppler frequency estimated by \ourname{} on each bistatic pair can be used as the input for velocity vector estimation algorithms.}

\begin{figure}
    \centering
    \tikzset{antenna/.style={insert path={-- coordinate (ant#1) ++(0,0.25) -- +(135:0.25) + (0,0) -- +(45:0.25)}}}
\tikzset{station/.style={naming,draw,shape=isosceles triangle,shape border rotate=90, minimum width=10mm, minimum height=10mm,outer sep=0pt,inner sep=3pt}}
\tikzset{mobile/.style={naming,draw,shape=rectangle,minimum width=12mm,minimum height=6mm, outer sep=0pt,inner sep=3pt}}
\tikzset{naming/.style={align=center,font=\small}}
\pgfdeclarelayer{bg} 
\pgfsetlayers{bg,main}
\newcommand{\BS}[1]{%
\begin{tikzpicture}
\node[] (base) {};
\draw[thick, line join=bevel] (base.north) -- ($(base.south east)-(-.5em,2em)$) node [midway] (mid_east) {} node [near end] (end_east) {};
\draw[thick, line join=bevel] (base.north) -- ($(base.south west)-(.5em,2em)$) node [midway] (mid_west) {} node [near end] (end_west) {};
\draw[thick, line join=bevel] (base.north) -- ($(base.south)-(0em,2.5em)$) node [midway] (mid_center) {} node [near end] (end_center) {};
\draw[thick, line join=bevel] (mid_east.center) -- (mid_center.center) -- (mid_west.center);
\draw[thick, line join=bevel] (end_east.center) -- (end_center.center) -- (end_west.center);

\draw[thick, line cap=rect] ([yshift=0pt]base.north) [antenna=1];
\end{tikzpicture}
}

\newcommand{\phone}[1]{%
\begin{tikzpicture}
\draw[rounded corners=2pt, fill=black] (0em,0em) rectangle (1.5em,2.5em) {};
\draw[rounded corners=1pt, fill=white] (.2em,.6em) rectangle (1.3em,2.3em) {};
\draw[fill=white] (.75em,.3em) circle (2pt) {};
\end{tikzpicture}
}

\newcommand{\UE}[1]{%
\begin{tikzpicture}[every node/.append style={rectangle,minimum width=0pt}]
\node[mobile] (box) {#1};

\draw ([xshift=.25cm] box.south west) circle (4pt)
      ([xshift=-.25cm]box.south east) circle (4pt);

\fill ([xshift=.25cm] box.south west) circle (1pt)
      ([xshift=-.25cm]box.south east) circle (1pt);

\draw (box.north) [antenna=1];
\end{tikzpicture}
}

\newcommand{\man}[1]{%
\begin{tikzpicture}
\node[circle,fill,minimum size=5mm] (head) {};
\node[rounded corners=1pt,minimum height=1.3cm,minimum width=0.4cm,fill,below = 1pt of head] (body) {};
\draw[line width=1mm,round cap-round cap] ([shift={(2pt,-1pt)}]body.north east) --++(-90:6mm);
\draw[line width=1mm,round cap-round cap] ([shift={(-2pt,-1pt)}]body.north west)--++(-90:6mm);
\draw[line width=.8mm,white,-round cap] (body.south) --++(90:5.5mm);
\end{tikzpicture}
}

\newcommand{\bisector}[8][]{%
    \path[#1] let
        \p1 = ($(#3)!1cm!(#2)$),
        \p2 = ($(#3)!1cm!(#4)$),
        \p3 = ($(\p1) + (\p2) - (#3)$)
    in
        ($(#3)!#6!(\p3)$) -- ($(\p3)!#5!(#3)$) node [] (b) {};
    \pic["$\gamma_{q}$", text=color3, draw=color3, -, thick, angle radius=1.5em, angle eccentricity=.9, pic text options={shift={(12pt,5.5pt)}}] {angle = b--#8--#7};
    }
    
\begin{tikzpicture}
    \tikzstyle{disp} = [draw, fill=white, circle, minimum size=0.1em]
    \tikzstyle{target} = [fill, circle] 
    \definecolor{orange}{RGB}{255,158,20}
    \definecolor{green}{RGB}{78,211,78}
    \definecolor{cyan}{RGB}{0,204,204}
    \definecolor{color1}{HTML}{003f5c}
    \definecolor{color2}{HTML}{bc5090}
    \definecolor{color3}{HTML}{ffa600}
    \definecolor{darkslategray583162}{RGB}{58,31,62}
    \definecolor{burlywood231175145}{RGB}{231,175,145}
    
    \begin{scope}
        \node (bs) {\BS{RX}};
        \node [] at(bs.north) (rx) {};
        \node [] at (bs.west) {RX};
        \node [above=1em of rx, scale=0.4, xshift=8em](target) {\man{Target}};
        \node [left= 0.1em of target] at (target.west) {Target};
        \node [right=17em of rx, anchor=north, yshift=-2em, scale=0.8](ue) {\phone{TX}};
        \node [] at ($(ue.west)-(.5em,0em)$) {TX};
        \node [] at($(ue.east)+(-.1em,1.2em)$) (tx) {};
        
        \draw [-, ultra thick] ($(target)+(8.5em,-1em)$) -- node[yshift=-.5em](middle1){} ($(target)+(10.5em,-1.3em)$);
        \node at ($(middle1)+(0em,1.2em)$) (static_path) {Static path};
        
        \node [darkslategray583162](v) at ($(tx)+(2em,4em)$) {$v^{\rm tx}$};

        \node [darkslategray583162](v_t) at ($(target)+(4em,.5em)$) {$v_{\rm q}$};
        
        \bisector[draw,densely dotted, semithick]{bs.north}{target}{tx}{.65}{.9}{v_t}{target}
        \draw [draw=darkslategray583162,->,thick] (target) -- (v_t);
        
        \draw [-] (bs.north) -- node[yshift=.5em] () {LoS} (tx.center);
        \node at ($(tx)!-5em!(rx)$)(tempL){};
        \draw [-, dashed] (tx.center) -- (tempL);

        \draw [-] (bs.north) -- (target);
        \draw [-] (target) -- (tx.center);
        \node at ($(tx)!-5em!(target)$)(tempT){};
        \draw [-, dashed] (tx.center) -- (tempT);

        \draw [-] (bs.north) -- ($(middle1)+(0,.5em)$);
        \draw [-] ($(middle1)+(0,.5em)$) -- node[](tempmid1){} (tx.center);
        \node at ($(tx)!-5em!(tempmid1)$)(temp1){};
        \draw [-, dashed] (tx.center) -- (temp1);

        \pic["$\alpha_{\rm t}$", text=color3, draw=color3, -, thick, angle radius=5.2em, angle eccentricity=.9, pic text options={shift={(-13pt,2pt)}}] {angle = target--tx--rx};
        \pic["$\xi_{\rm t}$", text=color3, draw=color3, -, thick, angle radius=2.1em, angle eccentricity=.9, pic text options={shift={(22pt, -2pt)}}] {angle = v--tx--target};
        \pic[draw=color3, -, thick, angle radius=1.95em, angle eccentricity=.9] {angle = v--tx--target};

        \pic["$\alpha_s$", text=color2, draw=color2, -, thick, angle radius=3.5em, angle eccentricity=.9, pic text options={shift={(-12pt,8pt)}}] {angle = tempmid1--tx--rx};
        \pic["$\xi_s$", text=color2, draw=color2, -, thick, angle radius=2.7em, angle eccentricity=.9, pic text options={shift={(1pt,10pt)}}] {angle = v--tx--tempmid1};
        \pic[ draw=color2, -, thick, angle radius=2.55em, angle eccentricity=.9] {angle = v--tx--tempmid1};


        \pic["$\eta$", xshift=-2em, draw, text=color1, draw=color1, -, thick, angle radius=1.3em, angle eccentricity=.9, pic text options={shift={(15pt,-5pt)}}] {angle= v--tx--rx};

        \draw [draw=darkslategray583162,->,thick] (tx.center) -- (v);

        \pic["$\beta_{q}$", text=color1, draw=color1, -, thick, angle radius=2em, angle eccentricity=.9, pic text options={shift={(20pt,4pt)}}] {angle = rx--target--tx};
    \end{scope}

\end{tikzpicture}
    \vspace{-.6cm}
    \caption{Reference scenario and multipath geometry. The time dependency is omitted for better visualization.}
    \label{fig:geometric_disp}
\end{figure}

\subsubsection{Continuous-time channel model} Consider the continuous-time channel between the TX and the RX and denote by $\tau_q(t)$, $\alpha_q(t)$, and $A_q(t)$ the propagation delay, the \ac{aod}, and the complex coefficient of the \mbox{$q$-th} path, respectively. 
$A_q(t)$ accounts for the propagation loss, and the scatterer's \ac{rcs}~\cite{richards2010principles}. If the devices are equipped with multiple antennas, $A_q(t)$ also includes the combined effect of the TX and RX beamforming vectors. 

As shown in \fig{fig:geometric_disp}, we consider the \ac{aod} to be measured relatively to the \ac{los} line connecting the TX and the RX. This assumes that the orientation of the RX with respect to the TX is known and has been compensated for. This could be done by obtaining the \ac{aod} or \ac{aoa} estimate of the \ac{los} path, for example, during beam-alignment protocols for communication. We use this technique in our experimental evaluation in \secref{sec:exp-results}. 
If available, an alternative is to use onboard sensors like an \ac{imu} to estimate the device orientation.
 
The offsets caused by the asynchrony of TX and RX, i.e., \ac{to}, \ac{cfo}, and \ac{po} are denoted by $\tau_{\rm o}(t)$, $f_{\rm o}(t)$, and $\psi_{{\rm o}}(t)$, respectively. 
Denoting by $\delta({\tau})$ the Dirac delta function, the \ac{cir} at time $t$ and delay $\tau$ is
\begin{equation}\label{eq:cir-continuous}
    h(t,\tau) = e^{j\psi_{{\rm o}}(t)}\sum_{q=1}^{Q}{A_q(t) e^{j\vartheta_q(t)}
    \delta(\tau-\tau_q(t)-\tau_{\rm o}(t)}),
\end{equation}
where \mbox{$\vartheta_q(t) = 2\pi (f_{{\rm D},q}(t)+f_{\rm o}(t)+f_{{\rm D},q}^{\mathrm{tx}}(t))t$}. Note that $f_{{\rm D},q}(t)$ is added to the \ac{cfo} and the Doppler shift introduced by the transmitter.

We stress that even though \ourname{} requires estimating the \ac{aod} (if the TX moves) or \ac{aoa} (if the RX moves) of the reflections, it does not require having multiple antennas on the \ac{isac} devices. Indeed, our approach is agnostic to the number of available antennas or radio frequency chains at the TX or RX, and can be implemented on cost-effective phased-array systems. In this case, the \ac{aod} or \ac{aoa} can be obtained through correlation-based algorithms such as the ones in~\cite{garcia2023scalable, pegoraro2024rapid}.

\subsubsection{\ac{cir} estimate model} \label{sec:cir-est-model}

In \ac{sc} systems, e.g., IEEE~802.11ay, the RX estimates the \ac{cir} directly using cross-correlation of the received signal with a known pilot sequence such as complementary Golay sequences which exhibit perfect autocorrelation property with no frequency shift. 
In \ac{ofdm} systems, e.g., 5G-NR, an estimate of the \ac{cfr} is computed at the receiver by exploiting a known preamble. The \ac{cir} can then be obtained via \ac{idft} of the estimated \ac{cfr}.
\rev{We denote by $G$ the \ac{snr} gain given by coherent processing of several pilot symbols during channel estimation.
This is equal to the length of the pilot sequence if standard matched filtering or frequency-domain symbol division are used, otherwise it can be estimated empirically.
$G$ will be used in the model of the noise distribution on the resulting \ac{cir} in \secref{sec:phase-meas}.
In this paper, we do not use the random data part of the communication packets to estimate the channel. Nevertheless, \ourname{} could be applied to the data symbols without modification, after demodulating and decoding them.}

As a consequence of the finite bandwidth of the \ac{isac} system, two or more scatterers may be closer than the delay resolution, given by $\Delta_{\tau} = 1/B$ where $B$ is the TX signal bandwidth. Hence, we consider a multipath channel including $M(t)\leq Q$ resolvable components, indexed by $m=1, \dots, M(t)$. The estimated \ac{cir} is discretized in the delay domain with granularity $\Delta_{\tau}$ and the discrete delay grid is given by $l\Delta_{\tau}, l=0, \dots, L-1$, where $L$ is the grid span. 

The \ac{cir} estimation is repeated across multiple time frames, indexed by $k$, at instants $t_k, k=0, \dots, K-1$, where $K$ is the number of frames and we consider $t_0 = 0$~s. 
We denote the inter-frame spacing by $T_k = t_k - t_{k-1}$. Note that the inter-frame spacing can be irregular, i.e., channel estimation does not need to happen on a regular sampling grid.

Using a common assumption in \ac{isac} and radar processing, we consider a short processing window of duration $t_{K-1}$~\cite{zhang2021overview}, where the parameters, $M(t)$, $\tau_m(t)$, $f_{{\rm D}, m}(t)$, $\alpha_m(t)$, $A_m(t), \eta(t),  \xi_m(t), v^{\rm tx}(t)$, and consequently the TX motion-induced Doppler shift $f_{{\rm D},m}^{\mathrm{tx}}(t)$ can be considered constant. 
Conversely, all the nuisance parameters, $\tau_{\rm o}(t)$, $f_{\rm o}(t)$, $\psi_{{\rm o}}(t)$, are time-varying within the window. Note that, although the \ac{cfo} is sometimes considered slowly time-varying, in \ac{isac} it can significantly drift during the processing window of $K$ frames. Moreover, in some systems, we need to consider the \textit{residual} \ac{cfo} after partial compensation by the receiver. The residual \ac{cfo} is fast time-varying, being the random residual of an imperfect estimation and compensation process~\cite{wu2024sensing}.

The noise-free estimated \ac{cir} at the $k$-th frame is
\begin{equation}
    \label{eq:cir-discrete}
    h[k,l] = e^{j\psi_{{\rm o}}(t_k)}\sum_{m=1}^{M}{A_{m}e^{j\vartheta_m[k]}\chi\left(l\Delta_{\tau}-\tau_m-\tau_{\rm o}(t_k)\right)},
\end{equation}
where \mbox{$\vartheta_m[k] = 2\pi (f_{{\rm D},m}+f_{\rm o}(t_k)+f_{{\rm D},m}^{\mathrm{tx}})t_k$}.
In \eq{eq:cir-discrete}, $\chi(l\Delta_{\tau})$ replaces the Dirac delta to account for non-ideal autocorrelation of the pilot sequence (in \ac{sc} systems) or the impact of the finite-length \ac{cfr} estimate (in \ac{ofdm} systems).

In our scenario, we consider that the $M$ propagation paths can be partitioned as follows: (i)~a \ac{los} path, which represents the direct propagation from the TX to the RX, (ii)~a single moving target path, (iii)~\mbox{$S$} \textit{static} scatterers paths for which \mbox{$f_{{\rm D}, m}=0$}. Point (ii) is not a requirement for \ourname{}, but simplifies the derivation. If multiple targets are present, \ourname{} should be applied on each corresponding peak in the \ac{cir}.
Note that assumption (i) simplifies the description of the algorithm, but is not strictly required. In fact, \ourname{} is still applicable when the \ac{los} is not available, by replacing the \ac{los} path with a reference static path caused by a strong scatterer.

In practice, TX could also measure $v^{\rm tx}$ using an onboard \ac{imu}, simplifying the derivation of the target's Doppler frequency.
However, the accuracy of such measurement is critical to the subsequent estimation process as discussed in~\cite{pegoraro2024jump}. Since low-cost sensors available on commercial cellular devices are often inaccurate, our approach considers $v^{\rm tx}$ to be unknown, i.e., it relies on the sole wireless signal. Still, our algorithm can readily incorporate a reliable estimate of $v^{\rm tx}$ from an \ac{imu}. This is demonstrated in \secref{sec:imu-exp-results}, where we show that \ourname{} is more accurate when only using the wireless signal, but external \ac{imu} measurements enhance its reliability when multipath components are momentarily undetected. 
Next, we focus on estimating the target Doppler frequency under the nuisance due to the \ac{cfo}, \ac{po}, and the TX motion. Hence, we assume that the RX can detect and separate the $M$ multipath components and extract their phase across time. To this end, the \ac{to} can be compensated for by obtaining \textit{relative} delay measurements with respect to the \ac{los}, as done in~\cite{pegoraro2024jump}. 

\subsection{Phase measurements}\label{sec:phase-meas}
In this section, we provide a model of the phase for each multipath component in \eq{eq:cir-discrete}. 

We group the \ac{cfo} and \ac{po} terms, which are the same on all propagation paths, into $\Psi_{{\rm o}}(t_k) = \psi_{{\rm o}}(t_k) + 2\pi f_{\rm o}(t_k)t_k$. Moreover, we use subscripts $\cdot_{\rm LoS}$ and $\cdot_{\rm t}$ to refer to quantities related to the \ac{los} and to the target-induced paths, respectively, as shown in~\fig{fig:geometric_disp}. For the paths caused by static scatterers, we use index $s=1, \dots, S$, where $S$ is the number of resolvable static scatterers detected in the processing window, with \mbox{$S = M-2$}.
The phase of the \ac{los} is
\begin{equation}\label{eq:los-phase}
    \phi_{\rm LoS}[k]= \Psi_{{\rm o}}(t_k)+\angle{A_{{\rm LoS}}} + 2\pi  t_k\left(\frac{v^{\rm tx}}{\lambda} \cos \eta \right) \rev{+ w_{\rm LoS}[k]},
\end{equation}
where $\angle{\cdot}$ is the phase operator, \mbox{$\xi_{\rm LoS} = \eta$}, \rev{and $w_{\rm LoS}[k]$ is the noise variable on the phase of the \ac{los} at time $k$.}
The phase of the target-induced path is affected by both the Doppler shift caused by the target, $f_{\rm D, t}$, and by the TX motion, $v^{\rm tx}$, as
\begin{equation}\label{eq:target-phase}
    \phi_{\rm t}[k]=\Psi_{{\rm o}}(t_k)+ \angle{A_{{\rm t}}} + 2\pi  t_k\left(f_{\rm D, t} + \frac{v^{\rm tx}}{\lambda} \cos \xi_{\rm t} \right) \rev{+ w_{\rm t}[k]},
\end{equation}
where $\xi_{\rm t}$ is the angle between the segment connecting the target to the TX and the TX velocity vector\rev{, and $w_{\rm t}[k]$ is the noise variable on the phase of the target path at time $k$.} The phase of the $s$-th static multipath component is
\begin{equation}\label{eq:static-phase}
    \phi_{s}[k]=\Psi_{{\rm o}}(t_k)+ \angle{A_{s}} + 2\pi  t_k \left(\frac{v^{\rm tx}}{\lambda} \cos \xi_s \right)
    \rev{+ w_{s}[k]},
\end{equation}
\rev{where $w_s[k]$ is the noise variable on the phase of the $s$-th static path at time $k$.}
From the \ac{cir}, the RX measures $\texttt{mod}_{2\pi}(\phi_i[k])$ where $i\in\{{\rm t}, {\rm LoS}, s\}$, $k=0, \dots,K-1$, and $\texttt{mod}_{2\pi}(\cdot)$ is the modulo $2\pi$ division, whose result is in $[0, 2\pi)$.  

\rev{\subsection{Noise distribution on phase measurements}\label{sec:noise-pdf}

Before discussing the limitations of existing synchronization techniques, we analyze the distributions of the noise variables $w_{\rm LoS}[k]$, $w_{\rm t}[k]$, and $w_{s}[k]$.
We assume the noise on the received signal is distributed as a complex Gaussian with variance $\sigma_w^2$. 
We only focus on Gaussian noise because, as explained in \secref{sec:cfo-canc}, \ourname{} exactly compensates for hardware impairments induced phase shifts. Therefore, the Gaussian noise on the received signal is the main nuisance source for the phase measurements.
Assuming unit TX power, the corresponding \textit{communication} \ac{snr}, based on the \ac{los} propagation path between the TX and RX is $\mathrm{SNR} = |A_{\rm LoS}|^2 / \sigma_w^2$.
This is the \ac{snr} we use in our simulations in \secref{sec:num-results}.
Note that the communication \ac{snr} is much higher than the \ac{snr} on the target paths used for sensing, due to the much lower propagation loss affecting the \ac{los}.

The channel estimation process reduces the noise variance by aggregating multiple pilot symbols coherently. 
The noise variance on the \ac{cir} is $\sigma_{h}^2 = \sigma_w^2 / G$ where $G$ is the \ac{snr} gain due to channel estimation. 
The noise on the phase measurements in \eqs{eq:los-phase}{eq:static-phase} depends on the noise on the \ac{cir} through a non-linear relation.
Denote path $i$ in the estimated \ac{cir} in \eq{eq:cir-discrete} at frame $k$ by \mbox{$h_i[k, l_i] \approx A_i e^{j(\vartheta_i[k] + \psi_{\rm o}(t_k))}$}, with $i\in \{{\rm LoS}, {\rm t}, s\}$. $l_i$ is the index of the peak corresponding to the $i$-th path and $l_i \Delta_{\tau} \approx \tau_i$. 
The real and imaginary components of $h_i[k, l_i]$ are denoted by $\Re\{h_i\}$ and $\Im\{h_i\}$, respectively. 
The phase of the $i$-th \ac{cir} path is given by $\angle{h_i[k, l_i]} = \mathrm{arctan}(\Im\{h_i\} /  \Re\{h_i\})$. 
If the \ac{snr} on the \ac{cir}, defined as $\mathrm{SNR}_{h} = G\cdot \mathrm{SNR}$, is \textit{high}, we can linearize the $\mathrm{arctan}$ function and approximate the resulting noise random variable as Gaussian. 
This assumption is reasonable if a sufficiently long pilot sequence is used to estimate the channel.
In this case, the noise variance on the phase of path $i \in \{{\rm LoS}, {\rm t}, s\}$ is given by $\sigma^2_{\phi, i}= \sigma_h^2 / (2 |A_i|^2)$ (the derivation is given in Appendix~\ref{app:app-b}).
Therefore, the noise variables $w_{\rm LoS}[k]$, $w_{\rm t}[k]$, and $w_{s}[k]$, are distributed as $\mathcal{N}(0, \sigma^2_{\phi, i})$ with $i \in \{{\rm LoS}, {\rm t}, s\}$.

For low $\mathrm{SNR}_h$, two factors have to be accounted for: (i)~the non-linearity of the $\mathrm{arctan}$ in the phase expression, (ii)~the wrapping of the phase around $2\pi$. 
An approximation of the corresponding noise probability density function resembles a von Mises distribution and is given in~\cite{luo2020analysis}, but is outside of the scope of this paper.
In \secref{sec:noise-pdf-sim}, we show that the noise distribution in the high $\mathrm{SNR}_h$ assumption well represents realistic scenarios in our simulations and experiments. 
}

\subsection{Limitations of existing techniques}\label{sec:limitations}

Existing techniques for \ac{to} and \ac{cfo} compensation use a reference antenna or propagation path to cancel out the offsets in the \ac{cir}.
Using a reference propagation path as done by JUMP~\cite{pegoraro2024jump}, the phase offsets removal step is 
\begin{align}\label{eq:phase-t-jump}
    \tilde\phi_{\rm t}[k] &= \angle{\left(e^{j\phi_{\rm t}[k]} / e^{j\phi_{\rm LoS}[k]}\right)}\\
    & = \angle{A_{{\rm t}}} - \angle{A_{{\rm LoS}}} + 2\pi  t_k\left(f_{\rm D, t} + \frac{v^{\rm tx}}{\lambda}( \cos \xi_{\rm t} - \cos \eta)\right)\nonumber,
\end{align}
\rev{where we omitted noise in the second expression for compactness}.
In \eq{eq:phase-t-jump}, if the phase measurements are obtained with a regular interval $t_k = T$, one can estimate the target Doppler from a sequence of measurements using, e.g., \ac{dft} or \ac{music}. 
This yields the desired target Doppler frequency only if the TX is \textit{static}, i.e., $v^{\rm tx} = 0$. 
If the TX moves, i.e., $v_{\rm tx} \neq 0$, the target Doppler estimate will contain a bias equal to $v^{\rm tx}( \cos \xi_{\rm t} - \cos \eta) / \lambda$ (proportional to the TX speed), preventing the correct estimation of the target Doppler.
This is exemplified in \fig{fig:bias-doppler}, where we compare the Doppler frequency estimated by \ourname{} and the combination of JUMP~\cite{pegoraro2024jump} and \ac{music}~\cite{schmidt1986multiple} on our experimental setup described in \secref{sec:setup}.
In this case, the target and the TX are both moving with distinct velocities.
JUMP correctly compensates for the \ac{cfo} but can not eliminate the TX Doppler. 
Moreover, note that the latter is \textit{path-specific} due to its dependency on the angle $\xi_{\rm t}$. Hence, it is non-trivial to compensate for it using standard phase difference-based methods, since every path is differently affected by the TX movement. 

\rev{\fig{fig:bias-doppler} also shows the Doppler frequency estimated by Mobi2Sense~\cite{zhang2022mobi2sense} (yellow line).
This compensates for the TX device motion, but can not cancel out phase offsets since it is designed for monostatic radar.
As a result, the \ac{cfo} and \ac{po} dominate the phase of the \ac{cir} paths, and the Doppler estimate is completely corrupted. 
}
\begin{figure}
    \centering
    \begin{subfigure}{0.45\columnwidth}
        \centering
        \input{plot/exp_doppler_music}    
    \end{subfigure}
    \hspace{1.5em}
    \begin{subfigure}{0.45\columnwidth}
        \centering
        \input{plot/exp_doppler_music_zoom}
    \end{subfigure}
    \vspace{-5mm}
    \caption{\rev{Example estimated target Doppler frequency with \ourname{}, the combination of JUMP~\cite{pegoraro2024jump} (for \ac{to} and \ac{cfo} removal) and \ac{music}~\cite{schmidt1986multiple} (for Doppler estimation), and Mobi2Sense~\cite{zhang2022mobi2sense} in our experimental evaluation.}}
    \label{fig:bias-doppler}
\end{figure}

\section{Methodology}\label{sec:method}

\begin{figure*}[t!]
    \centering
     \includegraphics[width=0.8\linewidth]{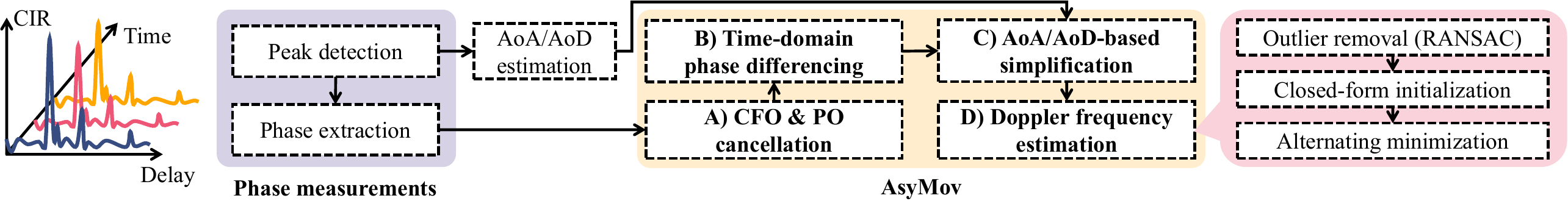}
    \caption{Block diagram of the proposed phase measurements steps and algorithm.}
    \label{fig:model}
    \vspace{-4mm}
\end{figure*}

In this section, we introduce \ourname{}, which estimates the bistatic Doppler frequency of the target from the phase measurements in \eqs{eq:los-phase}{eq:static-phase}. 
The phase measurements are extracted from the \ac{cir} estimate as described in the previous section and shown in the left part of~\fig{fig:model}.
The processing steps carried out by \ourname{} are shown in the yellow block in~\fig{fig:model} and can be summarized as follows.

\textit{A. \ac{cfo} and \ac{po} cancellation.} The phase offsets component, $\Psi_{{\rm o}}$, is common to all propagation paths. Hence, we can subtract the phase of the \ac{los} path from the other phase measurements, by \textit{canceling out} $\Psi_{{\rm o}}$. This step does not require a direct estimation of the phase offsets, enabling coherent processing of the multipath components with minimal complexity. This is further detailed in~\secref{sec:cfo-canc}.

\textit{B. Time-domain phase differencing.} By computing time-domain phase differences \textit{for each path}, we cancel out the path-specific phase terms due to the complex path amplitudes, $\angle{A_{i}}$, with $i\in\{{\rm t}, {\rm LoS}, s\}$, as detailed in~\secref{sec:phase-diff}. The resulting phase differences only depend on the Doppler shifts due to the TX and the target.

\textit{C. \ac{aod}-based simplification.} We leverage \ac{aod} estimation at the TX and the multipath geometry to make a key simplification in the phase measurements model, as described in~\secref{sec:aoa-simp}. This reduces the number of unknowns, thus enabling the estimation of the target Doppler frequency if at least $2$ static multipath components are detected.  

\textit{D. Doppler frequency estimation.} Using the simplified phase measurements model, we formulate the estimation of the target's Doppler frequency as a non-convex optimization problem across the multipath components (see \secref{sec:doppler-est}), that we solve using alternating minimization. To initialize the proposed algorithm, we derive and use the closed-form solution of the model using $S=2$ static paths.

\rev{The computational steps performed by \ourname{} are summarized in \alg{alg:asymov}, to which we refer in the detailed description provided in the next sections.}
\rev{Finally, we highlight the fact that a multi-target scenario would not increase the number of required static paths, as it would be sufficient to repeat \alg{alg:asymov} initializing it each time with one different target phase $\phi_{\rm t}$ for every available target in the environment.}

\vspace{-0.2cm}
\subsection{\ac{cfo} and \ac{po} cancellation}\label{sec:cfo-canc}

To cancel out the phase offsets term, $\Psi_{{\rm o}}$, we use \eq{eq:phase-t-jump} \rev{as shown in line~$2$ of \alg{alg:asymov}}, obtaining
\begin{align}
        & \tilde{\phi}_{\rm t}[k]= \Omega_{{\rm t}} + 2\pi t_k\left(f_{\rm D, t} + \frac{v^{\rm tx}}{\lambda} \left(\cos \xi_{\rm t} - \cos \eta \right) \right)  
    \rev{+ w'_{\rm t}[k]}
        , \label{eq:new-target-phase} \\ 
        & \tilde{\phi}_{s}[k]= \Omega_{s} + 2\pi t_k\left(\frac{v^{\rm tx}}{\lambda} \left(\cos \xi_s - \cos \eta\right)\right)
    \rev{+ w'_{s}[k]}, 
        \label{eq:new-static-phase}
\end{align}
where \mbox{$\Omega_{s} = \angle{A_{s}} - \angle{A_{{\rm LoS}}}$},  \mbox{$\Omega_{{\rm t}} = \angle{A_{{\rm t}}} - \angle{A_{{\rm LoS}}}$}.
\rev{The noise variables $ w'_{\rm t}[k]$ and $ w'_{s}[k]$ reflect the impact of the \ac{cfo} and \ac{po} removal and time-domain differencing operations. The phase offsets removal subtracts the phase of the \ac{los} path from the phases of the other paths, hence increasing the variance of the noise on phase $i \in \{{\rm t}, s\}$ to 
\begin{equation}
\label{eq:phase_noise}
    \sigma_{\tilde{\phi}, i}^2 = \sigma_{\phi, i}^2 + \sigma_{\phi, {\rm LoS}}^2 = \frac{\sigma_h^2}{2}\left(\frac{1}{|A_i|^2} + \frac{1}{|A_{\rm LoS}|^2}\right) = \frac{\sigma_h^2}{2}\kappa_i,
\end{equation}
where we defined $\kappa_i = |A_i|^{-2} + |A_{\rm LoS}|^{-2}$.
Hence, $w'_{i}[k]\sim \mathcal{N}(0, \sigma_{\tilde{\phi}, i}^2)$ with $i \in \{{\rm t}, s\}$.
This assumes that the noise on the phase of different paths is uncorrelated, which is true in \ac{sc} systems using ideal autocorrelation sequences (such as the Golay sequences used in our experiments) or \ac{ofdm} systems using full pilots in the frequency domain.
\rev{As a result of the phase differencing operation with the \ac{los} path, the noise variables on different paths become correlated, with covariance equal to $\sigma^2_{\phi, {\rm LoS}} = \sigma_w^2 / (2G|A_{\rm LoS}|^2)$}. 
}

Computing phase differences \textit{cancels out} \ac{cfo} and \ac{po} without estimating them, since they are common to all propagation paths.
Despite the absence of \ac{cfo} and \ac{po} in \eq{eq:new-target-phase}, the estimation of the target's Doppler frequency remains non-trivial due to the presence of the undesired frequency term $v^{\rm tx} \left(\cos \xi_{\rm t} - \cos \eta \right)/\lambda$  caused by the transmitter movement, as discussed in \secref{sec:limitations}. 

\begin{algorithm}[t!]
\small
	\caption{\ourname{} algorithm.}
	\label{alg:asymov}
	\begin{algorithmic}[1]
		\rev{\REQUIRE Phase measurements $\phi_{\rm LoS}[k]$, $\phi_{\rm t}[k]$, $\phi_{s}[k]$, $s=1, \dots, S$.
		\ENSURE Estimates of the parameters $\widehat{f}_{\rm D, t}, \widehat{v}^{\rm tx}, \widehat{\eta}$.
        \FOR{$k=0, \dots,K-1$}
        \STATE $\tilde\phi_{i}[k] = \angle{\left(e^{j\phi_{i}[k]} / e^{j\phi_{\rm LoS}[k]}\right)}$ (\ac{cfo} and \ac{po} cancellation) 
        \STATE $\Delta_i[k] = \angle{\left(e^{j\tilde{\phi}_i[k]} / e^{j\tilde{\phi}_i[k-1]}\right)}$ (time domain differencing)
        \STATE $\tilde{\Delta}_{i}[k] = \Delta_{i}[k] / (2\pi T_k)$, $i\in \{{\rm t}, s\}$ (normalization)
        \ENDFOR
        \STATE $\Bar{\mathbf{\Delta}} = \text{RANSAC}(\tilde{\mathbf{\Delta}})$ (outlier rejection and averaging)
        \STATE $\widehat{f}_{\rm D, t}, \widehat{v}^{\rm tx}, \widehat{\eta}$ $ \leftarrow $  \alg{alg:alt-min} with input $\Bar{\mathbf{\Delta}}$, $\alpha_{\rm t}$, $\{\alpha_1,\dots, \alpha_S\}$.}
	\end{algorithmic}
\end{algorithm}

\vspace{-0.2cm}
\subsection{Time-domain phase differencing}\label{sec:phase-diff}

After \ac{cfo} and \ac{po} cancellation, the RX computes first-order time-domain phase differences 
as 
\begin{equation}\label{eq:phase-td}
    \Delta_i[k] = \angle{\left(e^{j\tilde{\phi}_i[k]} / e^{j\tilde{\phi}_i[k-1]}\right)}
\end{equation}

with $i\in\{{\rm t}, s\}$ and $k = 1, \dots ,K-1$. \rev{This step corresponds to line~$3$ of \alg{alg:asymov}.} 
The phase differences are 
\begin{align}
    & \Delta_{\rm t}[k] = 2\pi T_k\left(f_{\rm D, t}+\frac{v^{\rm tx}}{\lambda}\left( \cos \xi_{\rm t} - \cos \eta\right) \right)
    \rev{+ w''_{\rm t}[k]}
    , \label{eq:phi-trg} \\
    & \Delta_s[k] = 2\pi T_k\left(\frac{v^{\rm tx}}{\lambda} \left(\cos \xi_s - \cos \eta \right)\right)
    \rev{+ w''_s[k]}
    , \label{eq:phi-m}  
\end{align}
where we recall that $T_k = t_k- t_{k-1}$. 
\rev{The time-domain phase differences cancel out the path-dependent constant phase terms  \mbox{$\Omega_{s}$}  and \mbox{$\Omega_{{\rm t}} $}, which would otherwise prevent a comparison of the phases of the multiple paths.}
\rev{The time-domain differencing operation in \secref{sec:phase-diff} doubles the noise variance with respect to \eq{eq:phase_noise}, assuming the noise is uncorrelated and has constant variance across time steps $k$.
Hence, the variance of the noise in \eq{eq:phi-m} and \eq{eq:phi-trg} is $\sigma_{\Delta, i}^2 = 2\sigma_{\tilde{\phi}, i}^2 $ and $w''_i[k] \sim \mathcal{N}(0, \sigma_{\Delta, i}^2)$, while the covariances among different paths are equal to $2\sigma^2_{\phi, {\rm LoS}}$.
Note that the temporal differencing also introduces correlation between the noise variables on \textit{adjacent} values of the differences, \rev{i.e., $w''_i[k], w''_i[k-1]$ and $w''_i[k+1], w''_i[k]$.
The covariance among adjacent differences is $\sigma_{2\Delta, i}=-\sigma_{\tilde{\phi}, i}^2 $, while the one for non-adjacent ones is $\sigma_{k\Delta, i}=0$ with $k=3, \dots, K-1$.}}

In \eq{eq:phi-trg} and \eq{eq:phi-m}, we assume that the inter-frame spacing for channel estimation, $T_k$, is sufficiently small so that the phase change between two subsequent frames is smaller than $\pi$, to avoid ambiguity. 
From \eq{eq:phi-trg}, it can be seen that the noise-free phase differences $\Delta_i$ are upper bounded by $ 2\pi T_k (3f_{\rm max})$, with $i\in\{{\rm t}, s\}$ and $f_{\rm max}$ being the maximum Doppler shift caused by the TX (or the target). $f_{\rm max}$ is a system design parameter that can be set depending on the specific scenario and application.
To fulfill the assumption, it is sufficient to impose \mbox{$|\Delta_i|<\pi$} which yields \mbox{$T_k<1/(6f_{\rm max})$}. The choice of $T_k$ is further discussed in \secref{sec:num-results}. 
\rev{Furthermore, to account for occasional fades or blockages that would cause noisy phase shifts, we implement the outlier detection strategy described in \secref{sec:doppler-est}.}

\vspace{-0.2cm}
\subsection{\ac{aod}-based simplification}\label{sec:aoa-simp}

By inspecting \fig{fig:geometric_disp}, a key simplification can be made in \eq{eq:phi-trg} and \eq{eq:phi-m} noticing that \mbox{$\cos \xi_i = \cos \left(\eta-\alpha_i \right)= \cos \left(\alpha_i-\eta\right)$}, with $i\in \{{\rm t}, s\}$.
This substitution removes the dependency on the unknown and path-dependent angle $\xi_i$. The new expressions of the phase differences are
\begin{align}
    & \Delta_{\rm t}[k] = 2\pi T_k\left(f_{\rm D, t}+\frac{v^{\rm tx}}{\lambda}\left( \cos (\eta - \alpha_{\rm t}) - \cos \eta\right) \right)
    \rev{+ w''_{\rm t}[k]}
    , \label{eq:phi-trg-simple} \\
    & \Delta_s[k] = 2\pi T_k\left(\frac{v^{\rm tx}}{\lambda} \left(\cos (\eta - \alpha_s) - \cos \eta \right)\right)
    \rev{+ w''_s[k]}
    , \label{eq:phi-m-simple}  
\end{align}
where the new dependency on $\alpha_i-\eta$, for $i\in \{{\rm t}, s\}$, is easier to handle since the RX estimates $\alpha_i$ and the unknown term $\eta$ is independent of the propagation paths.
\rev{
For convenience, we further reformulate \eq{eq:phi-trg-simple} and \eq{eq:phi-m-simple} as
\begin{align}
    & \tilde{\Delta}_{\rm t}[k] = f_{\rm D, t}+\frac{v^{\rm tx}}{\lambda}\left( \cos (\eta - \alpha_{\rm t}) - \cos \eta\right) \rev{+ \tilde{w}_{\rm t}[k]}
    , \label{eq:phi-trg-norm} \\
    & \tilde{\Delta}_s[k] = \frac{v^{\rm tx}}{\lambda} \left(\cos (\eta - \alpha_s) - \cos \eta \right) \rev{+ \tilde{w}_s[k]}
    , \label{eq:phi-m-norm}  
\end{align}
where $\tilde{\Delta}_{i}[k] = \Delta_{i}[k] / (2\pi T_k)$, for $i\in \{{\rm t}, s\}$ (line~$4$ in \alg{alg:asymov}).}
\rev{Dividing the phase measurements by the inter-frame spacing makes the noise variance depend on the specific frame~$k$. 
Hence, the variances of the noise variables $\tilde{w}_{\rm t}[k]$ and $ \tilde{w}_s[k]$, termed $\sigma_{i, k}^2$, and their covariances, termed $\sigma_{i\ell, k}$, for $i, \ell \in \{{\rm t}, s\}$ are respectively
\begin{equation}
    \label{eq:sigma_noise}
    \sigma_{i,k}^2 =\frac{\sigma_w^2 \kappa_i}{4\pi^2GT_k^2}, \quad \sigma_{i\ell, k} = \frac{\sigma_w^2}{4\pi^2 G T_k^2 |A_{\rm LoS}|^2}.
\end{equation}
As discussed in \secref{sec:noise-pdf}, \eq{eq:sigma_noise} is based on the assumption that the \ac{snr} on the \ac{cir} is high.
To validate the use of \eq{eq:sigma_noise} in the derivation of the \ac{crb}, in \secref{sec:noise-pdf-sim} we performed simulations to assess that \eq{eq:sigma_noise} models the actual noise variance on the input measurements to \ourname{}.

}

Note that, since the RX estimates $\alpha_s$ and $\alpha_{\rm t}$, the number of unknowns is reduced from $S+4$, i.e., $f_{\rm D, t}, v^{\rm tx}, \eta, \xi_{\rm t}, \xi_1, \dots, \xi_S$, to $3$, i.e., $f_{\rm D, t}, v^{\rm tx}, \eta$.
\rev{This makes \eq{eq:phi-trg} and \eq{eq:phi-m}, for $s=1,\dots, S$, a set of $S+1$ equations with $3$ unknowns, which can be solved if the number of static multipath components satisfies $S \geq 2$, so that $S+1\geq 3$}. \rev{In the next section, we discuss the identifiability of our measurement model, then we detail our solver algorithm in \secref{sec:doppler-est}.}

\rev{
\subsection{Identifiability of the measurement model}\label{sec:identifiability}

To prove the identifiability of our model, we define the following vector quantities: the phase differences vector at the $k$-th frame, $\tilde{\mathbf{\Delta}}[k] = [\tilde{\Delta}_{\rm t}[k], \tilde{\Delta}_{1}[k], \dots, \tilde{\Delta}_{S}[k]]^{\mathsf{T}}$, vector $\bm{\nu} = [f_{\rm D, t}, v^{\rm tx}]^{\mathsf{T}}$,
and the $(S+1) \times 2$ matrix 
\begin{equation}\label{eq:gamma}
\displaystyle
    \bm{\Gamma} = \begin{bmatrix}
        1 & \lambda^{-1}\left(\cos (\eta - \alpha_{\rm t}) - \cos \eta\right)\\
        0 &  \lambda^{-1}\left(\cos (\eta - \alpha_{\rm 1}) - \cos \eta\right)\\
        \vdots & \vdots \\
        0 & \lambda^{-1}\left(\cos (\eta - \alpha_{S}) - \cos \eta\right)
    \end{bmatrix}.
\end{equation}

The measurement model can then be written as
\begin{equation}\label{eq:delta-model}
    \tilde{\mathbf{\Delta}}[k] = \mathbf{\Gamma}\bm{\nu} + \mathbf{w}[k],
\end{equation}
where $\mathbf{w}[k]$ is a complex-valued Gaussian noise vector with covariance matrix $\mathbf{C}_k$ whose elements on the \textit{diagonal} are the variances $\sigma_{i,k}^2$, while the off-diagonal elements are equal to $\sigma_{i\ell, k}$, as derived in \eq{eq:sigma_noise}.

\eq{eq:delta-model} is a system of $S+1$ equations in $3$ unknowns. 
To prove the identifiability of $f_{\rm D, t}, v^{\rm tx}, \eta$, we verify whether in the absence of noise, \eq{eq:delta-model} provides a unique set of measurements for each value assumed by the parameters. 
If $\eta$ and $v^{\rm tx}$ are uniquely retrieved from equations $2, \dots, S+1$, i.e., only those pertaining to the static multipath components, then $f_{\rm D, t}$ can be uniquely obtained from the first equation.
Therefore, we analyze the identifiability of $\eta$ and $v^{\rm tx}$ from equations $2, \dots, S+1$.
These equations are of the form of \eq{eq:phi-m-norm} and each of them differs due to the angles $\alpha_s$.
The coefficient $v^{\rm tx} / \lambda$ is common to all equations, so $v^{\rm tx}$ can be uniquely identified if $\eta$ can.
To ensure $\eta$ is obtained without ambiguity, we must have \textit{at least} $2$ equations among the $S$ equations involving static multipath components in \eq{eq:phi-m-norm} that do not degenerate into giving no information on $\eta$.
This happens if there exists \textit{at least} a pair of angles $\alpha_{s_1}, \alpha_{s_2}$, with indices $s_1, s_2$, for which
\begin{itemize}
    \item Condition 1: $\alpha_{s_1} \neq \alpha_{s_2}$
    \item  Condition 2: $\alpha_{s_1} \neq 2\eta$, $\alpha_{s_2} \neq 2\eta$.
\end{itemize}
Note that violating either of the above two conditions means that one equation becomes uninformative on $\eta$.
Since both $\eta$ and $\alpha_s$ lie in the interval $[-\pi, \pi]$ by construction, we do not account for multiples of $2\pi$ in the above conditions.
Our measurement model is hence identifiable except for a set of zero measure given by the values that violate the above conditions.
Note that it is sufficient to have $2$ angles $\alpha_s$ that satisfy the conditions. The latter \textit{do not need to be} verified for all $s$ to guarantee identifiability.
Therefore, the more static multipath components are available, the lower the probability of encountering a pathological case for which the model is non-identifiable.
In the solution algorithm detailed in the next section, we check that matrix $\mathbf{\Gamma}$ does not incur one of such cases during the iterations, hence avoiding ambiguity.
}

\subsection{Doppler frequency estimation}
\label{sec:doppler-est}
\rev{To increase the robustness of \ourname{} to outliers in the phase measurements, we pre-process them using the \ac{ransac}~\cite{fischler1981random} algorithm.
We use \ac{ransac} to obtain a robust estimate of the mean phases within the processing window, $\Bar{\Delta}_{i}, i\in \{{\rm t}, s\}$, since phase differences of the same path are assumed constant during the time interval $KT$. 
The mean phases are collected in vector $\Bar{\mathbf{\Delta}} = [\Bar{\Delta}_{\rm t}, \Bar{\Delta}_{1}, \cdots, \Bar{\Delta}_S]$ which contains the average values of $\tilde{\Delta}_i[k]$ for $k=1, \dots, K-1$, $i=t,1,\cdots,S$, excluding the samples classified as outliers (line~$6$ in \alg{alg:asymov}).}
Then, we formulate the following non-convex optimization problem
\begin{align}\label{eq:min-prob}
    \argmin_{f_{\rm D, t}, \eta, v^{\rm tx}} &\left\|\Bar{\mathbf{\Delta}} - \mathbf{\Gamma}\bm{\nu}\right\|_2^2
    \\
    \text{subject to}&\,\, v^{\rm tx} \geq 0, \,\, \eta \in [0, 2\pi).
\end{align}
In \eq{eq:min-prob}, the variables $f_{\rm D, t}, v^{\rm tx}$ only appear in vector~$\bm{\nu}$. 
Conversely, $\eta$ only appears in~$\mathbf{\Gamma}$.
Our solution algorithm leverages such decoupled structure to alternatively optimize over the pair of variables $f_{\rm D, t}, v^{\rm tx}$ and then over~$\eta$. 
We provide a thorough description of the algorithm next.

\subsubsection{Alternating minimization solution}
\label{sec:minimization}

Assume that an initial estimate of $\eta$ is available, denoted by $\widehat{\eta}$. Considering the latter a fixed constant, the problem in \eq{eq:min-prob} becomes a standard \ac{ls} problem with a non-negativity constraint on $v^{\rm tx}$.
\rev{Denote by $\bm{\Gamma}(\widehat{\eta})$ the matrix $\mathbf{\Gamma}$ in \eq{eq:gamma} computed using $\eta = \widehat{\eta}$.
By checking the rows of $\bm{\Gamma}(\widehat{\eta})$, one can determine whether the current estimate, $\widehat{\eta}$, violates the identifiability conditions discussed in the previous section.
Specifically, if $\bm{\Gamma}(\widehat{\eta})$ has $2$ or more non-zero equal rows, this means Condition~$1$ is violated and only one of the equal rows should be retained.
Instead, if $\bm{\Gamma}(\widehat{\eta})$ contains any number of all-zero rows, Condition~$2$ is violated for such equations and such rows should be removed from the matrix.
In the following, we assume that this verification of the identifiability conditions is performed at each iteration of the algorithm when a new $\widehat{\eta}$ is obtained.
The problem is solvable with a unique solution as long as $S\geq2$ static multipath components are available with angles that do not violate Conditions~$1$ and $2$.
}

The \ac{ls} problem is solved in closed form by computing
\begin{equation}\label{eq:nu-prime-sol}
    \widehat{\bm{\nu}} = \Pi \left(\mathbf{\Gamma}^{\dag}(\widehat{\eta})\Bar{\mathbf{\Delta}}\right),
\end{equation}
where the symbol $^{\dag}$ denotes the pseudo-inverse of a matrix, $\widehat{\bm{\nu}} = [\widehat{f}_{\rm D, t}, \widehat{v}^{\rm tx}]^{\mathsf{T}}$ contains the estimates of the target Doppler frequency and TX movement speed, and $\Pi\left(\bm{\nu}\right) = [f_{\rm D, t}, \max(0, v^{\rm tx})]^{\mathsf{T}}$ projects the \ac{ls} solution onto the feasible set with $v^{\rm tx} \geq 0$. The complexity of solving~\eq{eq:nu-prime-sol} is $\mathcal{O}(S^2)$ since the dimension of $\bm{\nu}$ is fixed to $2$.

Conversely, when variables $f_{\rm D, t}, v^{\rm tx}$ are fixed, the problem reduces to finding the $\eta$ that minimizes \eq{eq:min-prob}.
Optimization over $\eta$ can be efficiently carried out by a grid search in the bounded interval $[0, 2\pi)$. 
We construct grid \mbox{$\mathcal{G} = \left\{0, 2\pi /N_{\eta}, \dots, 2\pi (N_{\eta} - 1) / N_{\eta}\right\}$} where $N_{\eta}$ is the number of grid candidates and we solve  
\begin{equation}\label{eq:eta-grid}
\widehat{\eta} = \argmin_{\eta \in \mathcal{G}} \left\| \Bar{\mathbf{\Delta}} - \bm{\Gamma}\widehat{\bm{\nu}}  \right\|^2_2.
\end{equation}
This approach has a complexity $\mathcal{O}(N_{\eta} S)$, which gives a total complexity of $\mathcal{O}(S^2 + N_{\eta} S)$. This is computationally affordable, since $S$ is typically low in practice and $\eta$ has bounded support, so the grid can be made fine-grained without excessively increasing its size. 
While alternative methods based on gradient descent could converge to local stationary points of the non-convex function, grid search guarantees that our solution is as close to the optimal value as determined by the grid step $2\pi /N_{\eta}$, which can be made arbitrarily small.

The above discussion motivates \ourname{}'s alternating minimization approach to solving the original problem in \eq{eq:min-prob}. \rev{The minimization proceeds as detailed in \alg{alg:alt-min}, which is used in the last computational step of \ourname{}, as shown in line~$7$ of \alg{alg:asymov}.
In \alg{alg:alt-min}, $^{(n)}$ denotes quantities at the $n$-th iteration and $\epsilon$ is a small positive constant.}

\rev{
The proposed approach converges to an approximation of a local minimum of the cost function following the result in~\cite{yang2019inexact}, since: (i)~the cost function is non-increasing across iterations, as proved in Appendix~\ref{app:opt_proof}, (ii)~the optimization step to obtain $\hat{\bm{\nu}}$ is solved \textit{optimally} with \acs{ls},  (iii)~the optimization step to obtain $\hat{\eta}$ approximates the optimal value with an error of at most $2\pi / N_{\eta}$ (the grid step), and (iv)~the function $\mathbf{\Gamma}(\eta)$ is infinitely differentiable (smooth) and its derivative is
bounded, being the sum of two sine functions.
If $N_{\eta}$ is sufficiently large, the algorithm is an instance of inexact (due to the grid search) block-coordinate descent with tight error bounds, and is guaranteed to converge to a local minimum of the cost function~\cite{yang2019inexact}.
Notably, thanks to our initialization method based on a closed-form solution (\secref{sec:closed-form}), which already provides a quite accurate estimate of the globally optimal parameters, local convergence is sufficient in practice to obtain accurate Doppler estimation and occurs in a few steps. 
}
The overall complexity of the minimization is $\mathcal{O}(N_{\rm it}(S^2 + N_{\eta} S))$, where $N_{\rm it}$ is the number of iterations.
Below, we provide a method to initialize $\widehat{\eta}$.

\begin{algorithm}[t!]
\setstretch{1.2}
\small
	\caption{Alternating minimization algorithm.}
	\label{alg:alt-min}
	\begin{algorithmic}[1]
		\REQUIRE Phase measurements $\Bar{\mathbf{\Delta}}$, angles $\alpha_{\rm t}$, $\{\alpha_1,\dots, \alpha_S\}$.
		\ENSURE Estimates of the parameters $\widehat{f}_{\rm D, t}, \widehat{v}^{\rm tx}, \widehat{\eta}$.
        \STATE Set $n=0$, initialize $\widehat{\eta}^{(0)}$ using \eq{eq:eta-hat} and $\widehat{\eta}^{(-1)} = \infty$
        \WHILE{$|\widehat{\eta}^{(n)} - \widehat{\eta}^{(n-1)}| > \epsilon$ }
        \STATE Fix $\widehat{\eta}^{(n)}$ and obtain $\widehat{f}_{\rm D, t}^{(n+1)}, (\widehat{v}^{\rm tx})^{(n+1)}$ using \eq{eq:nu-prime-sol}
        \STATE Fix $\widehat{f}_{\rm D, t}^{(n+1)}, (\widehat{v}^{\rm tx})^{(n+1)}$ and obtain $\widehat{\eta}^{(n+1)}$ using \eq{eq:eta-grid}
        \STATE $n \leftarrow n+1$
        \ENDWHILE
	\end{algorithmic}
\end{algorithm}

\subsubsection{Closed form initialization of $\eta$} \label{sec:closed-form}

Consider \eq{eq:phi-m-norm} with $2$ phase measurements from static paths, i.e., $s\in \{1,2\}$. We denote by $\bar{\Delta}_{\rm t}$, $\bar{\Delta}_{\rm 1}$, and $\bar{\Delta}_{\rm 2}$ the time-averaged phase differences for the sensing path, and static paths $1$ and $2$. Using also \eq{eq:phi-trg-norm}, a system with $3$ equations in $3$ unknowns is attained, which can be solved to find an initialization of $\widehat{\eta}$. We provide the solution in Appendix~\ref{app:system-sol}, giving the full expressions of $\widehat{\eta}$ in \eq{eq:eta-hat}. 
\rev{As discussed in \secref{sec:identifiability}, the solution requires the following conditions to be met: (i)~$\alpha_i \neq 0, i \in \{1,2,\mathrm{t}\}$, (ii)~$\alpha_{i} \neq \alpha_{\ell}, i \neq \ell \in \{1,2,\mathrm{t}\}$, and (iii)~$\alpha_{i} \neq 2\widehat{\eta}, i \in \{1,2\}$. }
Note that violating conditions (i)~and (ii)~ correspond to degenerate scenarios in which the \ac{los} path is not available, or two static multipath components have the same \ac{aod}.  Condition (iii)~instead is violated if the \ac{aod} of one of the multipath components is equal to $2\widehat{\eta}$.

\section{Analysis and numerical simulations}\label{sec:num-results}
In this section, we derive the \ac{crb} on the estimation of the parameters $f_{{\rm D, t}},\eta,  v^{\rm tx}$, based on the measurement model in \eq{eq:phi-trg-norm} and \eq{eq:phi-m-norm}. Then, we describe our simulations and the obtained numerical results.
The \ac{crb} derivation differs from that of existing works for bi-static asynchronous \ac{isac}, e.g.,  \cite{wang2024windowing, wang2024clutter}, that do not consider the motion of TX and RX, since it is based on our new model in \eq{eq:delta-model}, which accounts for the inter-dependencies among paths introduced by the phase differencing operations.
\vspace{-2pt}

\subsection{Performance limits analysis}\label{sec:crb}

To assess bounds on the performance of the proposed Doppler frequency estimation method, we utilize the \ac{crb}~\cite{kay1993statistical}, which represents a lower bound on the error variance of the parameter estimates. To this end, we define the vector containing all parameters $\boldsymbol{\theta} = [f_{\rm D, t}, \eta, v^{\rm tx}]^{\mathsf{T}}$. In the \ac{crb} derivation, we assume no prior knowledge of the parameters $\boldsymbol{\theta}$ is available, and we consider each path's \ac{aod} to be known exactly since the latter is an input to our algorithm and must be obtained by other means. Moreover, to simplify the derivation and make the expressions more compact, we restrict to the case $S=2$, i.e., two static paths are available, and we consider only a single time frame~$k$.
\rev{Extending the bound to an arbitrary $K$ requires taking into account the covariances among adjacent temporal phase differences $\sigma_{2\Delta, i}$, given in \secref{sec:phase-diff}, and is omitted for conciseness.}

\rev{The likelihood of the vector measurement model in \eq{eq:delta-model} is a multivariate Gaussian
\begin{equation}\label{eq:likelihood}
    p(\tilde{\mathbf{\Delta}}[k]; \boldsymbol{\theta}) \sim \mathcal{N}\left( \mathbf{\Gamma}\bm{\nu}, \mathbf{C}_k\right)
\end{equation}
where the parameters $\bm{\theta}$ are contained in $\mathbf{\Gamma}\bm{\nu}$. $\mathbf{C}_k$ has been defined in \secref{sec:identifiability}, and contains the variances and covariances from \eq{eq:sigma_noise}.
Element $i,\ell$ of the \ac{fim}, $\mathbf{J}$, is computed as in chapter $3.9$, Eq.~$3.31$ of~\cite{kay1993statistical}
\begin{equation}\label{eq:fim}
    [\mathbf{J}]_{i, \ell} = \left[\frac{\partial \mathbf{\Gamma}\bm{\nu}}{\partial \theta_i}\right]^{\top} \mathbf{C}_k^{-1} \left[\frac{\partial \mathbf{\Gamma}\bm{\nu}}{\partial \theta_\ell}\right].
\end{equation}}
\rev{We omit the full expression of the \ac{fim} for conciseness, and directly obtain the \ac{crb} by inverting it and taking the first element on the diagonal, $[\mathbf{J}^{-1}]_{1,1}$. The \ac{crb} on the Doppler frequency of the target is
\begin{equation}\label{eq:crlb}
[\mathbf{J}^{-1}]_{1,1} = \frac{\sigma_w^2\zeta_{{\rm t}, 1, 2}}{128 \pi^2 G T_k^2 \sin^2 \left(\frac{\alpha_1}{2}\right)\sin^2\left(\frac{\alpha_1 - \alpha_2}{2}\right)\sin^2\left(\frac{\alpha_2}{2}\right)},
\end{equation}
where $\zeta_{{\rm t}, 1, 2}$  exhibits a complex dependency on $\kappa_{\rm t}, \kappa_1, \kappa_2$, and the \acp{aod}, $\alpha_{\rm t}, \alpha_1, \alpha_2$. Its full expression is given in Appendix~\ref{app:fim-crlb}, in \eq{eq:zeta}.
}
We observe that $[\mathbf{J}^{-1}]_{1,1} \rightarrow +\infty$ when $\alpha_1,\alpha_2 \rightarrow 2n\pi, n\in \mathbb{N}_0 $ and $\alpha_1 \rightarrow \alpha_2$. This makes two of the equations in \eq{eq:delta-model} equal, making the system underdetermined.
Note that \eq{eq:crlb} does not contain $v^{\rm tx}$ and $\eta$, making it independent of the TX velocity vector. This means that the magnitude or direction of the TX velocity does not influence the error bound on the Doppler frequency.

\subsection{Simulation setup}\label{sec:sim-setup}

To validate \ourname{} we perform simulations for $5$~GHz, $28$~GHz, and $60$~GHz carrier frequencies $f_c$, as representative of, e.g., IEEE~802.11ax, \ac{fr2} \mbox{5G-NR}, and IEEE~802.11ay systems, respectively. 
The coordinates of the target and static objects are randomly generated in a $d \times d$ empty mobility area, and the real \ac{cir} is obtained using \eq{eq:cir-continuous}. We simulate the transmission of the waveform used to perform channel estimation. For $f_c=60$~GHz, an IEEE~802.11ay TRN field made of complementary Golay sequences is transmitted, while for $f_c=5$~GHz and $f_c=28$~GHz, we employ a random sequence of \ac{ofdm} symbols modulated with BPSK. At the receiver, we add \ac{cfo}, \ac{po}, and channel noise according to different \ac{snr} values. The received signal is sampled with period $1/B$ and the \ac{cir} is estimated. After the peak detection operation, the phases of the \ac{cir} peaks and the noisy \ac{aod} estimates, for each available path, are used as input to our algorithm.
All the parameters are summarized in \tab{tab:init_param}. The maximum TX speed, $v_{\rm max}$, differs for each carrier frequency to reflect the typical speed of the corresponding use case, i.e., IEEE~802.11ax, \mbox{5G-NR} and IEEE~802.11ay. Finally, to consistently compare various simulation configurations, we set the inter-frame spacing $T_k$ to a fixed value $T_k = T=1/(6f_{\rm max})$, which is the maximum time gap allowed in the measurements to avoid phase ambiguity, as discussed in~\secref{sec:phase-diff}. 
We remark that the selected $T$ is compatible with existing hardware and typical packet transmission times in wireless networks. 
\rev{The residual \ac{cfo} is modeled as $f_{\rm o}(kT) \sim \mathcal{N}(0,\sigma^2_{\rm o})$ which depends on time index $k$ because the phase drift accumulates at each new transmission. $\sigma^2_{\rm o}$ is chosen such that the frequency shift in $1$~ms is between $\pm 0.1$ parts-per-million~(ppm) of the carrier frequency $f_c$~\cite{3GPP_cfo}.
For the \ac{po} we assume the most stochastic scenario modeled as $\psi_{\rm o}(kT) \sim \mathcal{U}(0, 2\pi)$.
Note that our method cancels out the \ac{cfo} and \ac{po} \textit{exactly}, hence it is not significantly affected by their magnitude.}   
The error on the \ac{aod} estimates is modeled as an additive Gaussian random variable with distribution $\mathcal{N}(0,\sigma_\alpha^2)$.

\begin{figure*} 
    \centering
    \subcaptionbox{\mbox{$\acs{snr}=\SI{0}{\decibel}$, $\sigma_\alpha=5$°, and $KT=\SI{16}{\milli\second}$.}\protect\label{fig:boxplot_path}}[0.29\textwidth]{\begin{tikzpicture}
\definecolor{black28}{RGB}{28,28,28}
\definecolor{color3}{HTML}{bc5090}
\definecolor{burlywood231175145}{RGB}{231,175,145}
\definecolor{darkgray176}{RGB}{176,176,176}
\definecolor{darkslategray583162}{RGB}{58,31,62}
\definecolor{color2}{HTML}{ffa600}
\definecolor{color1}{HTML}{003f5c}
\matrix[draw, fill=white, ampersand replacement=\&, 
inner sep=2pt, 
    ] at (5.3,2.75) {  
        \node [shape=rectangle, draw=black ,fill=color1, label=right:\footnotesize{$f_{\rm c}=\SI{60}{\giga\hertz}$}] {};\&
        \node [shape=rectangle, draw=black ,fill=color3, label=right:\footnotesize{$f_{\rm c}=\SI{28}{\giga\hertz}$}] {}; \&
        \node [shape=rectangle, draw=black ,fill=color2, label=right:\footnotesize{$f_{\rm c}=\SI{5}{\giga\hertz}$}] {};\&
        \node [shape=rectangle, draw=black ,fill=white, pattern=north west lines, pattern color= black, label=right:\footnotesize{JM}] {};\&
        \node [shape=diamond, scale=.6, draw=black ,fill=black, label=right:\footnotesize{\ourname{}}] {};\&
        \node [shape=circle, scale=.7, draw=black ,fill=black, label=right:\footnotesize{JM}] {};\\
    };
  \begin{axis}
    [
    boxplot/draw direction=y,
    ylabel={Error $\varepsilon_{f_{\rm D, t}}$},
    xlabel={No. static paths $S$},
    ylabel shift=-5 pt,
    ymin=-0.011, ymax=0.27,
    ytick={0,0.05,0.1,0.15,0.2,0.25,0.3,0.35,0.4,0.45,0.5},
    yticklabels={$0$,,$0.1$,,$0.2$,,$0.3$,,$0.4$,,$0.5$},
    yticklabel style = {/pgf/number format/fixed},
    cycle list={{color1},{color3},{color2}},
    xtick={1,2,3,4},
    ymajorgrids,
    xticklabels={$2$, $4$, $6$, $8$},
    xlabel style={font=\footnotesize}, ylabel style={font=\footnotesize}, ticklabel style={font=\footnotesize},
    /pgfplots/boxplot/box extend=0.25,
    height=4cm,
    width=5.5cm,
    xmin=0.25,
    xmax=4.75,
    tickpos=left,
    ]
\addplot+[
fill, draw=black,
boxplot prepared={
	median=0.0217881961868356,
	upper quartile=0.0720364244487098,
	lower quartile=0.0080768498566264,
	upper whisker=0.167696904202991,
	lower whisker=6.29203536962354e-06,
	draw position=0.7
},
] coordinates {};

\addplot+[
fill, draw=black,
boxplot prepared={
	median=0.0406720438031957,
	upper quartile=0.115922632352021,
	lower quartile=0.0151282379882445,
	upper whisker=0.267043656177837,
	lower whisker=5.0678727588245e-06,
	draw position=1
},
] coordinates {};

\addplot+[
fill, draw=black,
boxplot prepared={
	median=0.023711747488531,
	upper quartile=0.0467696151318186,
	lower quartile=0.0106712613576449,
	upper whisker=0.100911255680748,
	lower whisker=1.80625298492444e-06,
	draw position=1.3
},
] coordinates {};

\addplot+[
fill, draw=black,
boxplot prepared={
	median=0.0192619155077619,
	upper quartile=0.0609583780577954,
	lower quartile=0.00726067152772785,
	upper whisker=0.141503159210245,
	lower whisker=2.58324650174922e-06,
	draw position=1.7
},
] coordinates {};

\addplot+[
fill, draw=black,
boxplot prepared={
	median=0.0340604271812231,
	upper quartile=0.100636968745714,
	lower quartile=0.0135247360066007,
	upper whisker=0.231103065593007,
	lower whisker=3.2680679562568e-06,
	draw position=2
},
] coordinates {};

\addplot+[
fill, draw=black,
boxplot prepared={
	median=0.020142840669711,
	upper quartile=0.0389312968226864,
	lower quartile=0.00932355168778696,
	upper whisker=0.0832856560063894,
	lower whisker=1.58094873171027e-06,
	draw position=2.3
},
] coordinates {};

\addplot+[
fill, draw=black,
boxplot prepared={
	median=0.0175950928955592,
	upper quartile=0.0553932498975595,
	lower quartile=0.00684635696640779,
	upper whisker=0.128136686149308,
	lower whisker=2.90491478978199e-06,
	draw position=2.7
},
] coordinates {};

\addplot+[
fill, draw=black,
boxplot prepared={
	median=0.0319330147943253,
	upper quartile=0.0929746982315427,
	lower quartile=0.0130084654733979,
	upper whisker=0.212917778023623,
	lower whisker=2.60378381830639e-08,
	draw position=3
},
] coordinates {};

\addplot+[
fill, draw=black,
boxplot prepared={
	median=0.0190467095323598,
	upper quartile=0.0355898272839043,
	lower quartile=0.00874387611991555,
	upper whisker=0.0757477738290354,
	lower whisker=9.69856784617988e-07,
	draw position=3.3
},
] coordinates {};

\addplot+[
fill, draw=black,
boxplot prepared={
	median=0.0175743727256008,
	upper quartile=0.0540797504787786,
	lower quartile=0.00684587746779399,
	upper whisker=0.124725852550749,
	lower whisker=1.6070455270364e-06,
	draw position=3.7
},
] coordinates {};

\addplot+[
fill, draw=black,
boxplot prepared={
	median=0.0299077783775768,
	upper quartile=0.0894695446492537,
	lower quartile=0.0125327684648711,
	upper whisker=0.204618490518458,
	lower whisker=2.59768967411557e-06,
	draw position=4
},
] coordinates {};

\addplot+[
fill, draw=black,
boxplot prepared={
	median=0.018701954511472,
	upper quartile=0.0346154899517717,
	lower quartile=0.00850352357155533,
	upper whisker=0.0736726348935232,
	lower whisker=1.14405906078488e-06,
	draw position=4.3
},
] coordinates {};
\end{axis}

    \begin{axis}[
        xtick = {1,2,3},
        ymin = 2,
        ymax = 16,
        xmin=0.25,
        xmax=4.75,
        hide x axis,
        hide y axis,
        height=4cm,
        width=5.5cm
    ]
        \addplot [draw=black, dashed, mark options={solid, fill=black}, mark=diamond*] 
        coordinates {
            (1, 4.67445)
            (2, 8.40604)
            (3, 11.05151)
            (4, 12.580123)
        };
    \end{axis}

    \begin{axis}[
        xmin=0.25,
        xmax=4.75,
        ymin = 2,
        ymax = 16,
        hide x axis,
        axis y line*=right,
        ylabel={\footnotesize{Time [ms]}},
        ylabel shift = -5pt,
        height=4cm,
        width=5.5cm,
        ylabel style={font=\footnotesize},
        yticklabel style = {/pgf/number format/fixed, ticklabel style={font=\footnotesize}},
    ]
  \end{axis}
  
\end{tikzpicture}}\hspace{5pt}
    \subcaptionbox{$\acs{snr}=\SI{5}{\decibel}$, $\sigma_\alpha=5$°, and $S=2$.\protect\label{fig:boxplot_K_music}}[0.4\textwidth]{\input{plot/boxplot_music_int}}\subcaptionbox{$f_{\rm c}=\SI{60}{\giga\hertz}$, $KT=\SI{16}{\milli\second}$, and $S=2$.\protect\label{fig:boxplot_aoa_music}}[0.29\textwidth]{\begin{tikzpicture}
    \definecolor{black28}{RGB}{28,28,28}
    \definecolor{color2}{HTML}{ffa600}
    \definecolor{burlywood231175145}{RGB}{231,175,145}
    \definecolor{darkgray176}{RGB}{176,176,176}
    \definecolor{darkslategray583162}{RGB}{58,31,62}
    \definecolor{color3}{HTML}{bc5090}
    \definecolor{color1}{HTML}{003f5c}
    \definecolor{color4}{HTML}{ff7c43}
    \definecolor{color5}{HTML}{ff7c43}
    
    \matrix[draw,  inner sep=2pt, fill=white, column sep=0em, ampersand replacement=\&] at (1.8,2.73) {  
        \node [shape=rectangle, draw=black, fill=white, pattern=north west lines, pattern color=color4, label=right:\footnotesize{JM}] {}; \&
        \node [shape=rectangle, draw=black ,fill=color1, label=right:\footnotesize{$\sigma_\alpha=1$°}] {}; \&
        \node [shape=rectangle, draw=black ,fill=color3, label=right:\footnotesize{$\sigma_\alpha=3$°}] {}; \&
        \node [shape=rectangle, draw=black ,fill=color2, label=right:\footnotesize{$\sigma_\alpha=5$°}] {}; \\
    };
    \tikzset{every mark/.append style={scale=0.9}}
    
      \begin{axis}
        [
        boxplot/draw direction=y,
        ylabel={Error $\varepsilon_{f_{\rm D, t}}$},
        ylabel shift = -5 pt,
        xlabel={SNR [dB]},
        ymin=-0.011, ymax=.29,
        yticklabel style = {/pgf/number format/fixed},
        cycle list={{color4},{color1},{color3},{color2}},
        xtick={1,2,3,4},
        ymajorgrids,
        xticklabels={$0$, $5$, $10$, $20$},
        xlabel style={font=\footnotesize}, ylabel style={font=\footnotesize}, ticklabel style={font=\footnotesize},
        /pgfplots/boxplot/box extend=0.125,
        height=4cm,
        width=5.5cm,
        ]
		\addplot+[
		fill=white, pattern=north west lines, 
        pattern color=color4, draw=black,
		boxplot prepared={
			median=0.264558649135578,
			upper quartile=0.462297732689348,
			lower quartile=0.128502880415872,
			upper whisker=0.896900745205922,
			lower whisker=0.000150890523746336,
			draw position=0.7
		},
		] coordinates {};

        \addplot+[
    	fill, draw=black,
    	boxplot prepared={
    		median=0.0154576306992317,
    		upper quartile=0.0686669131923394,
    		lower quartile=0.00539352872180648,
    		upper whisker=0.163500826009092,
    		lower whisker=7.0050210773741e-07,
    		draw position=0.9
    	},
    	] coordinates {};

        \addplot+[
    	fill, draw=black,
    	boxplot prepared={
    		median=0.0155451234745592,
    		upper quartile=0.0682233042649768,
    		lower quartile=0.00547051049702431,
    		upper whisker=0.162325083363619,
    		lower whisker=3.73393479349936e-07,
    		draw position=1.1
    	},
    	] coordinates {};
    		
    	\addplot+[
    	fill, draw=black,
    	boxplot prepared={
    		median=0.01672361308297,
    		upper quartile=0.0696783429756769,
    		lower quartile=0.00586347937895033,
    		upper whisker=0.165379007099074,
    		lower whisker=5.30904771497057e-06,
    		draw position=1.3
    	},
    	] coordinates {};
    
        \addplot+[
        fill=white, pattern=north west lines, 
        pattern color=color4, draw=black,
        boxplot prepared={
            median=0.239298814213977,
            upper quartile=0.414410413326405,
            lower quartile=0.116304501960479,
            upper whisker=0.798115396553603,
            lower whisker=1.89703753545132e-05,
            draw position=1.7
        },
        ] coordinates {};
    
        \addplot+[
    	fill, draw=black,
    	boxplot prepared={
    		median=0.00550940749797819,
    		upper quartile=0.0150083569393184,
    		lower quartile=0.00239475234014292,
    		upper whisker=0.0339286925042906,
    		lower whisker=9.84102025433151e-07,
    		draw position=1.9
    	},
    	] coordinates {};
    
        \addplot+[
    	fill, draw=black,
    	boxplot prepared={
    		median=0.00608029616055286,
    		upper quartile=0.0169595261585608,
    		lower quartile=0.00261099271130958,
    		upper whisker=0.0384229413417199,
    		lower whisker=6.37243670873998e-07,
    		draw position=2.1
    	},
    	] coordinates {};
    
    	\addplot+[
    	fill, draw=black,
    	boxplot prepared={
    		median=0.0070367689299547,
    		upper quartile=0.0194341803501735,
    		lower quartile=0.0030400799993659,
    		upper whisker=0.0440209102889859,
    		lower whisker=4.76333333926194e-06,
    		draw position=2.3
    	},
    	] coordinates {};
        
        \addplot+[
        fill=white, pattern=north west lines, 
        pattern color=color4, draw=black,
        boxplot prepared={
            median=0.189461332795162,
            upper quartile=0.332447799139231,
            lower quartile=0.0915771064039566,
            upper whisker=0.634365968174339,
            lower whisker=5.91906431295777e-05,
            draw position=2.7
        },
        ] coordinates {};
    
        \addplot+[
    	fill, draw=black,
    	boxplot prepared={
    		median=0.00366952283422067,
    		upper quartile=0.00951430308180164,
    		lower quartile=0.0015405223683254,
    		upper whisker=0.0214593412499629,
    		lower whisker=3.82655024734711e-07,
    		draw position=2.9
    	},
    	] coordinates {};
    
        \addplot+[
    	fill, draw=black,
    	boxplot prepared={
    		median=0.00448092889292549,
    		upper quartile=0.0112847390944868,
    		lower quartile=0.00190838358082867,
    		upper whisker=0.0252834232452499,
    		lower whisker=1.02475369679937e-06,
    		draw position=3.1
    	},
    	] coordinates {};
    
    	\addplot+[
    	fill, draw=black,
    	boxplot prepared={
    		median=0.00550534213180466,
    		upper quartile=0.0140270477564603,
    		lower quartile=0.00234640682271393,
    		upper whisker=0.0315242729144843,
    		lower whisker=1.9994498381282e-07,
    		draw position=3.3
    	},
    	] coordinates {};
    
        \addplot+[
        fill=white, pattern=north west lines, 
        pattern color=color4, draw=black,
        boxplot prepared={
            median=0.161738487026758,
            upper quartile=0.276983955078314,
            lower quartile=0.0767970570063543,
            upper whisker=0.534173987885451,
            lower whisker=3.32550280162903e-05,
            draw position=3.7
        },
        ] coordinates {};
    
        \addplot+[
    	fill, draw=black,
    	boxplot prepared={
    		median=0.00234545075188944,
    		upper quartile=0.00650597654644193,
    		lower quartile=0.000889572422121709,
    		upper whisker=0.0149065369894751,
    		lower whisker=5.62471484023397e-07,
    		draw position=3.9
    	},
    	] coordinates {};
    
        \addplot+[
    	fill, draw=black,
    	boxplot prepared={
    		median=0.0034330608628395,
    		upper quartile=0.00837384244146687,
    		lower quartile=0.00140028815248993,
    		upper whisker=0.0188339812680375,
    		lower whisker=1.58738690882127e-07,
    		draw position=4.1
    	},
    	] coordinates {};

    	\addplot+[
    	fill, draw=black,
    	boxplot prepared={
    		median=0.00467106542427099,
    		upper quartile=0.0110272106448026,
    		lower quartile=0.00182340526897922,
    		upper whisker=0.0247990181291291,
    		lower whisker=1.22537528435736e-07,
    		draw position=4.3
    	},
    	] coordinates {};
	\end{axis}

    \begin{axis}
        [
        boxplot/draw direction=y,
        ymin=-0.011, ymax=.29,
        yticklabel style = {/pgf/number format/fixed},
        xticklabel = \empty,
        yticklabel = \empty,
        xlabel style={font=\footnotesize}, ylabel style={font=\footnotesize}, ticklabel style={font=\footnotesize},
        /pgfplots/boxplot/box extend=0.125,
        height=4cm,
        width=5.5cm,
        ticks=none
        ]
    \end{axis}
\end{tikzpicture}}
    \caption{ Target Doppler frequency estimation error varying the number of available static scatterers, the duration of the processing window, the \ac{snr}, and the \acp{aoa} measurements error, with $f_{\rm c}=\SI{60}{\giga\hertz}$, $f_{\rm c}=\SI{28}{\giga\hertz}$, and $f_{\rm c}=\SI{5}{\giga\hertz}$.}
    \vspace{-4mm}
    \label{fig:boxplots}
\end{figure*}

\begin{table}[t!]
\footnotesize
    \centering
    \caption{Simulation parameters used in the numerical results. $\mathcal{U}(a, b)$ is the uniform distribution in the interval $[a, b]$. Values in curly brackets refer to parameters used for $60$, $28$, and $5$~GHz carriers.}
    \begin{tabular}{lcc}
       \toprule
       Target Doppler frequency [Hz] & $f_{\rm D, t}$     & $\pm\mathcal{U}(f_{\rm min}, f_{\rm max})$ \\
       Transmitter velocity [m/s] & $v^{\rm tx}$       & $\mathcal{U}(v_{\rm min}, v_{\rm max})$ \\
       Min. TX/target Doppler freq. [Hz]& $f_{\rm min}$ & $100$\\
       Min. TX/target velocity [m/s] & $v_{\rm min}$ & $0.5$ \\
       Max. TX/target Doppler freq. [kHz]& $f_{\rm max}$ & \{$1,0.93,0.3$\}\\
       Max. TX/target velocity [m/s] & $v_{\rm max}$ & \{$5,10,20$\} \\ 
       Mobility area side [m] & $d$ & \{$20,50,100$\} \\
       Channel estimation period [ms] & $T$ & \{$0.166,0.178,0.5$\}\\
       Bandwidth [GHz] & $B$ & \{$1.76, 0.4, 0.16$\}\\
       Subcarrier spacing [kHz]& $-$ & \{$\:\_\:,120,78.125$\} \\  
       \ac{cfo} time difference stand. dev. [MHz]& $\sigma_{\rm o}$ & \{$0.22,0.12,0.02$\}\\
       \bottomrule
    \end{tabular}
    \label{tab:init_param}
\end{table}

\rev{\subsection{Noise distribution model and CRB validation}\label{sec:noise-pdf-sim}

In \fig{fig:phase_error_dist}, we compare the empirical distribution of the input measurements to \alg{alg:alt-min} obtained from our simulations with the theoretical one under the assumption of high \ac{snr} on the \ac{cir}. 
The theoretical distribution is Gaussian with zero mean and variance $\sigma_{i, k}^2$, $i\in \{{\rm t}, s\}$, which is given in \eq{eq:sigma_noise}.
Note that the input measurements to \alg{alg:alt-min} are no longer phases due to the division by $2\pi T_k$, and their unit of measure is s$^{-1}$.
The empirical distribution is obtained from $1000$ simulations with an \ac{snr} of $5$~dB on the RX signal, and the IEEE~802.11ay scenario, with $60$~GHz  carrier frequency.
Since the channel estimation process in IEEE~802.11ay involves combining multiple repetitions of complementary Golay sequences~\cite{pegoraro2024rapid}, the \ac{snr} gain of the channel estimation process is not straightforward to obtain.
Therefore, we empirically computed the value of $G$, estimating $\mathrm{SNR}_h$ by comparing the estimated and the ground truth, noiseless \ac{cir}.
\fig{fig:phase_error_target} shows that the empirical noise distribution matches well with the theoretical one, validating our theoretical analysis of the noise distribution and the subsequent \ac{crb}. 
In \fig{fig:crb-comparison}, we show the \ourname{} \ac{rmse} and the related square-root \ac{crb}, $\sqrt{[\mathbf{J}]^{-1}_{1,1}}$, when transmitting at $f_c=28$~GHz with an aggregation window of $K=2$ samples. The \ac{rmse} of \ourname{} does not reach the \ac{crb}, meaning it is not efficient, but scales similarly when increasing the \ac{snr}, decreasing with the same slope. 
}

\begin{figure}
    \centering
    \begin{subfigure}{0.49\columnwidth}
        \centering
        \input{plot/phase_error_distribution_target}
        \caption{}
\label{fig:phase_error_target}
    \end{subfigure}
    \begin{subfigure}{0.49\columnwidth}
        \centering
\begin{tikzpicture}


\definecolor{darkgrey176}{RGB}{176,176,176}
\definecolor{darkorange25512714}{RGB}{255,127,14}
\definecolor{lightgrey204}{RGB}{204,204,204}
\definecolor{steelblue31119180}{RGB}{31,119,180}
\definecolor{color1}{HTML}{003f5c}
\definecolor{color2}{HTML}{bc5090}

\matrix[draw, fill=white, ampersand replacement=\&, inner sep=1pt, column sep=0.5mm] at (1.45,2.99) {  
    \node[label=right:\scriptsize{\ourname{}}] 
    {
        \begin{tikzpicture}
            \draw[color=color1, line width=1.5pt] (0,0) -- (0.4,0);
        \end{tikzpicture}
    }; \&

    \node[label=right:\scriptsize{$\sqrt{{\rm CRB}}$}] 
    {
        \begin{tikzpicture}
            \draw[dashed, color=color2, line width=1.5pt] (0,0) -- (0.4,0);
        \end{tikzpicture}
    };\\
};

\begin{axis}[
legend cell align={left},
legend style={fill opacity=0.8, draw opacity=1, text opacity=1, draw=lightgrey204, font=\footnotesize},
log basis y={10},
tick align=inside,
tick pos=left,
x grid style={darkgrey176},
xlabel={SNR [dB]},
grid=both,
xmin=-6.25, xmax=21.25,
xtick style={color=black},
y grid style={darkgrey176},
ylabel={RMSE [Hz]},
ymajorgrids,
ymin=168.601777105478, ymax=10000,
ymode=log,
minor y tick num=9, 
ytick style={color=black},
major grid style={black},
minor grid style={black},
xlabel style={font=\footnotesize}, ylabel style={font=\footnotesize}, ticklabel style={font=\footnotesize},
height=4cm,
width=4.5cm,
]
\addplot [semithick, color1, line width=1.5pt]
table {%
-5 8217.37000807465
-4 7380.52655814698
-3 6726.71647106745
-2 5999.08557979134
-1 5156.76780213356
0 4612.59468916251
1 4016.5377422371
2 3526.44141381512
3 2948.19932752305
4 2591.89262411312
5 2347.74441309431
6 2023.29932326386
7 1830.83624543732
8 1623.95440480694
9 1448.67265044553
10 1313.57406847419
11 1125.19149745989
12 1023.50724977692
13 881.418771091026
14 804.373305268538
15 729.04514952952
16 647.142225231764
17 566.579945047249
18 482.256196164901
19 442.277179448213
20 399.95425723298
};
\addplot [thick, color2, line width=1.5pt, dashed]
table {%
-5 3604.06623199403
-4 3222.40714727028
-3 2864.50830809331
-2 2572.12858007583
-1 2296.9788979699
0 2032.76289811607
1 1801.05092380106
2 1622.83569405917
3 1441.58770663403
4 1265.62333781965
5 1138.58635156833
6 1026.78269078755
7 900.156803363042
8 801.532374356761
9 717.107448069842
10 644.056552465687
11 573.773311611409
12 508.065302845377
13 454.699745447797
14 402.820821439759
15 362.873815837472
16 323.801257882749
17 287.895066188633
18 253.637686996272
19 227.879854323328
20 203.467998839599
};
\end{axis}

\end{tikzpicture}
        \vspace{-3mm}
        \caption{}
\label{fig:crb-comparison}
    \end{subfigure}
    \caption{\rev{In (a), the empirical measurements noise distribution (histogram) compared to the theoretical one, i.e., a Gaussian with zero mean and variance $\sigma_{i, k}^2$ from \eq{eq:sigma_noise} (purple dashed line), with $f_{\rm c} = 60$~GHz, SNR$=20$~dB, $\sigma_{\alpha}=5$°, $KT=16$~ms, and $S=2$. In (b), the Doppler estimation \ac{rmse} of \ourname{} compared to the square-root\ac{crb} with $f_c=28$~GHz and $K=2$.}}
    \label{fig:phase_error_dist}
\end{figure}

\subsection{State-of-the-art comparison}
\label{sec:jump+music}
To validate \ourname{}, we compare it with Doppler estimation algorithms used in asynchronous bistatic \ac{isac} systems. 
 As stated in \secref{sec:related}, to the best of our knowledge, no other \ac{isac} method tackles Doppler estimation with asynchronous and \textit{moving} devices. Therefore, we define a benchmark algorithm that applies the technique proposed in JUMP~\cite{pegoraro2024jump} to remove the phase offsets, followed by the \ac{music} algorithm to estimate the target Doppler frequency~\cite{schmidt1986multiple}. We call this benchmark \ac{jm}.
Since \ac{jm} does not account for the TX movement, it is biased by its Doppler frequency shift as discussed in \secref{sec:limitations}. The computational complexity of \ac{jm} is $\mathcal{O}(K^3)$ due to the eigenvalue decomposition of the covariance matrix of the channel measurements in \ac{music}.

\subsection{Doppler frequency estimation performance}\label{sec:doppler-freq-perf}

\rev{Since the possible Doppler frequencies vary for each considered $f_c$, to provide a fair comparison within different setups, we evaluate the performance of \ourname{} against \ac{jm} in terms of normalized absolute estimation error, defined as $\varepsilon_{f_{\rm D, t}} = |f_{\rm D, t} - \widehat{f}_{\rm D, t}| / |f_{\rm D, t}|$.}

\subsubsection{Number of static scatterers}
\label{sec:npath}
The number of available static scatterers $S$ is of prime importance for the proposed algorithm, as detailed in \secref{sec:doppler-est}. 
\rev{A minimum of $S=2$ static paths are needed for the algorithm to admit a solution, whereas $S>2$ increases its robustness to noise and reduces the probability of encountering one of the pathological cases in which the \ac{aod} of the paths violate the identifiability conditions in~\secref{sec:identifiability}.  }

This is confirmed by the results in \fig{fig:boxplot_path}, it is shown that the median estimation error and its spread are reduced if more static scatterers are available. For every considered carrier frequency, the median error lies below $5$\% of the actual Doppler frequency, even with \mbox{${\rm SNR}=0$~dB}. We recall that each scatterer adds one equation to the problem in \eq{eq:min-prob} without increasing the number of unknowns. This improves robustness at the cost of higher complexity. In \fig{fig:boxplot_path}, black diamonds represent the average time (across all $f_{\rm c}$) to solve the minimization problem on an Intel Xeon Gold 6252N CPU. 

\subsubsection{Processing window duration}
\label{sec:res_interval}
In \fig{fig:boxplot_K_music}, we evaluate the impact of averaging phase measurements over a longer aggregation window. We underline that, since $T$ differs for each carrier frequency, for a fixed observation interval $KT$, every considered $f_{\rm c}$ entails a different number of collected phases and \ac{aod} estimates. Increasing the number of considered frames $K$, while keeping $T$ constant, clearly improves robustness to noise. However, the window duration can not be made arbitrarily large but has to be tuned depending on the dynamicity of the environment to ensure that the channel parameters remain constant through the whole observation interval $KT$.
With $KT=\SI{32}{\milli\second}$, our method provides a median error below $1$\% of the true Doppler frequency for every considered $f_{\rm c}$.
On the other hand, \ac{jm} is less dependent on an increasing number of considered frames because the error is dominated by the bias described in \secref{sec:jump+music}. 
In addition, \fig{fig:boxplot_K_music} provides a comparison between the time needed to estimate the target Doppler frequency by \ourname{} and \ac{jm}, represented by black diamonds and circles, respectively. As described in \secref{sec:minimization}, \ourname{}'s computational complexity is independent of $K$, while for \ac{jm} it is $\mathcal{O}(K^3)$. Even for the smallest $K$, \ourname{} is over $100$ times faster than \ac{jm}.

\subsubsection{Measurements error}
\label{sec:res_snr}
In \fig{fig:boxplot_aoa_music}, we show the Doppler frequency estimation error as a function of the \ac{snr}, and of the noise affecting \ac{aod} measurements ($\sigma_{\alpha}$).
Lower \ac{snr} values greatly increase the estimation error spread, despite its median remaining below $0.02$. 
\fig{fig:boxplot_aoa_music} also shows that $\widehat{f}_{\rm D,t}$ is slightly affected by the \ac{aod} error, which means that the \ac{snr} level has a much stronger impact. Even though \ac{jm} improves with higher \ac{snr} levels, \ourname{} still outperforms it due to its capability of separating the target and the device Doppler.

Although our work focuses on Doppler frequency estimation, \ourname{} also estimates $\eta$ and $v^{\rm tx}$. In \fig{fig:speed_eta}, we show the normalized estimation errors for $\eta$ and $v^{\rm tx}$, which are respectively evaluated as \mbox{$\varepsilon_\eta = |\eta - \widehat{\eta}| / |\eta|$} and \mbox{$\varepsilon_{v^{\rm tx}}=|v^{\rm tx} - \widehat{v}^{\rm tx}| / |v^{\rm tx}|$}, where $\widehat{\eta}$ and $\widehat{v}^{\rm tx}$ represent the estimated TX motion direction angle and the estimated TX speed. 

\subsubsection{CIR measurement interval analysis} 
\label{sec:res_T}
As discussed in \secref{sec:phase-diff}, $T$ has to be sufficiently small to prevent ambiguity in the phase measurements. However, decreasing $T$ makes the phase differences more sensitive to noise due to the structure of \eq{eq:sigma_noise}. So, a suitable tradeoff has to be identified.  
The impact of an increasing $T$ on the estimation error due to the occurrence of phase ambiguities is evaluated in \fig{fig:average_error_sim}. Here, $T$ starts from the maximum value for which phase ambiguities are prevented. As expected, increasing $T$ beyond this value causes a growth in the estimation error $\varepsilon_{f_{\rm D, t}}$. \fig{fig:average_error_sim} is obtained for $f_{\rm c}=60$~GHz by way of example, but similar results hold for other carrier frequencies. 

\begin{figure}[t!]
    \centering
    \begin{subfigure}{\columnwidth}
        \centering
        \hspace{1cm}
        \begin{tikzpicture}
\definecolor{color2}{HTML}{bc5090}
\definecolor{color3}{HTML}{ffa600}
\definecolor{color1}{HTML}{003f5c}
\definecolor{black28}{RGB}{28,28,28}
\definecolor{burlywood231175145}{RGB}{231,175,145}
\definecolor{darkgray176}{RGB}{176,176,176}
\definecolor{darkslategray583162}{RGB}{58,31,62}
\matrix[draw, fill=white, ampersand replacement=\&, 
inner sep=2pt, 
    ] at (1.6,2.8) {  
        \node [shape=rectangle, draw=black ,fill=color1, label=right:\scriptsize{$f_{\rm c}=\SI{60}{\giga\hertz}$}] {};\&
        \node [shape=rectangle, draw=black ,fill=color2, label=right:\scriptsize{$f_{\rm c}=\SI{28}{\giga\hertz}$}] {}; \&
        \node [shape=rectangle, draw=black ,fill=color3, label=right:\scriptsize{$f_{\rm c}=\SI{5}{\giga\hertz}$}] {};\\
    };
\end{tikzpicture}
        \vspace{0.05cm}
    \end{subfigure}
    \begin{subfigure}{0.49\columnwidth}
        \centering
        \begin{tikzpicture}
\definecolor{black28}{RGB}{28,28,28}
\definecolor{color2}{HTML}{bc5090}
\definecolor{burlywood231175145}{RGB}{231,175,145}
\definecolor{darkgray176}{RGB}{176,176,176}
\definecolor{darkslategray583162}{RGB}{58,31,62}
\definecolor{color3}{HTML}{ffa600}
\definecolor{color1}{HTML}{003f5c}
  \begin{axis}
    [
    boxplot/draw direction=y,
    ylabel={Error $\varepsilon_\eta$},
    xlabel shift = -1 pt,
    ylabel shift = -5 pt,
    xlabel={SNR [dB]},
    ymin=-0.08, ymax=0.85,
    ytick={0,0.25,0.5,0.75,1,1.25,1.5,1.75,2},
    yticklabels={$0$,$0.25$,$0.5$,$0.75$,$1$,,$1.5$,,$2$},
    yticklabel style = {/pgf/number format/fixed},
    cycle list={{color1},{color2},{color3}},
    xtick={1,2,3,4},
    ymajorgrids,
    xticklabels={ $0$, $5$, $10$, $20$},
    xlabel style={font=\footnotesize}, ylabel style={font=\footnotesize}, ticklabel style={font=\footnotesize},
    /pgfplots/boxplot/box extend=0.25,
    height=3.5cm,
    width=4.8cm,
    ]

    \addplot+[
	fill, draw=black,
	boxplot prepared={
		median=0.0743237695953792,
		upper quartile=0.206135291606201,
		lower quartile=0.0250469418675565,
		upper whisker=0.476396726237431,
		lower whisker=1.79397713176381e-05,
		draw position=0.7
	},
	] coordinates {};

	\addplot+[
	fill, draw=black,
	boxplot prepared={
		median=0.104718867510952,
		upper quartile=0.259587035126255,
		lower quartile=0.0360023918413062,
		upper whisker=0.594728137478109,
		lower whisker=1.96737785486505e-05,
		draw position=1
	},
	] coordinates {};

	\addplot+[
	fill, draw=black,
	boxplot prepared={
		median=0.0942580057880106,
		upper quartile=0.221637987450462,
		lower quartile=0.038496562844569,
		upper whisker=0.495060916712417,
		lower whisker=3.04886629104449e-05,
		draw position=1.3
	},
	] coordinates {};

	\addplot+[
	fill, draw=black,
	boxplot prepared={
		median=0.0450211177596119,
		upper quartile=0.125664403392813,
		lower quartile=0.0161710811733704,
		upper whisker=0.289850707488633,
		lower whisker=9.76869410612856e-06,
		draw position=1.7
	},
	] coordinates {};

	\addplot+[
	fill, draw=black,
	boxplot prepared={
		median=0.0371555785719232,
		upper quartile=0.0867804070636233,
		lower quartile=0.0148192363785734,
		upper whisker=0.19449449517215,
		lower whisker=1.07370627173695e-06,
		draw position=2
	},
	] coordinates {};

	\addplot+[
	fill, draw=black,
	boxplot prepared={
		median=0.0648416570197363,
		upper quartile=0.149927138937161,
		lower quartile=0.0267686447896065,
		upper whisker=0.334344758711311,
		lower whisker=6.92600327702014e-06,
		draw position=2.3
	},
	] coordinates {};

	\addplot+[
	fill, draw=black,
	boxplot prepared={
		median=0.0374447116175446,
		upper quartile=0.098088254183698,
		lower quartile=0.013801602610322,
		upper whisker=0.224253179369207,
		lower whisker=1.20176316812749e-06,
		draw position=2.7
	},
	] coordinates {};

	\addplot+[
	fill, draw=black,
	boxplot prepared={
		median=0.030360415596439,
		upper quartile=0.0719903182875164,
		lower quartile=0.0117818726118621,
		upper whisker=0.162137958831305,
		lower whisker=9.44550295797736e-06,
		draw position=3
	},
	] coordinates {};

	\addplot+[
	fill, draw=black,
	boxplot prepared={
		median=0.0505060459984735,
		upper quartile=0.116677070241504,
		lower quartile=0.0201921765211281,
		upper whisker=0.261360984284302,
		lower whisker=3.55916375963035e-07,
		draw position=3.3
	},
	] coordinates {};

	\addplot+[
	fill, draw=black,
	boxplot prepared={
		median=0.0322717892985155,
		upper quartile=0.0825059678033986,
		lower quartile=0.0119252386855119,
		upper whisker=0.188339290244024,
		lower whisker=3.12479776449224e-06,
		draw position=3.7
	},
	] coordinates {};

	\addplot+[
	fill, draw=black,
	boxplot prepared={
		median=0.0248033436345417,
		upper quartile=0.0607037407092333,
		lower quartile=0.00971105728880997,
		upper whisker=0.13704325543688,
		lower whisker=3.24329299468213e-06,
		draw position=4
	},
	] coordinates {};

	\addplot+[
	fill, draw=black,
	boxplot prepared={
		median=0.0414343347344155,
		upper quartile=0.0975664981202827,
		lower quartile=0.016023747407107,
		upper whisker=0.219831707819459,
		lower whisker=7.01273735497963e-06,
		draw position=4.3
	},
	] coordinates {};

    \end{axis}
\end{tikzpicture}
        \vspace{-.2cm}
        \caption{Normalized $\eta$ estimation error.}
        \label{fig:boxplot_eta}
    \end{subfigure}
    \begin{subfigure}{0.49\columnwidth}
        \centering
        \begin{tikzpicture}
\definecolor{color2}{HTML}{bc5090}
\definecolor{color3}{HTML}{ffa600}
\definecolor{color1}{HTML}{003f5c}
\definecolor{black28}{RGB}{28,28,28}
\definecolor{burlywood231175145}{RGB}{231,175,145}
\definecolor{darkgray176}{RGB}{176,176,176}
\definecolor{darkslategray583162}{RGB}{58,31,62}
  \begin{axis}
    [
    boxplot/draw direction=y,
    ylabel={Error $\varepsilon_{v^{\rm tx}}$},
    xlabel shift = -1 pt,
    ylabel shift = -5 pt,
    xlabel={SNR [dB]},
    ymin=-0.08, ymax=0.85,
    ytick={0,0.25,0.5,0.75,1,1.25,1.5,1.75,2},
    yticklabels={$0$,$0.25$,$0.5$,$0.75$,$1$,,$1.5$,,$2$},
    yticklabel style = {/pgf/number format/fixed},
    cycle list={{color1},{color2},{color3}},
    xtick={1,2,3,4},
    ymajorgrids,
    xticklabels={ $0$, $5$, $10$, $20$},
    xlabel style={font=\footnotesize}, ylabel style={font=\footnotesize}, ticklabel style={font=\footnotesize},
    /pgfplots/boxplot/box extend=0.25,
    height=3.5cm,
    width=4.8cm,
    ]

    \addplot+[
	fill, draw=black,
	boxplot prepared={
		median=0.188691693833259,
		upper quartile=0.520221750434927,
		lower quartile=0.0656593819994641,
		upper whisker=1.20158151962962,
		lower whisker=7.28762175074445e-06,
		draw position=0.7
	},
	] coordinates {};

	\addplot+[
	fill, draw=black,
	boxplot prepared={
		median=0.25891764859988,
		upper quartile=0.697278552993303,
		lower quartile=0.090058520010702,
		upper whisker=1.60568526689773,
		lower whisker=1.58326982264409e-05,
		draw position=1
	},
	] coordinates {};

	\addplot+[
	fill, draw=black,
	boxplot prepared={
		median=0.261218211827209,
		upper quartile=0.581835341270658,
		lower quartile=0.0981970376574996,
		upper whisker=1.30670208467602,
		lower whisker=3.23045143325525e-05,
		draw position=1.3
	},
	] coordinates {};

	\addplot+[
	fill, draw=black,
	boxplot prepared={
		median=0.116652361249008,
		upper quartile=0.292221594396714,
		lower quartile=0.0439963542902754,
		upper whisker=0.664555976993339,
		lower whisker=2.19136218210972e-05,
		draw position=1.7
	},
	] coordinates {};

	\addplot+[
	fill, draw=black,
	boxplot prepared={
		median=0.100717020816473,
		upper quartile=0.22236355985569,
		lower quartile=0.0390286392914756,
		upper whisker=0.497118654205084,
		lower whisker=5.54532322171233e-05,
		draw position=2
	},
	] coordinates {};

	\addplot+[
	fill, draw=black,
	boxplot prepared={
		median=0.179169489485891,
		upper quartile=0.385856003745102,
		lower quartile=0.0694741839614141,
		upper whisker=0.860030243207658,
		lower whisker=6.23301227537737e-06,
		draw position=2.3
	},
	] coordinates {};

	\addplot+[
	fill, draw=black,
	boxplot prepared={
		median=0.0982604703720517,
		upper quartile=0.239153555825274,
		lower quartile=0.0369097145560284,
		upper whisker=0.541272642526236,
		lower whisker=2.05017053564481e-05,
		draw position=2.7
	},
	] coordinates {};

	\addplot+[
	fill, draw=black,
	boxplot prepared={
		median=0.0802448453052423,
		upper quartile=0.1754563387893,
		lower quartile=0.0318808510116896,
		upper whisker=0.390177178492074,
		lower whisker=2.86110869132814e-06,
		draw position=3
	},
	] coordinates {};

	\addplot+[
	fill, draw=black,
	boxplot prepared={
		median=0.135548558145441,
		upper quartile=0.292951385964636,
		lower quartile=0.053401015624624,
		upper whisker=0.651956278718631,
		lower whisker=3.63832352054051e-05,
		draw position=3.3
	},
	] coordinates {};

	\addplot+[
	fill, draw=black,
	boxplot prepared={
		median=0.0836460210135478,
		upper quartile=0.202469361502368,
		lower quartile=0.0321212294719916,
		upper whisker=0.456956976195309,
		lower whisker=1.85228097169358e-05,
		draw position=3.7
	},
	] coordinates {};

	\addplot+[
	fill, draw=black,
	boxplot prepared={
		median=0.0658882441645434,
		upper quartile=0.147190639209893,
		lower quartile=0.0262510125694735,
		upper whisker=0.327799029499976,
		lower whisker=6.27494561863599e-06,
		draw position=4
	},
	] coordinates {};

	\addplot+[
	fill, draw=black,
	boxplot prepared={
		median=0.106344005316074,
		upper quartile=0.240011265894288,
		lower quartile=0.0435928831430895,
		upper whisker=0.534398187438831,
		lower whisker=8.40985740744362e-06,
		draw position=4.3
	},
	] coordinates {};
 	\end{axis}
\end{tikzpicture}
        \vspace{-.61cm}
        \caption{Normalized $v^{\rm tx}$ estimation error.}
        \label{fig:boxplot_speed}
    \end{subfigure}
    \caption{Varying \ac{snr} values, with $\sigma_\alpha=5$°, $KT=\SI{16}{\milli\second}$, and $S=2$.}
    \label{fig:speed_eta}
\end{figure}
\begin{figure}[t!]
    \centering
    \begin{subfigure}{0.49\columnwidth}
        \centering
\begin{tikzpicture}
\definecolor{color1}{HTML}{003f5c}
\definecolor{color2}{HTML}{ffa600}
\definecolor{darkgray176}{RGB}{176,176,176}
    
\begin{axis}[
tick align=inside,
tick pos=left,
x grid style={darkgray176},
xlabel={Frame period $T$ [ms]},
xmin=-0.5, xmax=13.5,
ylabel shift = -5pt,
xlabel shift = -1 pt,
xtick={0,3,6,9,12},
xticklabels={0.14,0.2,0.26,0.32,0.38},
y grid style={darkgray176},
ylabel={Error $\Bar{\varepsilon}_{f_{\rm D, t}}$},
ymin=-0.325494304485038, ymax=1.99,
ytick distance=0.5,
ymajorgrids,
xmajorgrids,
xlabel style={font=\footnotesize}, ylabel style={font=\footnotesize}, ticklabel style={font=\footnotesize},
height=3.5cm,
width=4.8cm,
]

\addplot [semithick, color1]
table {%
0 0.169739250198556
1 0.0688312172547798
2 0.120619228162803
3 0.241649032510679
4 0.356781679508124
5 0.471710415018214
6 0.599547140988192
7 0.719915550172667
8 0.827366622329246
9 0.96345929266734
10 1.07749981394765
11 1.19853194554308
12 1.32860819411986
13 1.43561752459084
};
\addplot [semithick, color1, mark=-, mark size=2, mark options={solid}, only marks]
table {%
  0 0.0968059651676219
  0 0.24267253522949
  1 -0.0270294620910524
  1 0.164691896600612
  2 -0.090935511545534
  2 0.332173967871141
  3 -0.0174165126590544
  3 0.500714577680412
  4 0.0891274928725986
  4 0.62443586614365
  5 0.167735115959908
  5 0.77568571407652
  6 0.297317432700673
  6 0.90177684927571
  7 0.387759388861561
  7 1.05207171148377
  8 0.49876895056644
  8 1.15596429409205
  9 0.579925428317281
  9 1.3469931570174
  10 0.674244160386695
  10 1.48075546750861
  11 0.774513285612406
  11 1.62255060547376
  12 0.893164788719605
  12 1.76405159952011
  13 0.967299787837005
  13 1.90393526134468
  14 1.08275607861099
  14 2.03255681383724
  15 1.16730769672619
  15 2.19292297383077
  16 1.26331192335482
  16 2.35239043830026
  17 1.34629662608994
  17 2.50280189407742
  18 1.45074096712282
  18 2.63749542144828
  19 1.52451259723257
  19 2.79887711813684
  20 1.65454605348539
  20 2.90584137821982
  21 1.72284127175951
  21 3.07910552711426
  22 1.84001943414416
  22 3.20763465821764
  23 1.9203836006736
  23 3.38205138613473
  24 2.01726292024765
  24 3.53089497772606
  25 2.11983101760106
  25 3.68260091187573
  26 2.17668173384077
  26 3.87577257553877
  27 2.23771735007056
  27 4.06811368119417
  28 2.18970095502019
  28 4.47211985183201
};

\addplot [semithick, color1]
table {%
  0 0.0968059651676219
  0 0.24267253522949
};

\addplot [semithick, color1]
table {%
  1 -0.0270294620910524
  1 0.164691896600612
};

\addplot [semithick, color1]
table {%
2 -0.090935511545534
2 0.332173967871141
};

\addplot [semithick, color1]
table {%
3 -0.0174165126590544
3 0.500714577680412
};

\addplot [semithick, color1]
table {%
4 0.0891274928725986
4 0.62443586614365
};

\addplot [semithick, color1]
table {%
5 0.167735115959908
5 0.77568571407652
};

\addplot [semithick, color1]
table {%
6 0.297317432700673
6 0.90177684927571
};

\addplot [semithick, color1]
table {%
7 0.387759388861561
7 1.05207171148377
};

\addplot [semithick, color1]
table {%
8 0.49876895056644
8 1.15596429409205
};

\addplot [semithick, color1]
table {%
9 0.579925428317281
9 1.3469931570174
};

\addplot [semithick, color1]
table {%
10 0.674244160386695
10 1.48075546750861
};

\addplot [semithick, color1]
table {%
11 0.774513285612406
11 1.62255060547376
};

\addplot [semithick, color1]
table {%
12 0.893164788719605
12 1.76405159952011
};

\addplot [semithick, color1]
table {%
13 0.967299787837005
13 1.90393526134468
};

\addplot [semithick, color1]
table {%
14 1.08275607861099
14 2.03255681383724
};

\addplot [semithick, color1]
table {%
15 1.16730769672619
15 2.19292297383077
};

\addplot [semithick, color1]
table {%
16 1.26331192335482
16 2.35239043830026
};

\addplot [semithick, color1]
table {%
17 1.34629662608994
17 2.50280189407742
};

\addplot [semithick, color1]
table {%
18 1.45074096712282
18 2.63749542144828
};

\addplot [semithick, color1]
table {%
19 1.52451259723257
19 2.79887711813684
};

\addplot [semithick, color1]
table {%
20 1.65454605348539
20 2.90584137821982
};

\addplot [semithick, color1]
table {%
21 1.72284127175951
21 3.07910552711426
};

\addplot [semithick, color1]
table {%
22 1.84001943414416
22 3.20763465821764
};

\addplot [semithick, color1]
table {%
23 1.9203836006736
23 3.38205138613473
};

\addplot [semithick, color1]
table {%
24 2.01726292024765
24 3.53089497772606
};

\addplot [semithick, color1]
table {%
25 2.11983101760106
25 3.68260091187573
};

\addplot [semithick, color1]
table {%
26 2.17668173384077
26 3.87577257553877
};

\addplot [semithick, color1]
table {%
27 2.23771735007056
27 4.06811368119417
};

\addplot [semithick, color1]
table {%
28 2.18970095502019
28 4.47211985183201
};

\addplot [draw=color1, fill=color1, mark=*, only marks, mark options={solid, scale=0.7},]
table{%
x  y
0 0.169739250198556
1 0.0688312172547798
2 0.120619228162803
3 0.241649032510679
4 0.356781679508124
5 0.471710415018214
6 0.599547140988192
7 0.719915550172667
8 0.827366622329246
9 0.96345929266734
10 1.07749981394765
11 1.19853194554308
12 1.32860819411986
13 1.43561752459084
14 1.55765644622411
15 1.68011533527848
16 1.80785118082754
17 1.92454926008368
18 2.04411819428555
19 2.1616948576847
20 2.2801937158526
21 2.40097339943689
22 2.5238270461809
23 2.65121749340416
24 2.77407894898685
25 2.90121596473839
26 3.02622715468977
27 3.15291551563236
28 3.3309104034261
};

\addplot[very thick, color2, densely dashed]
table{%
1.5 -0.3
1.5 1.9
};
\node [label={[label distance=-5pt]0:\scriptsize{$T_{\rm max}=0.17$~ms}}] at (20,200) {};

\end{axis}
\end{tikzpicture}
        \vspace{-15pt}
        \caption{Simulation with $f_c=\SI{60}{\giga\hertz}$.}
        \label{fig:average_error_sim}
    \end{subfigure}
    \begin{subfigure}{0.49\columnwidth}
        \centering
\begin{tikzpicture}

\definecolor{darkgray176}{RGB}{176,176,176}
\definecolor{color1}{HTML}{003f5c}
\definecolor{color2}{HTML}{ffa600}

\begin{axis}[
tick align=inside,
tick pos=left,
x grid style={darkgray176},
xlabel={Frame period $T$~[ms]},
xmin=-0.5, xmax=13.5,
ylabel shift = -5pt,
xlabel shift = -1 pt,
ytick distance = 0.25,
xticklabels={0.5,1,1.5,2,2.5,3,3.5},
xtick={1,3,5,7,9,11,13},
y grid style={darkgray176},
ymin=-0.0448528289181404, ymax=1.24,
ylabel={Error $\Bar{\varepsilon}_{f_{\rm D, t}}$},
ymajorgrids,
xmajorgrids,
xlabel style={font=\footnotesize}, ylabel style={font=\footnotesize}, ticklabel style={font=\footnotesize},
height=3.5cm,
width=4.8cm,
]
\addplot [
  mark=*,
  only marks,
  mark size=2,
  mark options={solid, scale=0.7},
  scatter,
  scatter/@post marker code/.code={%
  \endscope
},
  scatter/@pre marker code/.code={%
  \scope[draw=color1, fill=color1]%
},
  visualization depends on={value \thisrow{draw} \as \drawcolor},
  visualization depends on={value \thisrow{fill} \as \fillcolor}
]
table{%
x  y  draw  fill
0 0.150381163527936 31,119,180 31,119,180
1 0.175409929753223 31,119,180 31,119,180
2 0.165501663923811 31,119,180 31,119,180
3 0.206655664709038 31,119,180 31,119,180
4 0.239745886912562 31,119,180 31,119,180
5 0.246903992948611 31,119,180 31,119,180
6 0.353965014302772 31,119,180 31,119,180
7 0.286392497762199 31,119,180 31,119,180
8 0.286665827321328 31,119,180 31,119,180
9 0.28379128840386 31,119,180 31,119,180
10 0.413346737622012 31,119,180 31,119,180
11 0.508424477838199 31,119,180 31,119,180
12 0.512582294029633 31,119,180 31,119,180
13 0.666365015623369 31,119,180 31,119,180
};
\addplot [semithick, color1]
table {%
0 0.150381163527936
1 0.175409929753223
2 0.165501663923811
3 0.206655664709038
4 0.239745886912562
5 0.246903992948611
6 0.353965014302772
7 0.286392497762199
8 0.286665827321328
9 0.28379128840386
10 0.413346737622012
11 0.508424477838199
12 0.512582294029633
13 0.666365015623369
};
\addplot [semithick, color1, mark=-, mark size=2, mark options={solid}, only marks]
table {%
0 0.125048738986343
0 0.184432811612226
1 0.12246103793374
1 0.247986496991783
2 0.135300842910002
2 0.201963179048407
3 0.159425412412766
3 0.263610011200929
4 0.162398007820376
4 0.344749144215259
5 0.188163353174029
5 0.316266613939041
6 0.225170178068154
6 0.583246148181194
7 0.193676368474216
7 0.418822879987699
8 0.22867176525049
8 0.349060555062631
9 0.230907992157723
9 0.353123119791393
10 0.312193666985505
10 0.530095500426039
11 0.372069520325441
11 0.676249441912877
12 0.377261037175579
12 0.717620665338255
13 0.398494281035125
13 1.16836176620203
};

\addplot [semithick, color1]
table {%
0 0.125048738986343
0 0.184432811612226
};
\addplot [semithick, color1]
table {%
1 0.12246103793374
1 0.247986496991783
};
\addplot [semithick, color1]
table {%
2 0.135300842910002
2 0.201963179048407
};
\addplot [semithick, color1]
table {%
3 0.159425412412766
3 0.263610011200929
};
\addplot [semithick, color1]
table {%
4 0.162398007820376
4 0.344749144215259
};
\addplot [semithick, color1]
table {%
5 0.188163353174029
5 0.316266613939041
};
\addplot [semithick, color1]
table {%
6 0.225170178068154
6 0.583246148181194
};
\addplot [semithick, color1]
table {%
7 0.193676368474216
7 0.418822879987699
};
\addplot [semithick, color1]
table {%
8 0.22867176525049
8 0.349060555062631
};
\addplot [semithick, color1]
table {%
9 0.230907992157723
9 0.353123119791393
};
\addplot [semithick, color1]
table {%
10 0.312193666985505
10 0.530095500426039
};
\addplot [semithick, color1]
table {%
11 0.372069520325441
11 0.676249441912877
};
\addplot [semithick, color1]
table {%
12 0.377261037175579
12 0.717620665338255
};
\addplot [semithick, color1]
table {%
13 0.398494281035125
13 1.16836176620203
};
\addplot[very thick, color2, densely dashed]
table{%
2.2 -0.3
2.2 1.9
};
\node [label={[label distance=-5pt]0:\scriptsize{$T_{\rm max}=0.8$~ms}}] at (30,139) {};
\end{axis}

\end{tikzpicture}
        \vspace{-15pt}
        \caption{Experimental setup $1$.}
        \label{fig:average_error_exp}
    \end{subfigure}
    \caption{Average normalized Doppler frequency error.}
    \label{fig:average_error}
\end{figure}

\section{Experimental results}\label{sec:exp-results}
In this section, we present experimental results obtained by \ourname{} on an IEEE~802.11ay-based \ac{isac} testbed.

\begin{figure*}[t!] 
    \centering
    \subcaptionbox{Setup $1$ (quasi-mono-static), $S=2$.\protect\label{fig:setup-mono}}[0.32\textwidth]{\includegraphics[width=0.25\textwidth]{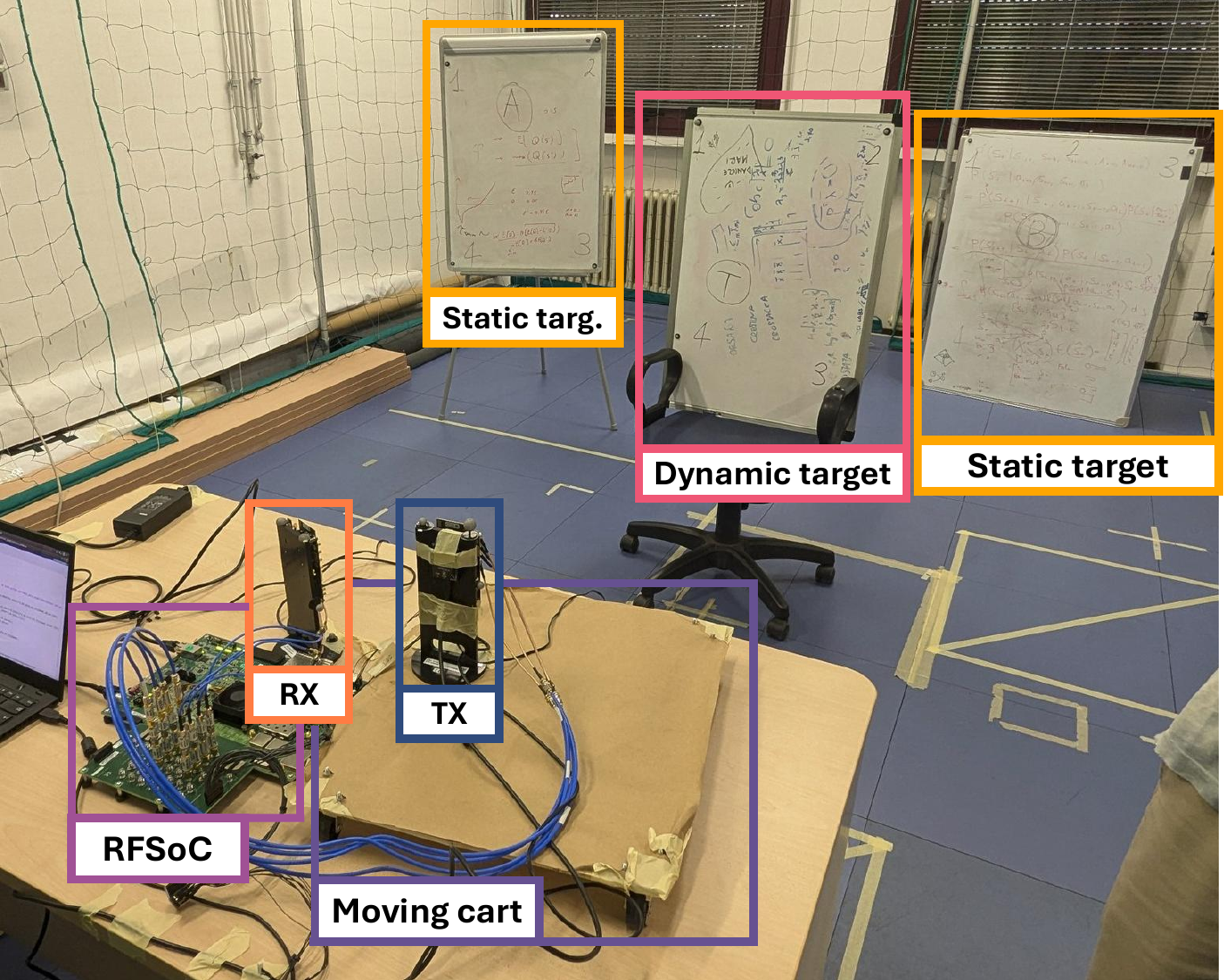}}
    \subcaptionbox{Setup $2$ (bi-static).\protect\label{fig:setup-bi}}[0.315\textwidth]{\includegraphics[width=0.245\textwidth]{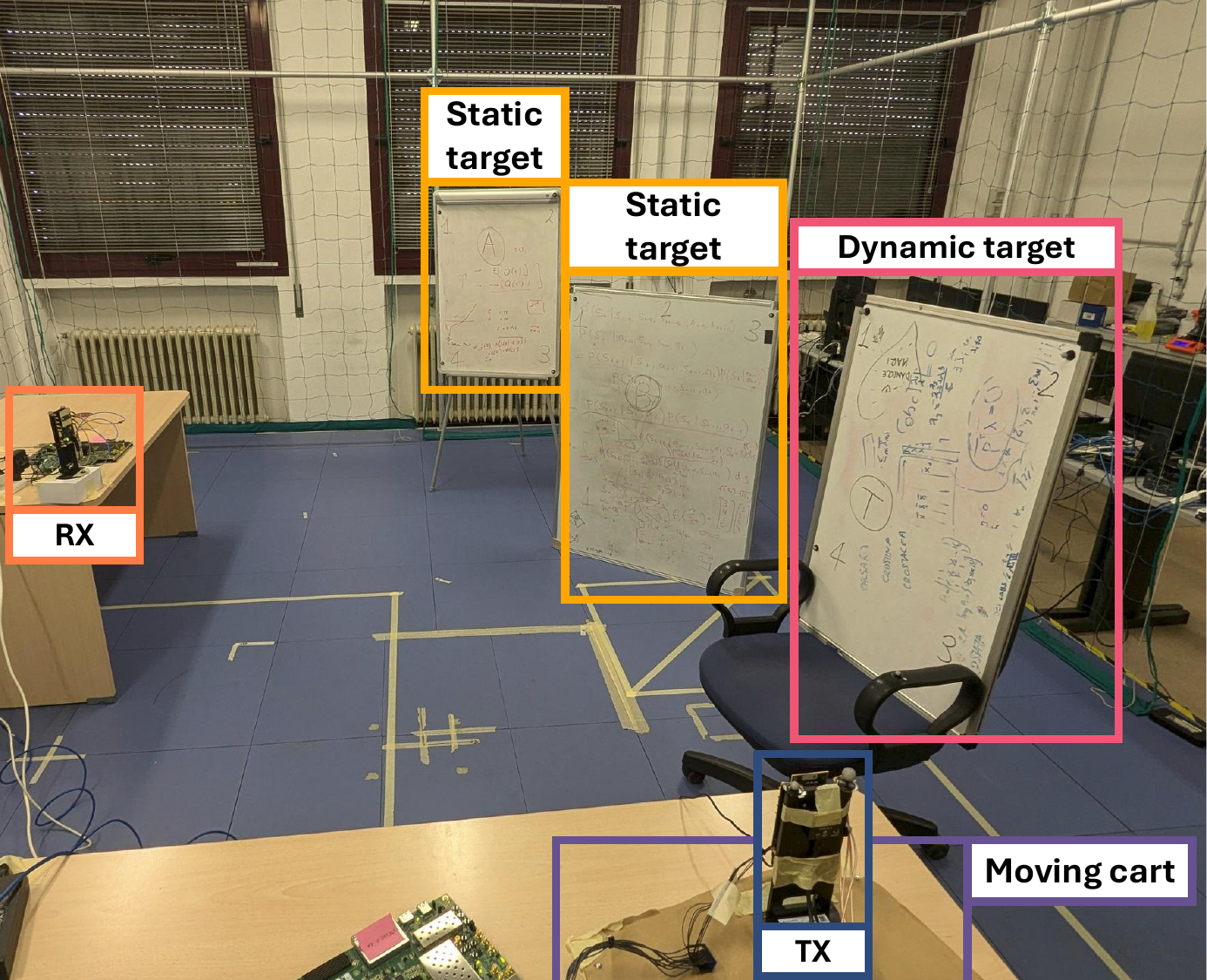}}
    \subcaptionbox{\centering{Setup $3$ \rev{(human target + moving interferers)}.}\protect\label{fig:setup-human}}[0.32\textwidth]{\includegraphics[width=0.235\textwidth]{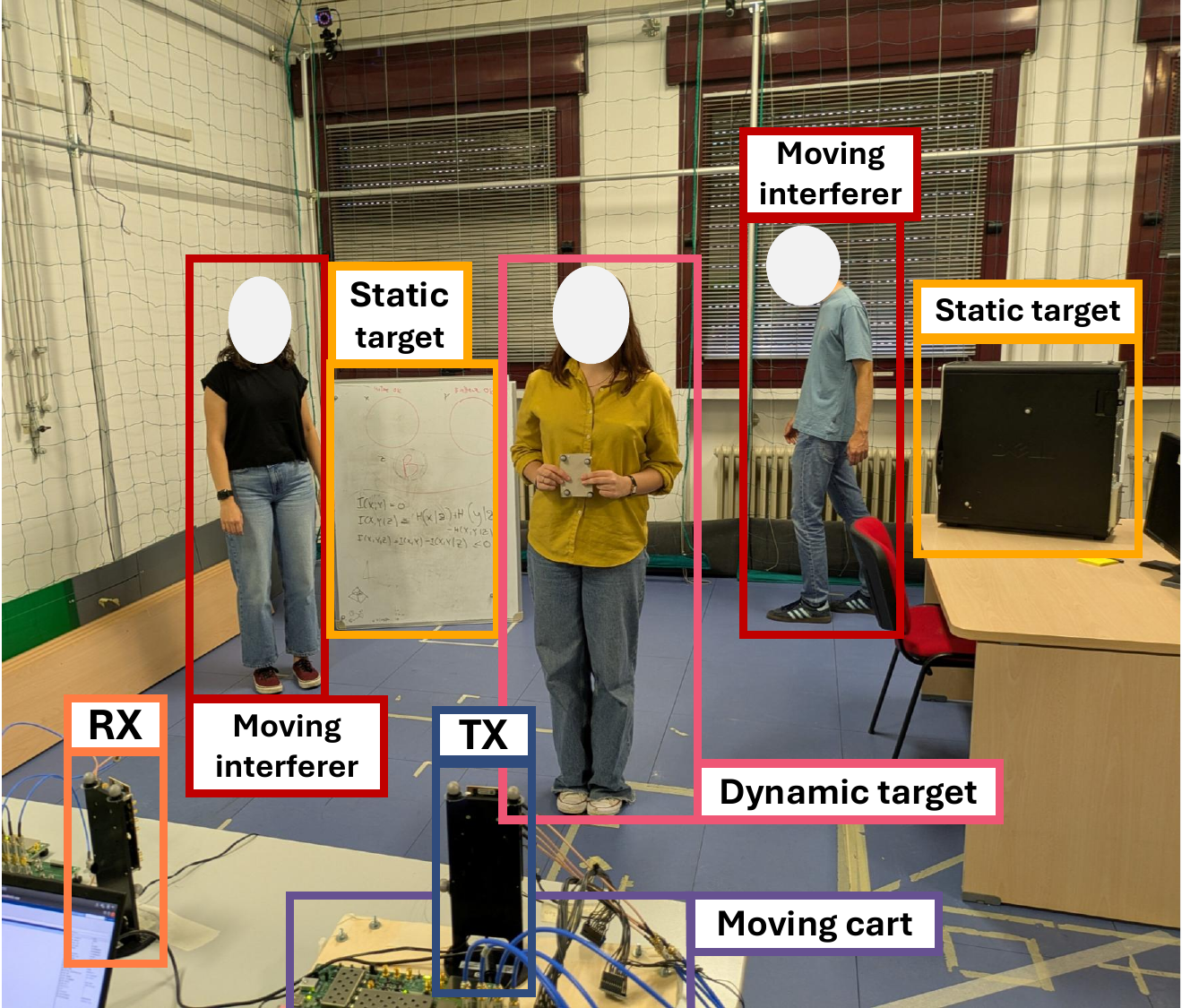}}
    \caption{Visualization of the experimental setup.}
    \label{fig:experimental-setup}
    \vspace{-4mm}
\end{figure*}

\subsection{\ac{isac} testbed}\label{sec:testbed}

Our testbed consists of two \ac{isac} nodes acting as TX and RX. 
\rev{Each node comprises a Xilinx \ac{rfsoc} for baseband signal processing and a \ac{mmwave} Sivers $16$- or $8$-elements phased array antennas operating at $f_c=60$~GHz or $f_c=28$~GHz, respectively}.   
The implementation is based on that presented in~\cite{pegoraro2024jump}. 
We refer to \cite{pegoraro2024jump, lacruz2021realtime} for additional details on the testbed setup.
The TX sends IEEE~802.11ay standard-compliant packets with $1.76$~GHz bandwidth, including an IEEE~802.11ay header and a TRN field. This field contains $12$ TRN units, each composed of $6$ complementary Golay sequences~\cite{802.11ay}. We use each TRN unit to estimate the channel for a different TX beam pattern, scanning the whole angular space in front of the TX. This allows us to estimate the \ac{aod} of each multipath component with respect to the antenna array orientation, using the algorithm in~\cite{pegoraro2024rapid}. 
The antenna beamwidth determines the resolution of the \ac{aod} estimation, which approximately equals $8^{\circ}$.
The RX performs downconversion, sampling, packet detection, and symbol-level synchronization. 
To perform \ac{cir} estimation, we compute the cross-correlation of the received signal with the known TRN units at the RX. Then, we apply peak detection to isolate significant multipath components and estimate \ac{aod} for each peak using the algorithm in~\cite{pegoraro2024rapid}. 
We compensate for the TX orientation by estimating the \ac{aod} of the \ac{los} path and removing it from the \ac{aod} of other paths.

\rev{This sensor has an acceleration measurement noise of $\pm 29.4$~mm/s$^2$ over $1$~ms, and a bias of $49$~mm/s$^2$. We experimentally characterized the acceleration noise distribution through an Allan deviation plot, verifying that it is dominated by white noise.}
The \ac{imu} is used to obtain measurements of the TX velocity vector according to a reference frame centered on the TX device.
Results obtained by integrating \ac{imu} measurements with our algorithm are presented in \secref{sec:imu-exp-results}. 

\subsection{Experimental setup}\label{sec:setup}

The experiments are performed in a $5\times 4$~m laboratory equipped with an infrared camera-based motion tracking system, as shown in~\fig{fig:experimental-setup}. 
The camera system allows obtaining the accurate positions of each object in the environment to be used as a ground-truth reference. For this, we place infrared markers on the TX, RX, static reflectors, and targets. 
To evaluate \ourname{} with mobile \ac{isac} nodes, we mount the TX on a wheeled cart, moving it during the measurements. \rev{\fig{fig:tx_pos} and \fig{fig:tx_velocity} describe the TX motion by plotting its position coordinates and its velocity components, respectively.}

\rev{
\begin{figure}
    \centering
    \begin{subfigure}{0.48\columnwidth}
        \centering
        \input{plot/coordinates_tx}
        \vspace{-15pt}
        \caption{TX position.}
        \label{fig:tx_pos}
    \end{subfigure}
    \begin{subfigure}{0.48\columnwidth}
        \centering
        \input{plot/speed_tx}
        \vspace{-15pt}
        \caption{TX velocity.}
        \label{fig:tx_velocity}
    \end{subfigure}
    \caption{\rev{Example TX position (a) and velocity (b) during movement. Blue and yellow colors refer to the $x$ and $y$ components, respectively.}}
    \label{fig:tx_trajectory}
\end{figure}
}

We consider $3$ different evaluation setups, as described next. \rev{In all of them, the default carrier frequency is $f_c=60$~GHz. In Setup $3$ additional results using $f_c=28$~GHz are presented.}

$\bullet$ \textit{Setup $1$ (SU$1$)}. As shown in \fig{fig:setup-mono}, the first setup includes quasi-monostatic TX and RX, placed at a distance of $40$~cm from one another and oriented towards the targets. 
In this setup, static and dynamic reflections are produced using metal whiteboards placed in the environment. The moving target whiteboard is mounted on a wheeled chair. 

$\bullet$ \textit{Setup $2$ (SU$2$)}. A more challenging evaluation is conducted in a bi-static setting, with widely separated TX and RX, as shown in~\fig{fig:setup-bi}. This experiment reproduces more realistic conditions where the \ac{los} path is weaker and the orientation of the \ac{isac} devices is unknown. 

$\bullet$ \textit{Setup $3$ (SU$3$)}. Similarly to SU$1$, this setup includes quasi-monostatic TX and RX, $50$~cm apart from one another and oriented towards the targets. However, as shown in \fig{fig:setup-human}, this is a more realistic setup where the target is human and the static paths are produced by common office furniture, including a desktop computer and a desk. \rev{In addition, $1$ or $2$ people are moving in the same room besides the target, acting as \textit{interferers}, to increase the similarity to a real office environment.} 
The target subject carries a plastic plate on which we place the infrared markers. 
\rev{In this setup, we collect measurements with both $f_c=60$~GHz and $f_c=28$~GHz.}

In all the considered experiments, the motion of the target and that of the TX follow \textit{uncoordinated} back and forth trajectories \rev{(the TX movement trajectory is shown in \fig{fig:tx_trajectory})}.
We do not use rails or stepped motors to simulate a realistic case where \ac{isac} devices may be handheld and targets follow irregular trajectories.
Unless specified, we consider the inter-frame duration $T_k$ to be constant and denote it by $T=0.254$~ms. 
In \secref{sec:rand_sampling}, we show that this simplification is not restrictive, since \ourname{} is robust to a random choice of $T_k$.
Furthermore, in the experiments, we re-analyze the choice of the inter-frame duration $T$ to avoid phase ambiguities. As shown in~\fig{fig:average_error_exp}, due to the lower velocities involved, \ourname{} achieves good performance on the experiments in Setup~$1$ for up to $T=0.8$~ms. The performance degradation is minimal up to $T=2.5$~ms due to the lower variations in Doppler frequencies observed in the experiments (as compared to the simulations), which leads to a lower probability of experiencing phase ambiguities.
\rev{Finally, we recall that the simulation framework is not able to model all the nuisance sources occurring in a real system, like multipath fading, hardware non-idealities besides \ac{po} and \ac{cfo}, and \ac{aod} estimation affected by non-Gaussian errors. 
All these aspects are instead present in our experiments and represent a more challenging validation scenario for \ourname{}, which explains the gap between simulation and experimental results.}

\subsection{Doppler frequency and TX speed estimation error}\label{sec:fd-exp-results}
\ourname{} is evaluated in terms of absolute estimation errors on the target Doppler frequency, the TX speed, and $\eta$, respectively defined as \mbox{$\widehat{\varepsilon}_{f_{\rm D,t}} = |f_{\rm D,t}-\widehat{f}_{\rm D,t}|$}, \mbox{$\widehat{\varepsilon}_{v^{\rm tx}}=|v^{\rm tx}-\widehat{v}^{\rm tx}|$}, \mbox{$\widehat{\varepsilon}_{\eta}=|\eta-\widehat{\eta}|$}. 
The ground truth $f_{\rm D,t}$ is obtained by plugging in \eq{eq:insta-doppler} the ground truth locations of the TX, the RX, and the target, obtained from the motion tracking system. 

\begin{figure}
    \centering
    \input{plot/exp_doppler}
    \vspace{-.5cm}
    \caption{Estimated target Doppler frequency, $\widehat{f}_{\rm D,t}$, with $S=2$ and $KT=16$~ms, for Setup $1$.}
    \label{fig:exp_doppler}
\end{figure}

\subsubsection{Robustness to irregular inter-frame spacing}
\label{sec:rand_sampling}

In \fig{fig:exp_doppler}, we show the ground truth and the estimated target Doppler frequency for a specific experiment in Setup~$1$. We also show the results obtained with a random inter-frame spacing $T_k$. This case is implemented by randomly discarding \ac{cir} estimates in the experiment -- each element is independently discarded with probability $0.5$. 
\ourname{} can accurately estimate the target Doppler frequency, with a marginal degradation despite a random $T_k$. Specifically, \ourname{} obtains Doppler frequency estimation errors of $3.03\pm3.61$~Hz and $5.07\pm6.66$~Hz with constant and random $T_k$, respectively (mean $\pm$ one standard deviation). The slight degradation of the latter (random) case is due to the fewer samples available in the observation window, as a consequence of the removal of \ac{cir} samples.

\subsubsection{State of the art comparison}
\label{sec:sota}
In \fig{fig:exp_setups}, we compare \ourname{} and \ac{jm} error results, $\widehat{\varepsilon}_{f_{\rm D,t}}$, in different setups using a mobile or static TX. We use $S=2$ and $KT=8$~ms or $KT=48$~ms.
 For both SU$1$ and SU$2$ (\fig{fig:exp_setups1} and \fig{fig:exp_setups2}), the error obtained by \ourname{} is only slightly higher with a moving TX compared to a static one. Conversely, \ac{jm} achieves similar performance only in the case of a static TX.
 When the TX moves, the bias described in \secref{sec:limitations} hinders a correct target Doppler estimation, leading to a $195\%$ and $45\%$ higher median error with respect to \ourname{} in SU$1$ and SU$2$, respectively. 
 \rev{Furthermore, in \fig{fig:exp_setup3} we show that \ourname{} is also effective in a realistic environment with a human target, office furniture, and other moving people acting as interferers. 
 As expected, the very low \ac{snr} on the target path due to the weak human body scattering at both $60$~GHz and $28$~GHz increases the error $\hat{\varepsilon}_{f_{\rm D}}$ with respect to Setups $1$ and $2$. 
 However, \ourname{} outperforms \ac{jm} also in this case, especially when the TX is moving. 
 \ac{jm} achieves a median error which is approximately $290\%$ higher than \ourname{}'s one at $60$~GHz (\fig{fig:exp_setups3}) and $56\%$ higher at $28$~GHz (\fig{fig:exp_setups4}). In the latter case, the performance gap is smaller because the TX motion bias described in \secref{sec:limitations} is inversely proportional to the wavelength, so \ac{jm} is less affected by the TX motion at lower frequencies when considering the absolute Doppler estimation error in Hz.}
 This demonstrates that only \ourname{} successfully copes with the motion of \ac{isac} devices.

\subsubsection{TX velocity estimation}
\label{sec:tx_velocity}
Besides the target Doppler, \ourname{} can additionally estimate the TX velocity.
In \fig{fig:exp_speed_eta}, we present both the velocity modulus and its direction. \fig{fig:exp_speedtx} shows that, with a sufficiently long window, i.e., $16$~ms or longer, \ourname{} achieves median errors lower than $5$~cm/s for $v^{\rm tx}$ in all the presented setups.
The error on the estimate of $\eta$ in \fig{fig:exp_eta}, gets lower than $15$° for Setups $1$ and $2$ with $KT\geq16$~ms. 
For Setup~$1$ and~$3$, $\widehat{\varepsilon}_{\eta}$ is affected by the difficulty of estimating the \ac{aod} for the \ac{los} path, since the RX is located at the very edge of the field of view of the TX antenna. 
Additionally, the performance in Setup~$3$ is further affected by lower accuracy of estimating the \acp{aod} for the weak paths scattered by a human body.

\rev{
\subsubsection{Real-time implementation}
\label{sec:real-time}
As we show in \fig{fig:boxplot_K_music}, \ourname{} execution time is from two to three orders of magnitude lower than \ac{jm}. 
To assess the applicability of \ourname{} in embedded devices, we tested it on a Nvidia Jetson TX2 board, measuring the time needed to estimate the Doppler frequency from the experimental measurements collected in Setup $3$. 
The algorithm is implemented in Python, so its efficiency could be further optimized by implementing it in a lower-level programming language. 
In \fig{fig:exp_setup3}, we show, as black diamonds, the average time needed to estimate the target Doppler frequency in one aggregation window for different values of the window $KT$. 
The execution time with $KT=48$~ms is below $100$~ms in the worst case, proving that \ourname{} obtains over $10$~Doppler frequency estimates per second on embedded hardware.
}

\begin{figure}[t]
    \centering
    \begin{subfigure}{\columnwidth}
        \centering
        \hspace{2pt}
        \begin{tikzpicture}[pics/legend entry/.style={code={%
        \draw[pic actions] 
        (0,0) -- (8pt,0);}}]
    \tikzset{
    legend square/.pic={
        \draw[pic actions] (0.5mm,-.75mm) rectangle (2mm,.75mm);
        }
        }
\definecolor{color4}{HTML}{ff764a}
\definecolor{color3}{HTML}{bc5090}
\definecolor{color1}{HTML}{374c80}
\definecolor{color2}{HTML}{ffa600}
\definecolor{black28}{RGB}{28,28,28}
\definecolor{burlywood231175145}{RGB}{231,175,145}
\definecolor{darkgray176}{RGB}{176,176,176}
\definecolor{darkslategray583162}{RGB}{58,31,62}
\matrix[draw, fill=white,inner sep=2pt,ampersand replacement=\&,column sep=5pt] at (0,0) { 
            \node [shape=rectangle, draw=black ,fill=color1, label=right:\scriptsize{\ourname{} Mob. TX}] {};\&
            \node [shape=rectangle, draw=black ,fill=color2, label=right:\scriptsize{\ac{jm} Mob. TX}] {};\&
            \node [shape=rectangle, draw=black ,fill=color3, label=right:\scriptsize{\ourname{} Stat. TX}] {};\&
            \node [shape=rectangle, draw=black ,fill=color4, label=right:\scriptsize{\ac{jm} Stat. TX}] {};\\
        };
\end{tikzpicture}
        \vspace{2pt}
    \end{subfigure}
    \begin{subfigure}{0.45\columnwidth}
        \centering
        \begin{tikzpicture}[pics/legend entry/.style={code={%
        \draw[pic actions] 
        (0,0) -- (8pt,0);}}]
\tikzset{
    legend square/.pic={
        \draw[pic actions] (0.5mm,-.75mm) rectangle (2mm,.75mm);
        }
        }
\definecolor{black28}{RGB}{28,28,28}
\definecolor{brown1814888}{RGB}{181,48,88}
\definecolor{burlywood231175145}{RGB}{231,175,145}
\definecolor{darkgray176}{RGB}{176,176,176}
\definecolor{darkslategray583162}{RGB}{58,31,62}
\definecolor{color4}{HTML}{ff764a}
\definecolor{color3}{HTML}{bc5090}
\definecolor{color1}{HTML}{374c80}
\definecolor{color2}{HTML}{ffa600}
  \begin{axis}
    [
    boxplot/draw direction=y,
    ylabel={$\hat{\varepsilon}_{f_{\rm D}}$ [Hz]},
    xlabel={Aggregation window $KT$ [ms]},
    xlabel shift=-5pt,
    ymin=-.5, ymax=19,
    ytick distance =5,
    yticklabel style = {/pgf/number format/fixed},
    cycle list={{color1},{color2},{color3},{color4}},
    xtick={1,2,3,4},
    ymajorgrids,
    xticklabels={$8$, $48$, $8$, $48$},
    xlabel style={font=\footnotesize}, ylabel style={font=\footnotesize}, ticklabel style={font=\footnotesize},
    ylabel shift=-6pt,
    /pgfplots/boxplot/box extend=0.125,
    height=3.5cm,
    width=4.2cm,
    xmin=0.25,
    xmax=2.75
    ]
    \addplot+[
    fill, draw=black,
    boxplot prepared={
    	median=2.7068846280999,
    	upper quartile=4.93999839558712,
    	lower quartile=1.2677988669615,
    	upper whisker=10.4299715272826,
    	lower whisker=0.00307303350037103,
    	draw position=.7
    },
    ] coordinates {};
    
    \addplot+[
    fill, draw=black,
    boxplot prepared={
    	median=7.48938259325776,
    	upper quartile=12.164487216693,
    	lower quartile=3.74663753672434,
    	upper whisker=24.5967620947449,
    	lower whisker=0.0030888937918121,
    	draw position=0.9
    },
    ] coordinates {};
    
    \addplot+[
    fill, draw=black,
    boxplot prepared={
    	median=1.99796066908952,
    	upper quartile=3.4842507726755,
    	lower quartile=0.90909092000344,
    	upper whisker=7.33293000186172,
    	lower whisker=0.000123196068159359,
    	draw position=1.1
    },
    ] coordinates {};
    
    \addplot+[
    fill, draw=black,
    boxplot prepared={
    	median=4.80180777268192,
    	upper quartile=8.01934578427462,
    	lower quartile=2.27185634046648,
    	upper whisker=16.6125795788688,
    	lower whisker=4.33560035872915e-05,
    	draw position=1.3
    },
    ] coordinates {};
    
    \addplot+[
    fill, draw=black,
    boxplot prepared={
    	median=2.42847965174581,
    	upper quartile=4.36828604810211,
    	lower quartile=1.1044266348951,
    	upper whisker=9.2296552884951,
    	lower whisker=0.000153268532329776,
    	draw position=1.7
    },
    ] coordinates {};
    
    \addplot+[
    fill, draw=black,
    boxplot prepared={
    	median=7.16628995606865,
    	upper quartile=11.2675368859339,
    	lower quartile=3.6478489823529,
    	upper whisker=22.6466038056729,
    	lower whisker=0.000704179568515428,
    	draw position=1.9
    },
    ] coordinates {};
    
    \addplot+[
    fill, draw=black,
    boxplot prepared={
    	median=1.62938587817928,
    	upper quartile=2.87880641784233,
    	lower quartile=0.773309871592755,
    	upper whisker=6.03286480396686,
    	lower whisker=0.00512511488260259,
    	draw position=2.1
    },
    ] coordinates {};

    \addplot+[
    fill, draw=black,
    boxplot prepared={
    	median=1.62531989099975,
    	upper quartile=2.9144719653848,
    	lower quartile=0.715647095434068,
    	upper whisker=6.18723781008143,
    	lower whisker=0.00146314442768869,
    	draw position=2.3
    },
    ] coordinates {};

\end{axis}

    
\end{tikzpicture}
        \vspace{-5pt}
        \caption{Setup~$1$.} \label{fig:exp_setups1}
    \end{subfigure}
    \hspace{1em}
    \begin{subfigure}{0.45\columnwidth}
        \centering
        \begin{tikzpicture}[pics/legend entry/.style={code={%
        \draw[pic actions] 
        (0,0) -- (8pt,0);}}]
\tikzset{
    legend square/.pic={
        \draw[pic actions] (0.5mm,-.75mm) rectangle (2mm,.75mm);
        }
        }
\definecolor{black28}{RGB}{28,28,28}
\definecolor{brown1814888}{RGB}{181,48,88}
\definecolor{burlywood231175145}{RGB}{231,175,145}
\definecolor{darkgray176}{RGB}{176,176,176}
\definecolor{darkslategray583162}{RGB}{58,31,62}
\definecolor{color4}{HTML}{ff764a}
\definecolor{color3}{HTML}{bc5090}
\definecolor{color1}{HTML}{374c80}
\definecolor{color2}{HTML}{ffa600}
  \begin{axis}
    [
    boxplot/draw direction=y,
    xlabel={Aggregation window $KT$ [ms]},
    xlabel shift=-5pt, 
    ymin=-.5, ymax=19,
    ytick distance =5,
    yticklabel style = {/pgf/number format/fixed},
    cycle list={{color1},{color2},{color3},{color4}},
    xtick={1,2,3,4},
    ymajorgrids,
    xticklabels={$8$, $48$, $8$, $48$},
    xlabel style={font=\footnotesize}, ylabel style={font=\footnotesize}, ticklabel style={font=\footnotesize},
    ylabel shift=-6pt,
    /pgfplots/boxplot/box extend=0.125,
    height=3.5cm,
    width=4.2cm,
    xmin=0.25,
    xmax=2.75
    ]
\addplot+[
fill, draw=black,
boxplot prepared={
	median=7.48741722790025,
	upper quartile=13.7959276683848,
	lower quartile=3.45722200536553,
	upper whisker=29.2721763784846,
	lower whisker=0.000268255213214275,
	draw position=.7
},
] coordinates {};

\addplot+[
fill, draw=black,
boxplot prepared={
	median=9.67473268444877,
	upper quartile=16.1434024748778,
	lower quartile=4.74242158521612,
	upper whisker=33.2047470373348,
	lower whisker=0.000739556917508821,
	draw position=0.9
},
] coordinates {};

\addplot+[
fill, draw=black,
boxplot prepared={
	median=5.83691501822577,
	upper quartile=10.8396163671031,
	lower quartile=2.63685591431652,
	upper whisker=23.0885379781168,
	lower whisker=0.00718033483639147,
	draw position=1.1
},
] coordinates {};

\addplot+[
fill, draw=black,
boxplot prepared={
	median=5.04063032924697,
	upper quartile=9.50666987164395,
	lower quartile=2.28987905216377,
	upper whisker=20.3189059107715,
	lower whisker=0.00204137333668797,
	draw position=1.3
},
] coordinates {};
    
\addplot+[
fill, draw=black,
boxplot prepared={
	median=5.64964441524712,
	upper quartile=10.1675092842348,
	lower quartile=2.69836429413165,
	upper whisker=21.321150820145,
	lower whisker=0.00184160177761328,
	draw position=1.7
},
] coordinates {};

\addplot+[
fill, draw=black,
boxplot prepared={
	median=8.2115261365403,
	upper quartile=13.8769958213102,
	lower quartile=4.15739416754889,
	upper whisker=28.3450895262758,
	lower whisker=0.0017337710443428,
	draw position=1.9
},
] coordinates {};

\addplot+[
fill, draw=black,
boxplot prepared={
	median=4.40706849833738,
	upper quartile=8.56275309505297,
	lower quartile=2.05338527970622,
	upper whisker=18.0536232251104,
	lower whisker=0.00190617865855813,
	draw position=2.1
},
] coordinates {};

\addplot+[
fill, draw=black,
boxplot prepared={
	median=4.18354730905973,
	upper quartile=8.07161696928158,
	lower quartile=1.90766975286174,
	upper whisker=17.3174481061954,
	lower whisker=0.00168688736108891,
	draw position=2.3
},
] coordinates {};

\end{axis}

    
\end{tikzpicture}
        \vspace{-17pt}
        \caption{Setup~$2$.} \label{fig:exp_setups2}
    \end{subfigure}
    \caption{Target Doppler frequency estimation error for different setups, with $S=2$ and varying aggregation window $KT$.}
    \label{fig:exp_setups}
    \vspace{-5pt}
\end{figure}

\begin{figure}[t]
    \centering
    \begin{subfigure}{\columnwidth}
        \centering
        \hspace{2pt}
        \begin{tikzpicture}[pics/legend entry/.style={code={%
        \draw[pic actions] 
        (0,0) -- (8pt,0);}}]
    \tikzset{
    legend square/.pic={
        \draw[pic actions] (0.5mm,-.75mm) rectangle (2mm,.75mm);
        }
        }
\definecolor{color4}{HTML}{ff764a}
\definecolor{color3}{HTML}{bc5090}
\definecolor{color1}{HTML}{374c80}
\definecolor{color2}{HTML}{ffa600}
\definecolor{black28}{RGB}{28,28,28}
\definecolor{burlywood231175145}{RGB}{231,175,145}
\definecolor{darkgray176}{RGB}{176,176,176}
\definecolor{darkslategray583162}{RGB}{58,31,62}
\matrix[draw, fill=white,inner sep=2pt,ampersand replacement=\&,column sep=5pt] at (0,0) { 
            \node [shape=rectangle, draw=black ,fill=color1, label=right:\scriptsize{\ourname{} Mob. TX}] {};\&
            \node [shape=rectangle, draw=black ,fill=color2, label=right:\scriptsize{\ac{jm} Mob. TX}] {};\&
            \node [shape=rectangle, draw=black ,fill=color3, label=right:\scriptsize{\ourname{} Stat. TX}] {};\&
            \node [shape=rectangle, draw=black ,fill=color4, label=right:\scriptsize{\ac{jm} Stat. TX}] {};\\
        };
\end{tikzpicture}
    \end{subfigure}
    \begin{subfigure}{0.45\columnwidth}
        \centering
        \begin{tikzpicture}[pics/legend entry/.style={code={%
        \draw[pic actions] 
        (0,0) -- (8pt,0);}}]
\tikzset{
    legend square/.pic={
        \draw[pic actions] (0.5mm,-.75mm) rectangle (2mm,.75mm);
        }
        }
\definecolor{black28}{RGB}{28,28,28}
\definecolor{brown1814888}{RGB}{181,48,88}
\definecolor{burlywood231175145}{RGB}{231,175,145}
\definecolor{darkgray176}{RGB}{176,176,176}
\definecolor{darkslategray583162}{RGB}{58,31,62}
\definecolor{color4}{HTML}{ff764a}
\definecolor{color3}{HTML}{bc5090}
\definecolor{color1}{HTML}{374c80}
\definecolor{color2}{HTML}{ffa600}
  \begin{axis}
    [
    boxplot/draw direction=y,
    ylabel={$\hat{\varepsilon}_{f_{\rm D}}$ [Hz]},
    xlabel={Aggregation window $KT$ [ms]},
    xlabel shift=-5pt,
    ymin=-2, ymax=65,
    ytick distance =10,
    yticklabel style = {/pgf/number format/fixed},
    cycle list={{color1},{color2},{color3},{color4}},
    xtick={1,2},
    ymajorgrids,
    xticklabels={$8$, $48$},
    xlabel style={font=\footnotesize}, ylabel style={font=\footnotesize}, ticklabel style={font=\footnotesize},
    ylabel shift=-5pt,
    /pgfplots/boxplot/box extend=0.125,
    height=3.5cm,
    width=4.2cm,
    xmin=0.25,
    xmax=2.75
    ]

    \addplot+[
    fill, draw=black,
    boxplot prepared={
    	median=11.1867272198156,
    	upper quartile=24.2245643576752,
    	lower quartile=4.7940997316024,
    	upper whisker=53.2903571934203,
    	lower whisker=0.000293122328253403,
    	draw position=.7
    },
    ] coordinates {};
    
    \addplot+[
    fill, draw=black,
    boxplot prepared={
    	median=36.5385824455988,
    	upper quartile=55.9593263257954,
    	lower quartile=18.7333279474288,
    	upper whisker=111.686748946818,
    	lower whisker=0.00571845097618962,
    	draw position=0.9
    },
    ] coordinates {};

    \addplot+[
    fill, draw=black,
    boxplot prepared={
    	median=11.5880122303476,
    	upper quartile=24.0011709901966,
    	lower quartile=4.92442669108856,
    	upper whisker=52.3514247328238,
    	lower whisker=0.00781180660692371,
    	draw position=1.1
    },
    ] coordinates {};
 
    \addplot+[
    fill, draw=black,
    boxplot prepared={
    	median=12.6711166384432,
    	upper quartile=31.4276328055631,
    	lower quartile=5.36366439539756,
    	upper whisker=70.4430565229151,
    	lower whisker=0.0076833397177154,
    	draw position=1.3
    },
    ] coordinates {};

    \addplot+[
    fill, draw=black,
    boxplot prepared={
    	median=9.25749899297111,
    	upper quartile=17.6440620323162,
    	lower quartile=4.23890550832164,
    	upper whisker=37.6511859909014,
    	lower whisker=0.00345396103722351,
    	draw position=1.7
    },
    ] coordinates {};

    \addplot+[
    fill, draw=black,
    boxplot prepared={
    	median=36.5980688653319,
    	upper quartile=53.6440069082684,
    	lower quartile=18.6717976831768,
    	upper whisker=106.058423748128,
    	lower whisker=0.0112539475570976,
    	draw position=1.9
    },
    ] coordinates {};

    \addplot+[
    fill, draw=black,
    boxplot prepared={
    	median=8.82706831950978,
    	upper quartile=17.8187013281574,
    	lower quartile=4.24128547621955,
    	upper whisker=38.1234519401831,
    	lower whisker=0.0224771433721287,
    	draw position=2.1
    },
    ] coordinates {};

    \addplot+[
    fill, draw=black,
    boxplot prepared={
    	median=11.7255650564585,
    	upper quartile=22.4885005615706,
    	lower quartile=5.38376244586185,
    	upper whisker=48.0884837771226,
    	lower whisker=0.00188573984138429,
    	draw position=2.3
    },
    ] coordinates {};

\end{axis}

\begin{axis}[
        xtick = {1,2},
        ytick = {20, 40, 60, 80, 100, 120},
        ymin=-4, ymax=130,
        xmin=0.25,
        xmax=2.75,
        hide x axis,
        axis y line*=right,
        height=3.5cm,
        width=4.2cm,
        y axis line style={draw=none},
        ylabel style={font=\footnotesize}, ticklabel style={font=\footnotesize},
        ylabel={Time [ms]},
        ylabel shift=-5pt,
    ]
        
        \addplot [draw=black, dashed, mark options={solid, fill=black}, mark=diamond*]
        coordinates {
            (1, 39.4012251170423)
            (2, 93.2799444294063)
        };
        
    \end{axis}
\end{tikzpicture}
        \vspace{-17pt}
        \caption{\rev{Setup~$3$ $f_{\rm c}=60$~GHz.}} \label{fig:exp_setups3}
    \end{subfigure}
    \hspace{1em}
    \begin{subfigure}{0.45\columnwidth}
        \centering
        \begin{tikzpicture}[pics/legend entry/.style={code={%
        \draw[pic actions] 
        (0,0) -- (8pt,0);}}]
\tikzset{
    legend square/.pic={
        \draw[pic actions] (0.5mm,-.75mm) rectangle (2mm,.75mm);
        }
        }
\definecolor{black28}{RGB}{28,28,28}
\definecolor{brown1814888}{RGB}{181,48,88}
\definecolor{burlywood231175145}{RGB}{231,175,145}
\definecolor{darkgray176}{RGB}{176,176,176}
\definecolor{darkslategray583162}{RGB}{58,31,62}
\definecolor{color4}{HTML}{ff764a}
\definecolor{color3}{HTML}{bc5090}
\definecolor{color1}{HTML}{374c80}
\definecolor{color2}{HTML}{ffa600}
  \begin{axis}
    [
    boxplot/draw direction=y,
    xlabel={Aggregation window $KT$ [ms]},
    xlabel shift=-5pt,
    ymin=-2, ymax=65,
    ytick distance =10,
    yticklabel style = {/pgf/number format/fixed},
    cycle list={{color1},{color2},{color3},{color4}},
    xtick={1,2},
    ymajorgrids,
    xticklabels={$8$, $48$},
    xlabel style={font=\footnotesize}, ylabel style={font=\footnotesize}, ticklabel style={font=\footnotesize},
    ylabel shift=-5pt,
    /pgfplots/boxplot/box extend=0.125,
    height=3.5cm,
    width=4.2cm,
    xmin=0.25,
    xmax=2.75
    ]

\addplot+[
fill, draw=black,
boxplot prepared={
	median=14.7395536347427,
	upper quartile=31.2509512179826,
	lower quartile=6.12567799785755,
	upper whisker=68.4126385286167,
	lower whisker=0.00109488932370994,
	draw position=0.7
},
] coordinates {};

\addplot+[
fill, draw=black,
boxplot prepared={
	median=21.0149203315561,
	upper quartile=35.4263147235676,
	lower quartile=10.5540929391967,
	upper whisker=72.7291462243304,
	lower whisker=0.0131501342922267,
	draw position=0.9
},
] coordinates {};

\addplot+[
fill, draw=black,
boxplot prepared={
	median=11.2400354023691,
	upper quartile=23.1873640554564,
	lower quartile=4.69092225289512,
	upper whisker=50.8498232082747,
	lower whisker=0.0239813510045508,
	draw position=1.1
},
] coordinates {};

\addplot+[
fill, draw=black,
boxplot prepared={
	median=13.1175666599621,
	upper quartile=28.9129659297882,
	lower quartile=5.60973328895988,
	upper whisker=63.6822365849501,
	lower whisker=0.000699039978943716,
	draw position=1.3
},
] coordinates {};

\addplot+[
fill, draw=black,
boxplot prepared={
	median=12.487041218634,
	upper quartile=24.784910745394,
	lower quartile=5.5217888554338,
	upper whisker=53.5548543043221,
	lower whisker=0.00164156055404874,
	draw position=1.7
},
] coordinates {};

\addplot+[
fill, draw=black,
boxplot prepared={
	median=19.326929575753,
	upper quartile=30.3741310059673,
	lower quartile=9.56825286331025,
	upper whisker=61.4980236142383,
	lower whisker=0.0128170659791067,
	draw position=1.9
},
] coordinates {};

\addplot+[
fill, draw=black,
boxplot prepared={
	median=7.80273309104836,
	upper quartile=15.4032467839405,
	lower quartile=3.10319623782737,
	upper whisker=33.1050758015489,
	lower whisker=0.000481561067257275,
	draw position=2.1
},
] coordinates {};

\addplot+[
fill, draw=black,
boxplot prepared={
	median=8.97511532566819,
	upper quartile=17.6053772763991,
	lower quartile=3.56095886086204,
	upper whisker=38.5734908431459,
	lower whisker=0.00365175412464502,
	draw position=2.3
},
] coordinates {};

\end{axis}

\begin{axis}[
        xtick = {1,2},
        ytick = {20, 40, 60, 80, 100, 120},
        ymin=-4, ymax=130,
        xmin=0.25,
        xmax=2.75,
        hide x axis,
        axis y line*=right,
        height=3.5cm,
        width=4.2cm,
        y axis line style={draw=none},
        ylabel style={font=\footnotesize}, ticklabel style={font=\footnotesize},
        ylabel={Time [ms]},
        ylabel shift=-5pt,
    ]
        
        \addplot [draw=black, dashed, mark options={solid, fill=black}, mark=diamond*]
        coordinates {
            (1, 23.165098067794)
            (2, 62.4304698531035)
        };
        
    \end{axis}
\end{tikzpicture}
        \vspace{-17pt}
        \caption{\rev{Setup~$3$ $f_{\rm c}=28$~GHz.}} \label{fig:exp_setups4}
    \end{subfigure}
    \caption{\rev{Target Doppler frequency estimation error and average computational time for different setups, with $S=2$.}}
    \label{fig:exp_setup3}
    \vspace{-5pt}
\end{figure}

\begin{figure}[t]
    \centering
    \begin{subfigure}{\columnwidth}
        \centering
        \hspace{30pt}
        \begin{tikzpicture}
    \definecolor{color2}{HTML}{bc5090}
    \definecolor{color4}{HTML}{ffa600}
    \definecolor{color3}{HTML}{ff764a}
    \definecolor{color1}{HTML}{003f5c}
    \definecolor{black28}{RGB}{28,28,28}
    \definecolor{burlywood231175145}{RGB}{231,175,145}
    \definecolor{darkgray176}{RGB}{176,176,176}
    \definecolor{darkslategray583162}{RGB}{58,31,62}

    \matrix[draw, fill=white,inner sep=2pt,ampersand replacement=\&,column sep=3pt ] at (60,100) { 
                    \node [shape=rectangle, draw=black ,fill=color1, label=right:\scriptsize{Setup $1$}] {};\&
                    \node [shape=rectangle, draw=black ,fill=color4, label=right:\scriptsize{Setup $2$}] {};\&
                    \node [shape=rectangle, draw=black ,fill=color2, label=right:\scriptsize{Setup $3$}] {};\\
                };
    
\end{tikzpicture}
        \vspace{0.03cm}
    \end{subfigure}
    \begin{subfigure}{0.48\columnwidth}
        \centering
        \begin{tikzpicture}
\definecolor{black28}{RGB}{28,28,28}
\definecolor{color2}{HTML}{ff764a}
\definecolor{color4}{HTML}{ffa600}
\definecolor{darkgray176}{RGB}{176,176,176}
\definecolor{darkslategray583162}{RGB}{58,31,62}
\definecolor{color3}{HTML}{bc5090}
\definecolor{color1}{HTML}{374c80}
  \begin{axis}
    [
    boxplot/draw direction=y,
    ylabel={$\hat{\varepsilon}_{\eta}$[°]},
    xlabel={Aggregation window $KT$ [ms]},
    ymin=-6, ymax=165,
    ytick distance =30,
    yticklabel style = {/pgf/number format/fixed},
    cycle list={{color1},{color4},{color3}},
    xtick={1,2,3,4,5,6},
    ymajorgrids,
    xticklabels={$2$, $4$, $8$, $16$, $32$, $48$},
    xlabel style={font=\footnotesize}, ylabel style={font=\footnotesize}, ticklabel style={font=\footnotesize},
    ylabel shift=-5pt,
    /pgfplots/boxplot/box extend=0.25,
    height=3.5cm,
    width=4.8cm,
    ]
\addplot+[
fill, draw=black,
boxplot prepared={
	median=55.4453344413898,
	upper quartile=166.790649774611,
	lower quartile=20.9067394527884,
	upper whisker=357.6662374971,
	lower whisker=0.0334309804090083,
	draw position=.7
},
] coordinates {};

\addplot+[
fill, draw=black,
boxplot prepared={
	median=27.3775507427976,
	upper quartile=162.424120020483,
	lower quartile=11.5858431438609,
	upper whisker=316.177721838647,
	lower whisker=0.00283662604561385,
	draw position=1
},
] coordinates {};

\addplot+[
fill, draw=black,
boxplot prepared={
	median=34.919147024245,
	upper quartile=166.135956824112,
	lower quartile=13.4864963458909,
	upper whisker=359.231866019386,
	lower whisker=0.00203525576921493,
	draw position=1.3
},
] coordinates {};

\addplot+[
fill, draw=black,
boxplot prepared={
	median=35.3415905581283,
	upper quartile=146.989120643099,
	lower quartile=15.1226997460039,
	upper whisker=344.456702408583,
	lower whisker=0.0128634500096894,
	draw position=1.7
},
] coordinates {};

\addplot+[
fill, draw=black,
boxplot prepared={
	median=16.9967072293393,
	upper quartile=129.033705437324,
	lower quartile=7.58301333621015,
	upper whisker=309.928671793964,
	lower whisker=0.00151178664023632,
	draw position=2
},
] coordinates {};

\addplot+[
fill, draw=black,
boxplot prepared={
	median=23.2645070576989,
	upper quartile=158.200372701568,
	lower quartile=9.47693361475414,
	upper whisker=359.52738737528,
	lower whisker=0.000281098561004001,
	draw position=2.3
},
] coordinates {};

\addplot+[
fill, draw=black,
boxplot prepared={
	median=20.3368641540084,
	upper quartile=93.3059389018657,
	lower quartile=8.47581688246473,
	upper whisker=220.375541900988,
	lower whisker=0.00304188194589017,
	draw position=2.7
},
] coordinates {};

\addplot+[
fill, draw=black,
boxplot prepared={
	median=10.1467456793201,
	upper quartile=26.0080116108923,
	lower quartile=4.73234354042113,
	upper whisker=57.8704443762097,
	lower whisker=0.00102419973646306,
	draw position=3
},
] coordinates {};

\addplot+[
fill, draw=black,
boxplot prepared={
	median=17.2240855847591,
	upper quartile=151.569461258626,
	lower quartile=8.03280199558846,
	upper whisker=359.400733658207,
	lower whisker=0.00132202091441513,
	draw position=3.3
},
] coordinates {};

\addplot+[
fill, draw=black,
boxplot prepared={
	median=10.5870277066505,
	upper quartile=32.8583191110348,
	lower quartile=4.79751512048132,
	upper whisker=74.4407947055204,
	lower whisker=0.0017953251478815,
	draw position=3.7
},
] coordinates {};

\addplot+[
fill, draw=black,
boxplot prepared={
	median=5.769980378649,
	upper quartile=12.4787235431088,
	lower quartile=2.62898384959714,
	upper whisker=27.2423320237097,
	lower whisker=0.000999375350090759,
	draw position=4
},
] coordinates {};

\addplot+[
fill, draw=black,
boxplot prepared={
	median=14.5185763590074,
	upper quartile=152.420758950175,
	lower quartile=7.42436151358283,
	upper whisker=359.325609150992,
	lower whisker=0.0141240134673808,
	draw position=4.3
},
] coordinates {};

\addplot+[
fill, draw=black,
boxplot prepared={
	median=6.57787817505971,
	upper quartile=17.703002614291,
	lower quartile=3.10092518273967,
	upper whisker=39.6021865149294,
	lower whisker=0.000452723712015768,
	draw position=4.7
},
] coordinates {};

\addplot+[
fill, draw=black,
boxplot prepared={
	median=3.47626081677885,
	upper quartile=7.26616226865958,
	lower quartile=1.54809515212365,
	upper whisker=15.8406356661478,
	lower whisker=0.00443878993701219,
	draw position=5
},
] coordinates {};

\addplot+[
fill, draw=black,
boxplot prepared={
	median=13.3413668406104,
	upper quartile=152.630236858692,
	lower quartile=7.17556165267304,
	upper whisker=358.975527435419,
	lower whisker=0.00486096997417462,
	draw position=5.3
},
] coordinates {};

\addplot+[
fill, draw=black,
boxplot prepared={
	median=5.72423126634186,
	upper quartile=16.1806595218756,
	lower quartile=2.59602753534611,
	upper whisker=36.512894147331,
	lower whisker=0.000249460037963445,
	draw position=5.7
},
] coordinates {};

\addplot+[
fill, draw=black,
boxplot prepared={
	median=2.85419969929993,
	upper quartile=5.75609683056362,
	lower quartile=1.29649438363315,
	upper whisker=12.4129436650478,
	lower whisker=2.19506497387556e-05,
	draw position=6
},
] coordinates {};

\addplot+[
fill, draw=black,
boxplot prepared={
	median=13.1649627450999,
	upper quartile=155.237246690313,
	lower quartile=7.10334412646486,
	upper whisker=359.215966868311,
	lower whisker=0.0161016828220397,
	draw position=6.3
},
] coordinates {};

\end{axis}
\end{tikzpicture}
        \vspace{-15pt}
        \caption{\rev{Absolute $\eta$ estimation error.}}
        \label{fig:exp_eta}
    \end{subfigure}
    \begin{subfigure}{0.48\columnwidth}
        \centering
        \begin{tikzpicture}
\definecolor{black28}{RGB}{28,28,28}
\definecolor{color2}{HTML}{ff764a}
\definecolor{color4}{HTML}{ffa600}
\definecolor{darkgray176}{RGB}{176,176,176}
\definecolor{darkslategray583162}{RGB}{58,31,62}
\definecolor{color3}{HTML}{bc5090}
\definecolor{color1}{HTML}{374c80}
  \begin{axis}
    [
    boxplot/draw direction=y,
    ylabel={$\hat{\varepsilon}_{v^{\rm tx}}$[m/s]},
    xlabel={Aggregation window $KT$ [ms]},
    ymin=-0.016, ymax=0.35,
    ytick distance =0.1,
    yticklabel style = {/pgf/number format/fixed},
    cycle list={{color1},{color4},{color3}},
    xtick={1,2,3,4,5,6},
    ymajorgrids,
    xticklabels={$2$,$4$, $8$, $16$, $32$, $48$},
    xlabel style={font=\footnotesize}, ylabel style={font=\footnotesize}, ticklabel style={font=\footnotesize},
    ylabel shift=-5pt,
    /pgfplots/boxplot/box extend=0.25,
    height=3.5cm,
    width=4.8cm,
    ]
\addplot+[
fill, draw=black,
boxplot prepared={
	median=0.112266230390616,
	upper quartile=0.210177731484034,
	lower quartile=0.0477458463938577,
	upper whisker=0.452134171807963,
	lower whisker=5.32860445297267e-06,
	draw position=.7
},
] coordinates {};

\addplot+[
fill, draw=black,
boxplot prepared={
	median=0.311989935166797,
	upper quartile=0.646014613823752,
	lower quartile=0.126589942251321,
	upper whisker=1.42408670355283,
	lower whisker=4.12673402535868e-05,
	draw position=1
},
] coordinates {};

\addplot+[
fill, draw=black,
boxplot prepared={
	median=0.0919720658320881,
	upper quartile=0.193488129674701,
	lower quartile=0.041815998693181,
	upper whisker=0.420993176603133,
	lower whisker=2.04198176840809e-05,
	draw position=1.3
},
] coordinates {};

\addplot+[
fill, draw=black,
boxplot prepared={
	median=0.0483938356618183,
	upper quartile=0.0989640967567948,
	lower quartile=0.0214541648748691,
	upper whisker=0.215145535399913,
	lower whisker=0.000171173663789687,
	draw position=1.7
},
] coordinates {};

\addplot+[
fill, draw=black,
boxplot prepared={
	median=0.156140424507425,
	upper quartile=0.298292901308507,
	lower quartile=0.0706746917706269,
	upper whisker=0.638831383528301,
	lower whisker=4.01853746072045e-05,
	draw position=2
},
] coordinates {};

\addplot+[
fill, draw=black,
boxplot prepared={
	median=0.0616872395608562,
	upper quartile=0.127608879271328,
	lower quartile=0.0283201497825213,
	upper whisker=0.27595979320996,
	lower whisker=2.48038356982505e-05,
	draw position=2.3
},
] coordinates {};

\addplot+[
fill, draw=black,
boxplot prepared={
	median=0.0288188650345279,
	upper quartile=0.0517055750562003,
	lower quartile=0.0134532061525396,
	upper whisker=0.108777081366336,
	lower whisker=4.99086099300244e-06,
	draw position=2.7
},
] coordinates {};

\addplot+[
fill, draw=black,
boxplot prepared={
	median=0.0949934509166375,
	upper quartile=0.167702800418776,
	lower quartile=0.0439543376287678,
	upper whisker=0.353093641708664,
	lower whisker=6.50514248021206e-06,
	draw position=3
},
] coordinates {};

\addplot+[
fill, draw=black,
boxplot prepared={
	median=0.0471057509517124,
	upper quartile=0.0965210708342651,
	lower quartile=0.0208634417128003,
	upper whisker=0.209623685486599,
	lower whisker=6.10907952280915e-06,
	draw position=3.3
},
] coordinates {};

\addplot+[
fill, draw=black,
boxplot prepared={
	median=0.0169396072696847,
	upper quartile=0.031798801476783,
	lower quartile=0.00835059914081039,
	upper whisker=0.0667077257966438,
	lower whisker=1.15781742398985e-05,
	draw position=3.7
},
] coordinates {};

\addplot+[
fill, draw=black,
boxplot prepared={
	median=0.054598944732893,
	upper quartile=0.0961355119624169,
	lower quartile=0.0261728194079169,
	upper whisker=0.201044937344008,
	lower whisker=3.41194155296098e-05,
	draw position=4
},
] coordinates {};

\addplot+[
fill, draw=black,
boxplot prepared={
	median=0.0395349206776009,
	upper quartile=0.081361566122735,
	lower quartile=0.0176170260066111,
	upper whisker=0.176971853735491,
	lower whisker=1.33223510315472e-05,
	draw position=4.3
},
] coordinates {};

\addplot+[
fill, draw=black,
boxplot prepared={
	median=0.0103715806430112,
	upper quartile=0.0190973462233101,
	lower quartile=0.0047085005760524,
	upper whisker=0.0403257661907905,
	lower whisker=3.33902699524435e-06,
	draw position=4.7
},
] coordinates {};

\addplot+[
fill, draw=black,
boxplot prepared={
	median=0.0335246735711243,
	upper quartile=0.0577128838253083,
	lower quartile=0.0160586034261336,
	upper whisker=0.119811858237185,
	lower whisker=5.06857647408765e-07,
	draw position=5
},
] coordinates {};

\addplot+[
fill, draw=black,
boxplot prepared={
	median=0.0363834391467363,
	upper quartile=0.0745211345800871,
	lower quartile=0.0154260751088201,
	upper whisker=0.163020740224482,
	lower whisker=5.07028664161607e-07,
	draw position=5.3
},
] coordinates {};

\addplot+[
fill, draw=black,
boxplot prepared={
	median=0.00819938670984689,
	upper quartile=0.0154000485167066,
	lower quartile=0.00390336213422432,
	upper whisker=0.0325509539457188,
	lower whisker=3.34325097911536e-07,
	draw position=5.7
},
] coordinates {};

\addplot+[
fill, draw=black,
boxplot prepared={
	median=0.0280634221706363,
	upper quartile=0.0499538628314583,
	lower quartile=0.0124165631950026,
	upper whisker=0.106211875638783,
	lower whisker=1.07112020414912e-05,
	draw position=6
},
] coordinates {};

\addplot+[
fill, draw=black,
boxplot prepared={
	median=0.0361067755429421,
	upper quartile=0.0715143839713568,
	lower quartile=0.0165042830863784,
	upper whisker=0.153920752728701,
	lower whisker=2.64194540120638e-06,
	draw position=6.3
},
] coordinates {};
    \end{axis}
\end{tikzpicture}
        \vspace{-15pt}
        \caption{\rev{Absolute $v_{\rm tx}$ estimation error.}}
        \label{fig:exp_speedtx}
    \end{subfigure}
    \caption{Varying the duration of the aggregation window, with $S=2$.}
    \label{fig:exp_speed_eta}
\end{figure}

 \begin{figure}[t!]
    \centering
    \input{plot/known_v}
    \caption{Target Doppler frequency with $S=2$ and $KT=16$~ms. }
    \label{fig:known_vtx}
\end{figure}

\subsection{Integration with \ac{imu}}\label{sec:imu-exp-results}

Finally, we test the integration of our model with the TX velocity measured by the onboard \ac{imu} sensor. This is done by obtaining an estimate of $v^{\rm tx}$ from the \ac{imu}, in the reference system of the TX. Then, this velocity is used as an additional input to \ourname{}, thus the number of unknowns is reduced to two and the system in \eq{eq:delta-model} can be solved using a single static reference path. 
\rev{We remark that, having also a reliable estimate of $\eta$, the number of unknowns would be reduced to one, solving the system in \eq{eq:delta-model} without any static reference path. However, the $\eta$ estimate obtained from the \ac{imu} implemented in our experimental setup is rather unreliable, and its use is likely to degrade the estimation performance; thus, we do not integrate it in our results.}

As shown in \fig{fig:known_vtx}, integrating the TX speed measurement from the \ac{imu} could enhance the reliability of \ourname{}.
In this experiment, the \ac{cir} peak corresponding to one of the required static paths is not detected for several subsequent frames in the interval $[1.5, 2.1]$~s. Within this time interval, \ourname{} lacks sufficient information (static paths) to estimate the target Doppler frequency and fails, as shown by the yellow curve. 
Conversely, when feeding the \ac{imu} measurement to \ourname{}, the Doppler frequency is estimated accurately even when not all the \ac{cir} peaks are detected (pink curve). 
Interestingly, when all the required \ac{cir} peaks are available, \ourname{} is more accurate \textit{without} using the \ac{imu} measurement. This is evident from the zoom plot in \fig{fig:known_vtx} and from \tab{tab:fd_est_vknown}, where we quantitatively evaluate the Doppler estimation error for three different time intervals in the experiment of \fig{fig:known_vtx}. While in the interval $[1.5, 2.1]$~s using the \ac{imu} gives a large performance gain, outside of this interval, estimating the Doppler solely based on the wireless channel cuts the estimation error by half with respect to using the \ac{imu}.
Based on these results, the \ac{imu} should only be used to compensate for the temporary lack of a sufficient number of static multipath components.

\begin{table}[t!]
\footnotesize
    \centering
    \caption{Average target Doppler frequency estimation error $\pm$ one standard deviation, with one undetected path  within $[1.5, 2.1]$~s.}
    \begin{tabular}{lccc}
       \toprule
        &\multicolumn{3}{c}{\textbf{Time interval}}\\ 
        \cmidrule(lr){2-4}
         \textbf{Error [Hz]} & $[0,1.5]$~s & $[1.5,2.1]$~s & $[2.1,7.3]$~s \\ 
        \midrule
        w/ IMU  & $5.32\pm4.04$ & $\boldsymbol{10.72\pm12.32}$ & $6.51\pm3.89$ \\
         w/o IMU & $\boldsymbol{2.06\pm2.24}$ & $168\pm518$ & $\boldsymbol{2.91\pm2.33}$ \\
       \bottomrule
    \end{tabular}
    \label{tab:fd_est_vknown}
\end{table}

\section{Concluding remarks}\label{sec:conclusion}

In this paper, we proposed \ourname{}, a method to estimate the Doppler frequency of a target in an \textit{asynchronous} \ac{isac} system featuring \textit{mobile} \ac{isac} devices. Our approach effectively disentangles the target's Doppler frequency from the phase offsets due to clock asynchrony and the Doppler caused by the device movement if at least $2$ static multipath components are available. By exploiting phase differences across multipath components, across time, and the multipath geometry, \ourname{} casts Doppler estimation as a non-linear least-squares problem that is solved through alternating minimization, attaining high-quality estimates for the target's Doppler frequency and the velocity of the moving \ac{isac} device. In contrast with traditional \ac{music}-based processing, \ourname{} handles channel estimates obtained at irregular time intervals. Moreover, it can be seamlessly integrated with measurements from an \ac{imu} sensor, positioned on the moving \ac{isac} device, to provide increased robustness as some of the static paths become temporarily unavailable.   
\ourname{} has been thoroughly assessed via theoretical analysis, numerical simulation, and experiments on moving object and human targets, showing comparable performance to static \ac{isac} devices. 

\appendices

\rev{
\section{} \label{app:opt_proof}
The proof that across the iterations \alg{alg:alt-min} the cost function in \eq{eq:min-prob} is non-increasing proceeds by induction.

$\bullet$ \textit{Base case ($n=0$):} Denote by $\hat{\eta}^{(0)}, \hat{f}_{\rm D,t}^{(0)},(\hat{v}^{\rm tx})^{(0)}$ as the closed form solution of \eq{eq:delta-model}, used as initial values for the unknowns. Using the notation $\mathbf{\Gamma}^{(n)} = \mathbf{\Gamma}(\eta^{(n)})$, we have
\begin{equation*}\label{eq:base-case}
\left\|\bar{\mathbf{\Delta}} - \mathbf{\Gamma}^{(0)}\bm{\nu}^{(0)}\right\|_2^2 \stackrel{(a)}{\geq} \left\|\bar{\mathbf{\Delta}} - \mathbf{\Gamma}^{(0)}\bm{\nu}^{(1)}\right\|_2^2 \stackrel{(b)}{\geq} \left\|\bar{\mathbf{\Delta}} - \mathbf{\Gamma}^{(1)}\bm{\nu}^{(1)}\right\|_2^2,
\end{equation*}
where $(a)$ is because $\hat{f}_{\rm D,t}^{(1)}$,$(\hat{v}^{\rm tx})^{(1)}$ are the optimal solutions of the \ac{ls} problem in \eq{eq:nu-prime-sol} with fixed $\hat{\eta}^{(0)}$, while $(b)$ follows from the definition of grid search in \eq{eq:eta-grid}.
The above chain of inequalities proves that in the first iteration, the cost function in \eq{eq:min-prob} does not decrease.

$\bullet$ \textit{Induction step:} Assume that the cost function does not decrease in the $n$-th step, i.e., $
     \left\|\bar{\mathbf{\Delta}} - \mathbf{\Gamma}^{(n)}\bm{\nu}^{(n)}\right\|_2^2 \geq \left\|\bar{\mathbf{\Delta}} - \mathbf{\Gamma}^{(n+1)}\bm{\nu}^{(n+1)}\right\|_2^2$.
Then, since $\hat{f}_{\rm D,t}^{(n+2)}$,$(\hat{v}^{\rm tx})^{(n+2)}$ are the optimal solutions of \eq{eq:nu-prime-sol} using $\eta^{(n+1)}$ and $\eta^{(n+2)}$ is obtained from the grid search in \eq{eq:eta-grid}, we have
\begin{equation}
    \begin{split}
        \left\|\bar{\mathbf{\Delta}} - \mathbf{\Gamma}^{(n+1)}\bm{\nu}^{(n+1)}\right\|_2^2 \geq & \left\|\bar{\mathbf{\Delta}} - \mathbf{\Gamma}^{(n+1)}\bm{\nu}^{(n+2)}\right\|_2^2 \geq \\\geq & \left\|\bar{\mathbf{\Delta}} - \mathbf{\Gamma}^{(n+2)}\bm{\nu}^{(n+2)}\right\|_2^2
    \end{split}
\end{equation}

$\bullet$ \textit{Conclusion:} Since both the base case and the induction step have been proved by mathematical induction, every iteration of \alg{alg:alt-min} does not increase the cost function. 
}

\vspace{-3mm}
\section{}\label{app:system-sol}

We first observe that \eq{eq:phi-m-norm} does not contain the Doppler frequency of the target. Hence, we use it with $s=1$ to obtain write $v^{\rm tx}$ as a function of $\eta$ (note that a similar expression is obtained using $s=2$)

\begin{equation}\label{eq:vel-func-eta}
    v^{\rm tx} = \frac{\lambda \bar{\Delta}_1}{\cos{(\eta-\alpha_1)}-\cos{\eta}}.
\end{equation}
Then, using the above result into \eq{eq:phi-m-norm} with $s=2$, and solving for $\eta$ we get 
\begin{equation}
    \label{eq:tan-eta}
    \tan \eta = \frac{ \bar{\Delta}_2 (\cos \alpha_1 - 1) - \bar{\Delta}_1 (\cos \alpha_2 - 1)}{\bar{\Delta}_1 \sin{\alpha_2} - \bar{\Delta}_2 \sin{\alpha_1}} \triangleq \xi.
\end{equation}
Inverting the above equation as $\widehat{\eta} = \arctan \xi$ would cause an ambiguity, since the codomain of the $\arctan$ function is $[-\pi/2,\pi/2]$. This would exclude from the solution the (valid) cases where $\eta \in [\pi/2,3\pi/2]$.
To resolve the ambiguity, we start by noticing that 
\begin{equation}\label{eq:deriv1}
    \bar{\Delta}_1 \sin{\alpha_2} - \bar{\Delta}_2 \sin{\alpha_1}  = \frac{v^{\rm tx}}{\lambda} u(\alpha_1, \alpha_2) \cos \eta,
\end{equation}
where $ u(\alpha_1, \alpha_2) = \sin \alpha_1 (1 - \cos \alpha_2) + \sin \alpha_2 (\cos \alpha_1 - 1)$.
\eq{eq:deriv1} is obtained after some manipulations by substituting the expressions of $\bar{\Delta}_1$ and $\bar{\Delta}_2$. The left-hand side of \eq{eq:deriv1} and $u(\alpha_1, \alpha_2)$ can be computed numerically using the phase measurements and the \ac{aod} estimates. Hence, we have that
\begin{equation}
    \mathrm{sign}\left(\cos \eta\right) = \mathrm{sign}\left(\frac{\bar{\Delta}_1 \sin{\alpha_2} - \bar{\Delta}_2 \sin{\alpha_1}}{u(\alpha_1, \alpha_2)} \right),
\end{equation}
since $v^{\rm tx}/\lambda$ is positive by definition. From the sign of $\cos \eta$ we can understand whether $\eta \in [-\pi/2,\pi/2]$ (if $\cos \eta$ is positive) or $\eta \in [\pi/2,3\pi/2]$ (if $\cos \eta$ is negative). This is then used to disambiguate \eq{eq:tan-eta} as follows, finding an initialization of~$\eta$
\begin{equation}\label{eq:eta-hat}
    \widehat{\eta}=
    \begin{cases}
   \arctan \xi
    \,\, &\mathrm{if} \, \cos \eta \geq 0,\\
    \pi + \arctan \xi  &\mathrm{if} \, \cos \eta < 0.
    \end{cases}
\end{equation}

\section{} \label{app:app-b}
The noise variance on the phase of path $i$ is~\cite{ku1966notes}
\begin{align}
     \sigma^2_{\phi, i} &= \left(\frac{\partial \angle{h_i[k, l_i]}}{\partial \Re\{h_i\}}\right)^2\frac{\sigma_h^2}{ 2} + \left(\frac{\partial \angle{h_i[k, l_i]}}{\partial \Im\{h_i\}}\right)^2\frac{\sigma_h^2}{ 2}
\end{align}
which gives $\sigma^2_{\phi, i}= \sigma_w^2/ (2 G|A_i|^2)$, meaning that the noise variance on the phase of a path is inversely proportional to the squared magnitude of that path.

\section{}\label{app:fim-crlb}

\rev{
Term $\zeta_{{\rm t}, 1, 2}$ in the \ac{crb} expression in \eq{eq:crlb} is given by
\begin{equation}\label{eq:zeta}
    \zeta_{{\rm t}, 1, 2} = -\rho_{{\rm t}, 1, 2} + \sum_{i \in \{{\rm t}, 1, 2\}}\kappa_i \left[3 + \gamma\left(\{{\rm t}, 1, 2\}\setminus i \right)\right],
\end{equation}
where ``$\setminus$" denotes the set subtraction operation, the expression of $\gamma$, function of a set of two indices $\{i,\ell\}$ is
\begin{equation}\label{eq:small_gamma}
        \gamma(\{i, \ell\}) = 2 \left[\sin \alpha_i - \sin (\alpha_i-\alpha_\ell) - \sin \alpha_\ell\right]^2,
\end{equation}
and $\rho_{{\rm t}, 1, 2}$ is given in \eq{eq:rho}.
}

\begin{figure*}[!t]
\normalsize
\setcounter{mytempeqncnt}{\value{equation}}
\vspace{-4mm}
\rev{\begin{align}\label{eq:rho}
\rho_{{\rm t}, 1, 2} = \frac{16}{|A_{\rm LoS}|^2}\Bigg[
\sin\!\left(\frac{\alpha_1}{2}\right)
\sin\!\left(\frac{\alpha_2}{2}\right)
\sin(\alpha_{\rm t})
\Bigg(
\sin\!\left(\frac{\alpha_1 + \alpha_2 - 3\alpha_{\rm t}}{2}\right)
- \sin\!\left(\frac{\alpha_1 - \alpha_2 - \alpha_{\rm t}}{2}\right)
- \sin\!\left(\frac{3\alpha_1 - \alpha_2 - \alpha_{\rm t}}{2}\right)\\\nonumber
 + \sin\!\left(\frac{\alpha_1 + \alpha_2 - \alpha_{\rm t}}{2}\right)
+ 2 \cos\!\left(\frac{\alpha_2}{2}\right)
  \sin\!\left(\frac{\alpha_1 - 2\alpha_2 + \alpha_{\rm t}}{2}\right)
\Bigg)
\Bigg]
\end{align}}
\vspace{-4mm}
\hrulefill
\end{figure*}

\bibliography{IEEEabrv.bib, references.bib}

@STRING{IEEE_J_VT         = "{IEEE} Trans. Veh. Technol."}

@STRING{IEEE_J_STSP       = "{IEEE} J. Sel. Topics Signal Process."}

@STRING{IEEE_J_SP         = "{IEEE} Trans. Signal Process."}

@STRING{IEEE_J_WCOM       = "{IEEE} Trans. Wireless Commun."}

@STRING{IEEE_J_WCOML      = "{IEEE} Wireless Commun. Lett."}

@STRING{IEEE_J_MC         = "{IEEE} Trans. Mobile Comput."}

@STRING{IEEE_J_GRS        = "{IEEE} Trans. Geosci. Remote Sens."}

@STRING{IEEE_J_SENSOR     = "{IEEE} Sensors J."}

@STRING{IEEE_O_ACC        = "{IEEE} Access"}

@STRING{IEEE_M_COM        = "{IEEE} Commun. Mag."}

@STRING{IEEE_O_CSTO       = "{IEEE} Commun. Surveys Tuts."}

@STRING{IEEE_M_SP         = "{IEEE} Signal Process. Mag."}

@STRING{IEEE_M_WC         = "{IEEE} Wireless Commun."}

@IEEEtranBSTCTL{IEEEexample:BSTcontrol,
CTLuse_forced_etal       = "yes",
CTLmax_names_forced_etal = "6",
CTLnames_show_etal       = "1" }

@ARTICLE{802.11ay,
author={{IEEE 802.11 working group}},
journal={IEEE P802.11ay/D3.0},
title={{Part 11: Wireless LAN MAC and PHY Specifications-Amendment: Enhanced Throughput for Operation in License-Exempt Bands Above 45 GHz}},
year={2019}
}

@inproceedings{pegoraro2022sparcs,
  title={{SPARCS: A Sparse Recovery Approach for Integrated Communication and Human Sensing in mmWave Systems}},
  author={Pegoraro, Jacopo and Lacruz, Jesus O and Rossi, Michele and Widmer, Joerg},
  booktitle={21st ACM/IEEE Int. Conf. on Information Processing in Sensor Networks (IPSN)},
  month={May},
  address={Milan, Italy},
  year={2022}
}

@article{zhang2021overview,
  title={An overview of signal processing techniques for joint communication and radar sensing},
  author={Zhang, J Andrew and Liu, Fan and Masouros, Christos and Heath, Robert W and Feng, Zhiyong and Zheng, Le and Petropulu, Athina},
  journal=IEEE_J_STSP,
  volume={15},
  number={6},
  pages={1295--1315},
  year={2021},
  publisher={IEEE}
}

@book{richards2010principles,
  title={Principles of modern radar},
  author={Richards, Mark A and Scheer, Jim and Holm, William A and Melvin, William L},
  year={2010},
  publisher={Scitech Publishing Inc.},
  address={ Raleigh, NC, USA}
}

@article{meneghello2022sharp,
  title={{SHARP: Environment and Person Independent Activity Recognition with Commodity IEEE 802.11 Access Points}},
  author={Meneghello, Francesca and Garlisi, Domenico and Dal Fabbro, Nicol{\`o} and Tinnirello, Ilenia and Rossi, Michele},
  journal=IEEE_J_MC,
pages={6160--6175},
volume={22},
number={10},
  year={2023},
  publisher={IEEE}
}

@ARTICLE{zhao2024multiple,
	author={Zhao, Jingbo and Lu, Zhaoming and Zhang, J. Andrew and Dong, Shixu and Zhou, Shuang},
	journal=IEEE_J_VT, 
	title={{Multiple-Target Doppler Frequency Estimation in ISAC with Clock Asynchronism}}, 
	year={2024},
	volume={73},
	number={1},
	pages={1382-1387},
	}

@ARTICLE{zhang2022integration,
	author={Zhang, J. Andrew and Wu, Kai and Huang, Xiaojing and Guo, Y. Jay and Zhang, Daqing and Heath, Robert W.},
	journal=IEEE_M_COM, 
	title={Integration of Radar Sensing into Communications with Asynchronous Transceivers}, 
	year={2022},
	volume={60},
	number={11},
	pages={106-112},
	}

@article{zhang2021enabling,
  title={Enabling joint communication and radar sensing in mobile networks—A survey},
  author={Zhang, J Andrew and Rahman, Md Lushanur and Wu, Kai and Huang, Xiaojing and Guo, Y Jay and Chen, Shanzhi and Yuan, Jinhong},
  journal=IEEE_O_CSTO,
  volume={24},
  number={1},
  pages={306--345},
  year={2021},
  publisher={IEEE}
}

@article{liu2023seventy,
  title={{Seventy years of radar and communications: The road from separation to integration}},
  author={Liu, Fan and Zheng, Le and Cui, Yuanhao and Masouros, Christos and Petropulu, Athina P and Griffiths, Hugh and Eldar, Yonina C},
  journal=IEEE_M_SP,
  volume={40},
  number={5},
  pages={106--121},
  year={2023},
  publisher={IEEE}
}

@ARTICLE{singh2021multi,
  author={Singh, Anuradha and Rehman, Saeed Ur and Yongchareon, Sira and Chong, Peter Han Joo},
  journal=IEEE_J_SENSOR, 
  title={{Multi-Resident Non-Contact Vital Sign Monitoring Using Radar: A Review}}, 
  year={2021},
  volume={21},
  number={4},
  pages={4061-4084}}

@ARTICLE{3GPP_cfo,
author={{3GPP}},
title={{Technical Specification Group Radio Access Network; User
Equipment (UE) radio transmission and reception; Part 2: Range 2
Standalone (Release 16)}},
journal={3rd Generation Partnership Project (3GPP),
techreport TR 38.101-2},
month={Mar.},
year={2020, version 16.3.0}
}

@ARTICLE{wu2024sensing,
  author={Wu, Kai and Pegoraro, Jacopo and Meneghello, Francesca and Zhang, J. Andrew and Lacruz, Jesus O. and Widmer, Joerg and Restuccia, Francesco and Rossi, Michele and Huang, Xiaojing and Zhang, Daqing and Caire, Giuseppe and Guo, Y. Jay},
  journal=IEEE_M_SP, 
  title={{Sensing in Bistatic ISAC Systems With Clock Asynchronism: A signal processing perspective}}, 
  year={2024},
  volume={41},
  number={5},
  pages={31-43}}

@inproceedings{zhang2022mobi2sense,
author = {Zhang, Fusang and Xiong, Jie and Chang, Zhaoxin and Ma, Junqi and Zhang, Daqing},
title = {{Mobi2Sense: empowering wireless sensing with mobility}},
publisher = {{ACM}},
address = {New York, NY, USA},
booktitle = {Proc. of the 28th Annual Int. Conf. on Mobile Computing And Networking},
pages = {268–281},
numpages = {14},
location = {Sydney, NSW, Australia},
series = {MobiCom '22},
year = {2022}
}

@ARTICLE{pegoraro2024jump,
  author={Pegoraro, Jacopo and Lacruz, Jesus O. and Azzino, Tommy and Mezzavilla, Marco and Rossi, Michele and Widmer, Joerg and Rangan, Sundeep},
  journal=IEEE_J_WCOM, 
  title={{JUMP: Joint Communication and Sensing With Unsynchronized Transceivers Made Practical}}, 
  year={2024},
  volume={23},
  number={8},
  pages={9759-9775},
 }

@book{willis2005bistatic,
  title={Bistatic radar},
  author={Willis, Nicholas J},
  volume={2},
  year={2005},
  publisher={SciTech Publishing}
}

@article{ku1966notes,
  title={{Notes on the use of propagation of error formulas}},
  author={Ku, Harry H},
  journal={Journal of Research of the National Bureau of Standards},
  volume={70},
  number={4},
pages={263--273},
  year={1966}
}

@article{kay1993statistical,
  title={Statistical signal processing: estimation theory},
  author={Kay, Steven M},
  journal={Prentice Hall},
  volume={1},
  pages={Chapter--3},
  year={1993}
}

@ARTICLE{ni2021uplink,
  author={Ni, Zhitong and Zhang, J. Andrew and Huang, Xiaojing and Yang, Kai and Yuan, Jinhong},
  journal=IEEE_J_SP, 
  title={{Uplink Sensing in Perceptive Mobile Networks With Asynchronous Transceivers}}, 
  year={2021},
  volume={69},
  number={},
  pages={1287-1300},
 }

@ARTICLE{li2022CSI,
  author={Li, Xinyu and Zhang, J. Andrew and Wu, Kai and Cui, Yuanhao and Jing, Xiaojun},
  journal=IEEE_J_SENSOR, 
  title={{CSI-Ratio-Based Doppler Frequency Estimation in Integrated Sensing and Communications}}, 
  year={2022},
  volume={22},
  number={21},
  pages={20886-20895},
 }

@article{zhang2020exploring,
  title={{Exploring LoRa for Long-range Through-wall Sensing}},
  author={Zhang, Fusang and Chang, Zhaoxin and Niu, Kai and Xiong, Jie and Jin, Beihong and Lv, Qin and Zhang, Daqing},
  journal={Proc. of the ACM on Interactive, Mobile, Wearable and Ubiquitous Technologies},
  volume={4},
  number={2},
  pages={1--27},
  year={2020},
  publisher={ACM New York, NY, USA}
}

@article{zeng2020multisense,
  title={{MultiSense: Enabling Multi-person Respiration Sensing with Commodity WiFi}},
  author={Zeng, Youwei and Wu, Dan and Xiong, Jie and Liu, Jinyi and Liu, Zhaopeng and Zhang, Daqing},
  journal={Proc. of the ACM on Interactive, Mobile, Wearable and Ubiquitous Technologies},
  volume={4},
  number={3},
  pages={1--29},
  year={2020},
  publisher={ACM New York, NY, USA}
}

@INPROCEEDINGS{yu2024efficient,
  author={Yu, Puxi and Ma, Dingyou and Zhang, Qixun and Feng, Zhiyong},
  booktitle={2024 2nd Int. Conf. On Mobile Internet, Cloud Computing and Information Security (MICCIS)}, 
  title={{Efficient Clock Offset Calibration Method for Uplink Sensing in Asynchronous ISAC System}}, 
  year={2024},
  volume={},
  number={},
  pages={43-47},
}

@INPROCEEDINGS{kanhere2021target,
  author={Kanhere, Ojas and Goyal, Sanjay and Beluri, Mihaela and Rappaport, Theodore S.},
  booktitle={2021 IEEE 93rd Vehicular Technology Conf. (VTC)}, 
  title={{Target Localization using Bistatic and Multistatic Radar with 5G NR Waveform}}, 
  year={2021},
  volume={},
  number={},
  pages={1-7}}

@ARTICLE{gao2023integrated,
  author={Gao, Zhen and Wan, Ziwei and Zheng, Dezhi and Tan, Shufeng and Masouros, Christos and Ng, Derrick Wing Kwan and Chen, Sheng},
  journal=IEEE_J_WCOM, 
  title={{Integrated Sensing and Communication With mmWave Massive MIMO: A Compressed Sampling Perspective}}, 
  year={2023},
  volume={22},
  number={3},
  pages={1745-1762},
 }

@INPROCEEDINGS{pucci2022performance,
  author={Pucci, Lorenzo and Matricardi, Elisabetta and Paolini, Enrico and Xu, Wen and Giorgetti, Andrea},
  booktitle={2022 IEEE Int. Conf. on Communications Workshops (ICC Workshops)}, 
  title={{Performance Analysis of a Bistatic Joint Sensing and Communication System}}, 
  year={2022},
  volume={},
  number={},
  pages={73-78},
}

@inproceedings{lacruz2021realtime,
title = {{A Real-Time Experimentation Platform for sub-6 GHz and Millimeter-Wave MIMO Systems}},
author = {Lacruz, Jesus O. and Ortiz, Rafael Ruiz and Widmer, Joerg},
booktitle={Proc. of the 19th ACM Annual Int. Conf. on Mobile Systems, Applications, and Services (MobiSys)},
year = {2021},
address = {Virtual Event, Wisconsin, USA},
pages = {427–439},
numpages = {13},
}

@ARTICLE{pegoraro2024rapid,
  author={Pegoraro, Jacopo and Lacruz, Jesus O. and Meneghello, Francesca and Bashirov, Enver and Rossi, Michele and Widmer, Joerg},
  journal=IEEE_J_MC, 
  title={{RAPID: Retrofitting IEEE 802.11ay Access Points for Indoor Human Detection and Sensing}}, 
  year={2024},
  volume={23},
  number={5},
  pages={4501-4519},
 }

@article{barneto2021full,
  title={{Full Duplex Radio/Radar Technology: The Enabler for Advanced Joint Communication and Sensing}},
  author={Barneto, Carlos Baquero and Liyanaarachchi, Sahan Damith and Heino, Mikko and Riihonen, Taneli and Valkama, Mikko},
  journal=IEEE_M_WC,
  volume={28},
  number={1},
  pages={82--88},
  year={2021},
  publisher={IEEE}
}

@inproceedings{heino2021design,
  title={{Design of Phased Array Architectures for Full-Duplex Joint Communications and Sensing}},
  author={Heino, Mikko and Barneto, Carlos Baquero and Riihonen, Taneli and Valkama, Mikko},
  booktitle={2021 15th European Conf. on Antennas and Propagation (EuCAP)},
  pages={1--5},
  year={2021},
  organization={IEEE}
}

@ARTICLE{ventura2024bistatic,
  author={Ventura, Gianmaria and Bhalli, Zaman and Rossi, Michele and Pegoraro, Jacopo},
  journal=IEEE_J_WCOML, 
  title={{Bistatic Doppler Frequency Estimation With Asynchronous Moving Devices for Integrated Sensing and Communications}}, 
  year={2024},
  volume={13},
  number={10},
  pages={2872-2876},
 
}

@ARTICLE{garcia2021ieee,
  author={Garcia-Rodriguez, Adrian and López-Pérez, David and Galati-Giordano, Lorenzo and Geraci, Giovanni},
  journal=IEEE_M_COM, 
  title={{IEEE 802.11be: Wi-Fi 7 Strikes Back}}, 
  year={2021},
  volume={59},
  number={4},
  pages={102-108},
 }

@article{liu2023towards,
author = {Liu, Jinyi and Li, Wenwei and Gu, Tao and Gao, Ruiyang and Chen, Bin and Zhang, Fusang and Wu, Dan and Zhang, Daqing},
title = {{Towards a Dynamic Fresnel Zone Model to WiFi-based Human Activity Recognition}},
year = {2023},
issue_date = {June 2023},
publisher = {{ACM}},
address = {New York, NY, USA},
volume = {7},
number = {2},
journal = {Proc. ACM Interact. Mob. Wearable Ubiquitous Technol.},
month = jun,
articleno = {65},
numpages = {24}
}

@inproceedings{chang2024msense,
author = {Chang, Zhaoxin and Zhang, Fusang and Xiong, Jie and Chen, Weiyan and Zhang, Daqing},
title = {{MSense: Boosting Wireless Sensing Capability Under Motion Interference}},
year = {2024},
isbn = {9798400704895},
publisher = {{ACM}},
address = {New York, NY, USA},
booktitle = {Proc. of the 30th Annual Int. Conf. on Mobile Computing and Networking},
pages = {108–123},
numpages = {16},
location = {Washington D.C., DC, USA},
series = {ACM MobiCom '24}
}

@article{vandersmissen2018indoor,
  title={{Indoor person identification using a low-power FMCW radar}},
  author={Vandersmissen, Baptist and Knudde, Nicolas and Jalalvand, Azarakhsh and Couckuyt, Ivo and Bourdoux, Andre and De Neve, Wesley and Dhaene, Tom},
  journal=IEEE_J_GRS,
  volume={56},
  number={7},
  pages={3941--3952},
  year={2018},
  publisher={IEEE}
}

@ARTICLE{chu2023deep,
  author={Chu, Yi and Cumanan, Kanapathippillai and Sankarpandi, Sathish K. and Smith, Stephen and Dobre, Octavia A.},
  journal=IEEE_O_ACC, 
  title={{Deep Learning-Based Fall Detection Using WiFi Channel State Information}}, 
  year={2023},
  volume={11},
  pages={83763-83780},
}

@ARTICLE{fang2021rotor,
  author={Fang, Xin and Xiao, Guoqing},
  journal=IEEE_J_SENSOR, 
  title={{Rotor Blades Micro-Doppler Feature Analysis and Extraction of Small Unmanned Rotorcraft}}, 
  year={2021},
  volume={21},
  number={3},
  pages={3592-3601},
 }

@ARTICLE{garcia2023scalable,
  author={Garcia, Dolores and Lacruz, Jesus O. and Mateo, Pablo Jiménez and Palacios, Joan and Ruiz, Rafael and Widmer, Joerg},
  journal=IEEE_J_MC, 
  title={{Scalable Phase-Coherent Beam-Training for Dense Millimeter-Wave Networks}}, 
  year={2023},
  volume={22},
  number={4},
  pages={2353-2369}
 }

@ARTICLE{wang2024windowing, author={Wang, Xiao-Yang and Yang, Shaoshi and Zhai, Hou-Yu and Masouros, Christos and Andrew Zhang, J.}, 
journal={IEEE Transactions on Communications}, 
title={{Windowing Optimization for Fingerprint-Spectrum-Based Passive Sensing in Perceptive Mobile Networks}}, 
year={2025}, 
volume={73}, 
number={2}, 
pages={1367-1382}, 
doi={10.1109/TCOMM.2024.3446610}}

@ARTICLE{wang2024clutter,
  author={Wang, Xiao-Yang and Yang, Shaoshi and Zhang, Jianhua and Masouros, Christos and Zhang, Ping},
  journal={IEEE Journal on Selected Areas in Communications}, 
  title={{Clutter Suppression, Time-Frequency Synchronization, and Sensing Parameter Association in Asynchronous Perceptive Vehicular Networks}}, 
  year={2024},
  volume={42},
  number={10},
  pages={2719-2736},
  doi={10.1109/JSAC.2024.3414581}}

@ARTICLE{schmidt1986multiple,
  author={Schmidt, R.},
  journal={IEEE Transactions on Antennas and Propagation}, 
  title={Multiple emitter location and signal parameter estimation}, 
  year={1986},
  volume={34},
  number={3},
  pages={276-280},
  doi={10.1109/TAP.1986.1143830}}

@ARTICLE{ma2021target,
  author={Ma, Hui and Antoniou, Michail and Stove, Andrew G. and Cherniakov, Mikhail},
  journal={IEEE Transactions on Aerospace and Electronic Systems}, 
  title={{Target Kinematic State Estimation With Passive Multistatic Radar}}, 
  year={2021},
  volume={57},
  number={4},
  pages={2121-2134},
  doi={10.1109/TAES.2021.3069283}}

@INPROCEEDINGS{luo2020analysis,
  author={Luo, Zhongtao and Zhan, Yanmei and Jonckheere, Edmond},
  booktitle={2020 IEEE/CIC International Conference on Communications in China (ICCC)}, 
  title={{Analysis on Functions and Characteristics of the Rician Phase Distribution}}, 
  year={2020},
  volume={},
  number={},
  pages={306-311},
  doi={10.1109/ICCC49849.2020.9238805}}

@article{fischler1981random,
  title={Random sample consensus: a paradigm for model fitting with applications to image analysis and automated cartography},
  author={Fischler, Martin A and Bolles, Robert C},
  journal={Communications of the ACM},
  volume={24},
  number={6},
  pages={381--395},
  year={1981},
  publisher={ACM New York, NY, USA}
}

@ARTICLE{yang2019inexact,
  author={Yang, Yang and Pesavento, Marius and Luo, Zhi-Quan and Ottersten, Bj\''{o}rn},
  journal={IEEE Transactions on Signal Processing}, 
  title={{Inexact Block Coordinate Descent Algorithms for Nonsmooth Nonconvex Optimization}}, 
  year={2019},
  volume={68},
  number={},
  pages={ 947 - 961},
  doi={10.1109/TSP.2019.2959240}}
\bibliographystyle{IEEEtran}
\end{document}